\documentclass{emulateapj}

\slugcomment{Accepted by ApJ}
\shortauthors{Kuraszkiewicz et al.}
\shorttitle{SEDs of Red 2MASS AGN.}

\def\etal   {{\it et~al.}}

\def\nh {${\rm N_H}$}
\def\lax    {${_<\atop^{\sim}}$}
\def\gax    {${_>\atop^{\sim}}$}
\def\chandra {{\it Chandra}}
\def\jmk {{J$-$K$_S$}}
\def\O3 {{[O\,{\small III}]}}
\def\O2 {{[O\,{\small II}]}}
\newcommand{\kms}{\ifmmode~{\rm km~s}^{-1}\else ~km~s$^{-1}~$\fi}
\def\plotfiddle#1#2#3#4#5#6#7{\centering \leavevmode
    \vbox to#2{\rule{0pt}{#2}}
    \includegraphics{#1}}

\begin{document}

\title{The Spectral Energy Distributions of Red 2MASS AGN.}
\author{Joanna Kuraszkiewicz\altaffilmark{1}, Belinda J. Wilkes\altaffilmark{1}, Gary Schmidt\altaffilmark{2},
Himel Ghosh\altaffilmark{3}, Paul S. Smith\altaffilmark{2}, Roc Cutri\altaffilmark{4}, Dean Hines\altaffilmark{5}, Eric
M. Huff\altaffilmark{2,6}, Jonathan  C. McDowell\altaffilmark{1} \& Brant Nelson\altaffilmark{4}}  
\altaffiltext{1}{Harvard-Smithsonian Center for Astrophysics, Cambridge, MA 02138}
\altaffiltext{2}{Steward Observatory, University of Arizona, Tucson, AZ 85721}
\altaffiltext{3}{Department of Astronomy, Ohio State University, Columbus, OH
  43210-1173} 
\altaffiltext{4}{IPAC, Caltech, MS 100-22, Pasadena, CA 91125}
\altaffiltext{5}{Space Science Institute, 4750 Walnut Street, Suite 205,
  Boulder, CO 80301}
\altaffiltext{6}{current address:  Department of Astronomy, University of
  California, Berkeley, Berkeley, CA 94720}

\begin{abstract}

We present infrared (IR) to X--ray spectral energy distributions
(SEDs) for 44 red AGN selected from the 2MASS survey on the basis of
their red \jmk\ color ($>2$~mag.) and later observed by \chandra . In
comparison with optically-, radio-, and X-ray selected AGN, their
median SEDs are red in the optical and near-IR with little/no blue
bump. Comparison of the various broad-band luminosity ratios show that
the main differences lie at the blue end of the optical and in the
near-IR to far-IR ratios (when available), with the red 2MASS AGN
being redder than the other samples. It thus seems that near-IR color
selection isolates the reddest subset of AGN that can be classified
optically. The shape of the SEDs is generally consistent with modest
absorption by gas (in the X-ray) and dust (in the optical-IR), as
demonstrated by comparing the optical and near-IR colors with a
reddened median SED and observed optical+near--IR to intrinsic X-ray
ratios. The levels of obscuration, estimated from X-rays, far-IR and
our detailed optical/near-IR color modeling are all consistent
implying \nh\ $\leq $few$\times 10^{22}$~cm$^{-2}$.  We present SED
models that show how the AGN optical/near-IR colors change due to
differing amounts of reddening, AGN to host galaxy ratio, redshift and
scattered light emission and apply them to the sources in the
sample. We find that the 2MASS AGN optical color, B$-$R, and to a
lesser extent the near-IR color, \jmk , are strongly affected by
reddening, host galaxy emission, redshift, and in few, highly
polarized objects, also by scattered AGN light ($<$2\% of intrinsic
AGN light in R band is scattered; this contribution becomes
significant as the direct AGN light is absorbed).  The lack of low
equivalent widths in the distribution of the [O\,{\small
III}]\,$\lambda5007$ emission line implies a predominance of inclined
objects in the red 2MASS sample.  The obscuration/inclination of the
AGN allows us to see weaker emission components which are generally
swamped by the AGN.

\end{abstract}


\section{Introduction}

The orientation dependence of the appearance of an AGN has been
well-known for many years. Radio observations clearly demonstrate
extended, often very large structures that appear different as a
function of their orientation (Barthel 1989) to our line of
sight. Optical polarization reveals the presence of AGN in scattered
light that are not visible directly due to high obscuration (Antonucci
\& Miller 1985). These results led to the development of unification
models which relate observationally different AGN and radio galaxies
to one another via viewing angle (Antonucci 1993). It is clear that
obtaining an unbiased view of the AGN population is a challenge and,
since most initial surveys were carried out in optical wavebands and
based on a search for blue sources, obscured and/or edge-on AGN have
been missed.  Even within the optically-selected subset of AGN,
radio$-$X-ray spectral energy distributions (SEDs) show a wide variety
of properties that affect selection (Elvis \etal\ 1994).

Evidence for a large, obscured subset of the AGN population which is
mostly ``missing" was demonstrated by modeling of the Cosmic X-ray
Background (CXRB, Comastri \etal\ 1995, Gilli, Risaliti \& Salvati
1999).  Initial reports of a red, obscured population missed by
traditional optical surveys was based on radio-selected AGN (Webster
\etal\ 1995, Kim \& Elvis 1999). Detailed study has shown that these
can be explained in terms of an additional, red non-thermal
synchrotron continuum component, linked to the radio emission (Francis
\etal\ 2001, but see Richards \etal\ 2003). The importance of this
particular set of red AGN to the population seems not to be large
(Boyle \& DiMatteo 1995).  Much effort has been made both to reduce
selection bias in sample selection and to understand the angular
dependence and thus the relation between those selected in different
wavebands.  The Sloan Digital Sky Survey (SDSS), while still an
optical survey, has developed sophisticated color selection techniques
which successfully reveal AGN with a variety of non-stellar colors,
many much redder than found in previous optically-selected samples of
AGN (Richards \etal\ 2003). The Two Micron All-Sky Survey (2MASS)
revealed a significant subset of predominantly broad-lined AGN through
their red near-infrared (NIR) colors. Their number density rivals that
of optically-selected AGN at low redshifts (Cutri \etal\ 2002).  In
addition, the typically high optical polarization (Smith \etal\ 2002,
2003) of the 2MASS AGN suggests substantial obscuration around the
nuclear energy source. \chandra\ observations show weak, hard X-ray
emission compared with normal, low-redshift AGN (Wilkes \etal\ 2002).
These properties suggest that they are AGN obscured at a level
intermediate between the well-studied, unobscured, broad-lined AGN
revealed by optical surveys (face-on in the Unified Scheme, Antonucci
\& Miller 1985) and the obscured narrow-lined AGN believed to be
viewed edge-on.

The advent of the Great Observatories: \chandra\ and {\it Spitzer} over the
past 5$-$10 years has facilitated many deeper, multi-wavelength
surveys that have extended sufficiently deeply into the population to
reveal large numbers of previously unknown AGN with properties very
different from the traditional broad emission line AGN (Alexander
\etal\ 2003, Polletta \etal\ 2006, 2007).  These new AGN also extend
the observed properties of AGN SEDs over a much wider range.  With a
large number of new candidate AGN covering a wide range of properties,
the quest to understand the AGN population and the nature and variety
of the structure of their central regions is only now beginning in
earnest.  The most recent version of CXRB models, based on results
from the current deep X-ray surveys (Gilli, Comastri \& Hasinger 2007)
includes a population of moderately obscured AGN, which may be
explained by a combination of the new AGN candidates, but still calls
for a significant highly-obscured population that has not yet been
found, although recent {\it Spitzer} and \chandra\ results are suggestive
(Daddi \etal\ 2008, Fiore \etal\ 2008).
                   
The importance of red AGN to the total population thus remains
uncertain.  At low redshift, the 2MASS red AGN may account for as much
as 20\% of the AGN population (Francis, Nelson \& Cutri 2004), and
likely represent a significant subset of the moderately-obscured AGN
required by current Cosmic X-ray Background (CXRB) models.  Many of
these red AGN, especially those with lower obscuration ($A_V$ \lax\
few) should be picked up by the SDSS. However, those AGN with higher
obscuration whose optical colors are dominated by the host galaxy will
not, since their colors will lie too close to the stellar locus to be
classified as a quasar\footnote{We have cross-correlated the 44 red
2MASS AGN in our sample with the SDSS dr6 database and found 23 2MASS
AGN in the area covered by the SDSS. Using the ugri color-color, low-z
QSO selection criteria from Richards \etal\ (2002) we find that 6 of
these sources would have been excluded: 3 (13\%) blue sources
(NLS1/BALQSOs with optical colors dominated by AGN emission) due to
$(u-g) < 0.5$ criterion and 3 (13\%) red sources (with optical colors
dominated by host galaxy and $A_V > 10$) whose optical colors fall too
close (within the $2\sigma-4\sigma$ error) to the stellar
locus. However the red 2MASS sample used here is not complete and these
conclusions should be treated with caution.}. The
high-redshift population corresponding the red 2MASS sources is not yet
known due to the combination of the bright magnitude limit of the
2MASS survey and the lower efficiency of the near-IR selection as the
optical emission shifts into the observed waveband.  They most likely
overlap with higher-redshift populations of red AGN being found in the
longer-wavelength {\it Spitzer} surveys (Treister \etal\ 2006, Lacy
\etal\ 2007).

As transition objects between unobscured, face-on and Compton-thick,
edge-on sources, the 2MASS red AGN provide a unique view of the AGN
central regions. The partial obscuration of the bright, direct AGN
light facilitates study of weaker components (Pounds, Wilkes \& Page
2005, Wilkes \etal\ 2008) and thus of the complex structure of
material close to the nucleus that not only re-processes the light,
but also may be integral to fueling the AGN itself, as well as the
powerful radio jets and outflowing material we observe on a wide
variety of spatial scales.

We have embarked on a multi-wavelength study of a bright subset of the
2MASS red AGN that have been observed by \chandra . In this paper, we
present our X-ray and optical data, collated with data from the
literature to generate IR-X-ray SEDs. The properties of the SEDs are
presented including previously unpublished optical spectra,
polarimetry and spectrophotometry and measurements of the continuum
and emission line properties.  These properties are compared with
those of more traditional AGN to identify similarities and differences
and therefore probe the structure and properties of the AGN.  We
conclude that, orientation-dependent obscuration, host galaxy
properties and scattering effects are factors in determining the SED
properties of these red AGN.  However, as we show in a companion paper
(Kuraszkiewicz \etal\ 2008), where we perform Principal Component
Analysis (PCA) on the SED and emission line properties, the
$L/L_{Edd}$ ratio is a dominant factor.

\section{The Sample}

The Two Micron All Sky Survey (2MASS; Skrutskie \etal\ 2006) is
yielding a catalog of near IR-selected AGN (Cutri \etal\ 2002) larger
and deeper than those discovered by IRAS (Soifer \etal\ 1984).
Spectroscopic follow-up of red candidates, selected to have \jmk$> 2$
from the high galactic latitude 2MASS Point Source Catalog, reveals
that $\sim$75\% are previously-unidentified emission-line AGN, with
$\sim$85\% showing broad optical emission lines (Type 1-1.9:
Seyfert~1, intermediate and QSO), and the remainder being narrow-line
objects (Type~2: Seyfert~2, QSO~2, and LINER; Cutri \etal\ 2002).
They span a redshift range $0.1<z<2.3$ with median $\sim$0.25.  The
inferred surface density is $\sim$0.5~deg$^{-2}$ brighter than
$K_s<14.5$~mag., higher than that of optically selected AGN at the
same IR magnitudes and indicating that 2MASS may reveal $>$25,000 such
objects over the sky (Cutri \etal , in preparation). Red \jmk\
selection is inhomogeneous with redshift, since the \jmk\ color is a
near-IR spectral index for $z<0.25$, an IR--to--optical index for
$0.25<z<1.2$, and optical spectral index for $z>1.2$.  Hence reddened
AGN (that show no 1~$\mu$m inflection) will be picked up by the red
\jmk\ selection at all redshifts, while the ``normal'' blue AGN, with
red near-IR colors, will be only picked up at low redshift (see also
Fig.~5 in Barkhouse \& Hall 2001, who find a large number of blue AGN
with \jmk\ $>2$ at $z<0.5$). In this paper we study a well-defined,
flux-limited, color-selected subset of 44 red 2MASS AGN (for a list of
objects see Table~1) selected to
have B$-$K$_S>4.3$ and K$_S<13.8$. This subsample is
representative\footnote{However, sources that fulfill the color
selection but are not detected at J are not included in the sample.}
of the low redshift red AGN population, with $0 < z < 0.37$, a full
range of spectral types (7 Type~1, 11 Type~2, and 26 intermediate type
sources), a wide range of observed $K_S$-band-to-X-ray(1keV) slopes:
1.1\lax $\alpha_{KX}$\lax 2, and a broad range of observed optical
polarization fraction at R band $0 < P(\%) < 13$.  This subset was
observed by \chandra\ and was shown to be relatively X-ray faint and
hard in comparison with optically selected broad-line AGN (Wilkes
\etal\ 2002).  Assuming the spectral hardness is due to absorption,
deduced equivalent hydrogen column densities are at the level of log
\nh\ $\sim 21-$23 and absorption-corrected X-ray fluxes are at the low
end of the expected range based on the assumption that the K magnitude
is intrinsic (Wilkes \etal\ 2002), even lower if the K magnitude is
also affected by absorption.

\section{Multi-wavelength observations}

\subsection{\chandra\  data}

\chandra\ ACIS-I/ACIS-S observations have been obtained for all 44
2MASS AGN in our sample.  The observations were designed to detect
each AGN based on its K$_S$ magnitude combined with the lowest known
X-ray to K flux ratio for an AGN at the time $\alpha_{KX}$=2 (MKN 231,
Turner 1999).  A wide range of net counts was found, from 3 sources
with few ($\sim$5) counts to several with $>$200 counts, implying a
range of $\times 100$ in observed X-ray to K flux across the sample
(Wilkes \etal\ 2002).

Spectral fits were made to the \chandra\ X-ray data to provide the
best estimate of the X-ray fluxes. For the higher count sources (\gax
80, referred to as ``{\it C}'' fits), a simple power-law plus
rest-frame absorption was fitted with both parameters free. For
sources with counts between $\sim 30$-$80$ the power-law was fixed to
$\Gamma=2$ but the \nh\ remained free (``{\it B}'' fits).  For the
lowest count sources, \lax 30 counts (``{\it A}'' fits), fits were
made using a power-law with $\Gamma = 2$ and \nh\ $=
7.6\times10^{21}$~cm$^{-2}$ (the median absorption from the {\it C}
fits). In a few cases where the {\it B} fits did not converge, {\it A}
fits were made despite higher counts. This fitting is described and
presented in more detail in a companion paper (Wilkes \etal\ 2008, in
preparation), where we present the X-ray fluxes used in the SEDs. When
available, we also included in the SEDs the X-ray fluxes from the WGA
Catalog of ROSAT point sources (White, Giommi \& Angelini 1995).

\subsection{HST Spectroscopy}

The ultraviolet (UV) spectrophotometry for 9 objects was obtained
using the Space Telescope Imaging Spectrograph (STIS) aboard the
Hubble Space Telescope (HST).  Only 5 objects had sufficient signal to
noise to be used here. Four objects were observed with the PRISM
grating, and one (1516+1900) was observed with the higher resolution
gratings G140L and G230L (see Table~2 for details). No
emission line analysis was done for these spectra, since so few
objects were detected.  For the purpose of including the HST spectra
in the SEDs the underlying continuum was fitted using the
IRAF~\footnote{IRAF (Image Reduction and Analysis Facility) is
distributed by the National Optical Astronomy Observatories, which are
operated by AURA, Inc., under cooperative agreement with the National
Science Foundation.}  ``continuum'' task. The fitted continuum was
then binned into broader wavelength bands to delineate the SEDs.

\subsection{Optical and IR Photometry}

The near-IR flux densities for the sample were compiled using the
J,H,K$_S$ magnitudes from the Two Micron All-Sky Survey (2MASS) Point
Source Catalog. The optical (B,R,I) photometry was taken from the
Palomar Digital Sky Survey (DPOSS~I and II) photographic plates and
retrieved through the SuperCOSMOS Sky Survey (Hambly \etal\ 2001a,b,c)
or/and the USNO--A2.0 Catalog (Monet \etal\ 1998). An 0.3~magnitude
uncertainty was adopted for SuperCOSMOS data and 0.4~mag. uncertainty
for USNO--A2 data. These uncertainties account for the photometric
accuracy of the plates and the different photographic emulsions used
at different epochs.  The 2MASS and DPOSS magnitudes for our red 2MASS
AGN sample objects are presented in Table~1 (note
that in most cases SuperCOSMOS R magnitudes were taken at two epochs
denoted here as R1 and R2).  A small number of AGN were also found in
the SDSS database. These objects are presented in
Table~3. In a few cases optical (B,R) photometry from
the DPOSSII plates differed substantially from the SDSS data (u,g,r,i
magnitudes were translated to B,V,R following Jester \etal\ 2005) or
from the shape of the optical spectrum described below. For such
objects (see Table~1) only SDSS photometry was used
for the SEDs, as it is a CCD-based photometry with typical accuracies
of 0.03~mag.

\subsection{Optical Spectrophotometry}

Optical spectroscopy of the red 2MASS AGN was obtained between 1998
and 2005, primarily with the Norris Spectrograph on the Palomar
200-inch Hale telescope and/or the Boller \& Chivens Spectrograph on
the 90-inch Bok telescope at Kitt Peak. Six spectra were taken with
the 2.3m telescope at the Siding Spring Observatory and four spectra
with the FAST Spectrograph on the 1.5m Tillinghast telescope, one
spectrum on the 6.5m MMT telescope on Mt. Hopkins, and one spectrum
with the Keck~II telescope on Mauna Kea. Details of the observations
are shown in Table~4 and the spectra are presented in
Fig.~\ref{fig:spectra_SEDs_all} (right column).  For the purpose of
including the optical spectra in the SEDs, we used (as before for the
HST spectra) the IRAF ``continuum'' task to obtain the underlying
continuum, which was then binned into broader wavelength bands to
delineate the SEDs.  Due to slit losses and observations made under
non-photometric conditions, some of our optical spectra had to be
grayshifted to match the optical photometry from DPOSS and/or SDSS
surveys (the factor by which the spectra were multiplied is also shown
in Table~4).  Where possible, the long wavelength end
of the grayshifted spectrum was also matched with a linear
interpolation between the J and I photometry (spectra ending at
wavelengths between the J and I filters) or between I and R photometry
(spectra ending at shorter wavelengths). In a few cases where the I
photometry did not fit the overall SED, the long wavelength end of the
spectrum was grayshifted to agree with a linear interpolation between
J and R photometry.

\subsection{Optical Polarimetry}

Optical polarimetry was obtained as an extension of the detailed
optical study of red 2MASS AGN by Smith \etal\ (2002, 2003), and
details of the observational procedures can be found in those
publications.  Briefly, $R$-band imaging polarimetry was generally
acquired for all sources using the CCD Imaging/Spectropolarimeter SPOL
(Schmidt, Stockman \& Smith 1992) at the 2.3~m Bok Telescope. This was
followed up by spectropolarimetry of the more strongly polarized
targets at the 6.5~m MMT and 2.3~m Bok Telescope to ascertain the
polarization characteristics of the continuum {\it vs.\/} broad and
narrow emission lines.  In a few cases only spectropolarimetry was
acquired and the quoted $R$-band values were derived from an
integration of the results over the filter passband.  The
spectropolarimetric observations also provide high quality,
low-resolution ($\sim$15\AA) total flux spectra, suitable for accurate
optical classification and the measurement of line strengths, host
galaxy contributions, and continuum slopes as well as suitable for
including in the SEDs.  The new polarimetry results are listed in
Table~5.  Figure~\ref{fig:specpol} shows the
spectropolarimetry data for those sources where it has not been
previously published. For those objects having linear polarization
previously measured by Smith \etal\ (2002, 2003) we recall these
values in column 4 of Table~1.

\subsection{IRAS photometry}

Twelve sources in our red 2MASS AGN sample were found to have IRAS
fluxes in the Faint Source Catalog version~2 (Moshir \etal\ 1990). The
12, 25, 60 and 100$\mu$m fluxes are presented in Table~6.

\section{IR to X-ray Spectral Energy Distributions (SEDs)}

The multi-wavelength data described in the Section~3 was combined to
generate SEDs of the red 2MASS objects, together with the far-IR to
X-ray data available in the NASA/IPAC Extragalactic Database (NED). A
summary of all references used in the compilation of the SEDs is
presented in Table~7.  The full SEDs (except for the
X-rays, which were corrected for \nh\ during X-ray spectral fitting
described in section 3.1) were then corrected for galactic extinction
using the Galactic neutral hydrogen column from Dickey and Lockman
(1990) and Stark \etal\ (1992) and assuming a fixed conversion of
$N(HI)/E(B-V)=5.0\times10^{21}$~cm$^2$mag$^{-1}$ (Burstein \& Heiles
1978).  After this, the data were shifted to the rest frame using a
cosmological model with $\Omega_{o} = 1$ and $H_{o}$ = 75
kms$^{-1}$Mpc$^{-1}$ (Mould \etal\ 2000).  No k-corrections and no
assumptions about the intrinsic spectrum were required, since we are
working with the complete spectral energy distributions.  No host
galaxy subtraction was made, since there is no consistent way to
estimate the strength of the host galaxy emission for the full
sample. The resulting, rest-frame far--IR to X-ray SEDs of the full
sample are presented in Fig.~\ref{fig:spectra_SEDs_all}.

\subsection{Comparison of the SEDs and broad-band optical and IR
  colors with other AGN samples}

In Fig.~\ref{fig:median}a we plot the median SED for our red 2MASS AGN
normalized at 1.5$\mu$m, together with the 68, 90, and 100
Kaplan-Meier percentile envelopes (Feigelson \& Nelson 1985, Isobe,
Feigelson, \& Nelson 1986), which take into account upper limits in
the data (mostly in IRAS data).  In Fig.~\ref{fig:median}b we compare
this median with the medians (redshift and host galaxy corrected) of
the optically and radio selected AGN from Elvis \etal\ (1994;
hereafter E94) and the hard-X-ray selected AGN from Kuraszkiewicz
\etal\ (2003; hereafter the HEAO sample).  All three samples have
similar redshift ranges $0 < z$\lax 0.37.  Optical selection will pick
mostly unobscured (\nh$ < 10^{21}$~cm$^{-2}$) AGN.  Hard-X-ray
selection, on the other hand, has the advantage of finding samples
that are more representative of the real AGN population, since it is
less biased by the affects of obscuration along the line of sight.  It
is apparent from the comparison of the medians of these three samples
that in the optical and UV, the 2MASS median occupies the redder
envelope of the HEAO (i.e.  representative) AGN sample, while the
optically selected AGN occupy the bluer envelope. The median 2MASS AGN
SED is also relatively bright in the IR (has a more pronounced IR bump
due to the \jmk\ $> 2$ selection) in comparison with the other two
samples.

Differences in the median SEDs are confirmed by comparing the
distributions of various (octave and decade) IR, optical, and UV
luminosity ratios in these samples. We find that the distributions of
the following luminosity ratios were significantly different ($>$99\%
in the two-tailed K-S test) when the 2MASS sample was compared with
the HEAO sample and the optically/radio selected E94 sample:
L(0.2-0.4$\mu$m)/L(0.4-0.8$\mu$m), L(0.2-0.4$\mu$m)/L(0.8-1.6$\mu$m),
L(0.4-0.8$\mu$m)/L(0.8-1.6$\mu$m) and
L(1-10$\mu$m)/L(10-100$\mu$m). These distributions are presented in
Fig.~\ref{fig:histogram_L}. Objects, that have HST spectra are marked
with ``x''. Note that, although the 3000\AA\ bump (Wills, Netzer \&
Wills 1985) lies in the 0.2-0.4$\mu$m range, our luminosity estimates
are not contaminated by this bump since our SEDs were compiled using
the underlying continuum fits to the optical/UV spectra (Sections~3.2,
3.4). Objects with high polarization values ($P>3$\%) are marked in
Fig.~\ref{fig:histogram_L} with ``p''.  If higher polarization in
these objects was due to the presence of larger amounts of dust in the
nucleus (as in the case of the highly polarized 2MASX
J10494334+5837501 - see Schmidt \etal\ 2007) we might expect to find
all highly polarized objects among the reddest AGN in our 2MASS
sample.  However these objects span a range of UV/opt and opt/IR
ratios (see Fig.~\ref{fig:histogram_L}) and do not tend to group
towards the reddest 2MASS objects. It is possible that the reddest
(most obscured) AGN do not show high polarization because scattered
nuclear light is diluted by host galaxy emission (Smith \etal\ 2002,
2003) and/or the optical/IR colors of the highly polarized sources are
made bluer by host galaxy emission and the scattered light itself,
which certainly is the case in our 2MASS AGN as discussed in
Section~5.2.

We present in Fig.~\ref{fig:opttwo_color} a comparison of the near-IR
selected 2MASS sample with the hard-X-ray (Kuraszkiewicz \etal\ 2003)
and optically and radio selected (E94) samples in a two color
(L(0.4-0.8$\mu$m)/L(0.8-1.6$\mu$m)
vs. L(0.2-0.4$\mu$m)/L(0.8-1.6$\mu$m)) diagram. The hard-X-ray
selected HEAO sample covers the entire range of parameter space
occupied by the other two samples. The 2MASS sample includes the
redder objects, and the optical/radio selected sample has bluer colors
with little overlap between the two. Clearly the hard-X-ray (2-10~keV)
selection of the HEAO sample is an efficient way to select AGN
regardless of their optical/IR properties.

\section{Analysis of the optical/IR colors}

\subsection{The 2MASS B$-$R and \jmk\ colors}

In Fig.~\ref{fig:two_color} we plot the observed \jmk\ versus B$-$R
colors of the red 2MASS AGN and compare them with the observed colors
of E94 AGN (crosses). The colors of both samples have not been
corrected for host galaxy emission.  The conversion from J$-$K to
\jmk\ colors of the E94 AGN was made using transformations from
Bessell (2005).  The 2MASS objects cover the B$-$R colors of the E94
sample and extend towards redder optical colors. The \jmk\ colors are
on average 1 magnitude redder than those of the optical/radio selected
AGN (E94) due to the red \jmk $> 2$ selection.

Let us assume that the optical/radio selected AGN from E94 are
representative of the unobscured AGN population. We show their median
colors (z=0 SED, individual SEDs comprising the median were corrected
for host galaxy emission) in Fig.~\ref{fig:two_color} by a filled
circle and denote by ``$A_V$=0''. The long-dash-short-dash line shows
the change in color when this median SED is obscured by Milky Way
dust, with extinctions ranging from $A_V$=0 to 4~magnitudes (filled
circles at the loci). The dotted line shows these colors for an AGN at
z=0.3, i.e. at a redshift close to the highest redshift in the 2MASS
sample (see Section~5.5 for a detailed discussion of the redshift
dependence of optical/IR colors). We see that the range of B$-$R
colors shown by the red 2MASS AGN translates into $A_V$ between 0 and
3~magnitudes, while the range of \jmk\ colors translates into $A_V$
between 0 to 10~mag. (0 to 7~mag. for an AGN with z=0.3).  Possible
explanations for finding lower extinction estimated from the B$-$R
color than the \jmk\ color include:

\begin{enumerate}
\item{optical colors of a highly reddened AGN are affected (bluened)
by significant host galaxy emission (for $A_V\geq 2$ reddened AGN
colors are redder than the host galaxy colors -- see discussion in
Section~5.2),}
\item{a significant scattered light component, as indicated by high
  polarization levels, bluens the optical colors,}
\item{large amounts of hot circumnuclear dust are present, which does
not obscure the optically emitting region, but does produce
stronger near-IR emission,} 
\item{the intrinsic SEDs for the reddest \jmk\ objects have a much
steeper/more pronounced intrinsic big blue bump (BBB) than the E94
median; applying the large $A_V$ inferred from the \jmk\ colors would
then produce bluer than expected B$-$R,}
\item{dust composition/grain size differs from the Milky Way dust
composition.}
\end{enumerate}

We will discuss points 1 and 2 in greater detail in Sections~5.2 and
5.3.  The possibility of hot dust in few objects will be discussed in
Section~8.1.  For option 4: to produce the reddest \jmk\ objects (\jmk
=3.5) from a standard E94 median AGN SED, extinction of
$A_V\sim$10~mag. is needed (if AGN is at z=0). To obtain a reddened
B$-$R=1.5~mag color observed in these objects, we would need to start
with a B$-$R$ = -1$~mag., much bluer than the B$-$R$ =
-0.2$~mag. expected from a pure standard accretion disk model, where
$F_{\nu} \propto \nu^{1/3}$ (Shakura \& Sunyaev 1973).  This is also
much bluer than the bluest observed colors of the optical/radio
selected QSOs from E94 (indicated by crosses in
Fig.~\ref{fig:two_color}) that extend to B$-$R$=0.4$~mag., and the
SDSS QSOs that extend to the expected, from accretion disk modeling,
limit of B$-$R$=-0.2$~mag. Therefore, we can rule out an extremely
blue BBB.

Czerny \etal\ (1995) calculated extinction curves for different grain
sizes and compositions including graphite, amorphous carbon, and a
mixture of carbon and silicate grains. All extinction curves are very
similar at optical wavelengths ($\lambda > 3000$\AA), but differ
significantly in the UV, where the Milky Way and pure graphite dust
show the lowest absorption in the UV. If the dust composition is
different from Milky Way or pure graphite dust, or the grains are
small, the dust will more easily absorb the UV, heat up and produce
stronger near-IR emission. Unfortunately, none of the reddest \jmk\
sources, have ultraviolet (HST or GALEX) measurements, so option 5 is
difficult to confirm/rule out.

\subsection{Effects of host galaxy on the B$-$R and \jmk\  colors}

Imaging of the 2MASS AGN at R band shows that the observed, nuclear to
host galaxy flux ratio is 10$\times$ lower when compared to the ratio
found in the normal/blue selected AGN (Hutchings \etal\ 2003). This
suggests that the 2MASS AGN have relatively weaker observed nuclear
flux and/or higher host galaxy contribution at optical wavelengths
than normal AGN. We test that this is indeed the case.  In
Fig.~\ref{fig:two_color_host} we plot the effects of host galaxy on
AGN optical/near-IR colors by adding to Fig.~\ref{fig:two_color}
curves showing the colors of a reddened AGN modified by the host
galaxy emission.  We use host galaxy template models (evolutionary
stellar population synthesis models along the Hubble morphological
sequence) from Buzzoni (2005) that take into account both the
morphological type of the galaxy and the age of the stars ranging from
1 to 15--Gyr.  We select their reddest host in B$-$R, which is an
elliptical galaxy with a 15--Gyr stellar population (E15) and the
bluest host, being an Sd galaxy with a 5--Gyr stellar population
(Sd05), and an itermediate color host: an Sa galaxy with a 5--Gyr
stellar population (Sa05).  These host galaxy templates were combined
with reddened AGN generated from the E94 median SED, using Milky Way
dust with $A_V$ ranging between 0 and 10~magnitudes in steps of
1~mag. This demonstrates the range of possible colors. In each case a
range of host galaxy strength relative to the intrinsic, {\it
unreddened} AGN is shown, normalized in the R band to be
2,5,10,20,30,40 times weaker than the AGN.  For example, an Sd05
galaxy with 5 times brighter intrinsic (unreddened) AGN than host
galaxy at R band is indicated by (Sd05;5). The same notation applies
to the AGN + elliptical E15 (reddest possible) host galaxy models.

By comparing the colors of a pure reddened AGN (long-dash--short-dash
line) with the colors of a reddened AGN+host galaxy in
Fig.~\ref{fig:two_color_host} we see that the optical B$-$R colors are
strongly affected by host galaxy emission.  For a high host galaxy
contribution (e.g. (E15;2), and (Sd05;5)) the B$-$R colors become
bluer and more consistent with the host galaxy colors (represented by
thick solid lines) at $A_V > 2$. For smaller host galaxy contribution
((E15;40), (Sa05;40) and (Sd05;40)) B$-$R colors become dominated by
host galaxy at $A_V > 4$. The \jmk\ colors, on the other hand, are
much less dependent on reddening and host galaxy. They reach a peak at
higher $A_V$ which value is dependent on the AGN/host galaxy
ratio. AGN with weaker host galaxies peak at redder \jmk\ color and
higher $A_V$ (compare for example curves (Sa05;40) and (Sd05,5) where
the peak is at $A_V=14$ and $A_V=8$ respectively).  The \jmk\ colors
then plummet towards bluer colors with increasing $A_V$, reaching the
host galaxy colors at $A_V \gg 20$.

The colors of the four reddest \jmk $>3$ sources can be modeled by a
highly reddened AGN with $A_V=11-12$~mag. and a relatively weak host
galaxy (e.g. (Sa05;40)). In this case the B$-$R colors are consistent
with the host galaxy colors, while the red \jmk\ colors are due to the
reddened AGN.

To better understand the dependence of the B$-$R and \jmk\ colors on
AGN reddening and host galaxy contribution we plot in
Fig.~\ref{fig:colors_SED} a reddened AGN SED (E94 median AGN SED
reddened by Milky Way dust - solid lines) and three (E15, Sd05 and
Sa05) host galaxies from Buzzoni (2005) normalized to be 5 times
(strong host galaxy contribution) and 40 times (weak host galaxy
contribution) weaker at R band than the AGN nucleus (dotted lines).
The above conclusions from the color-color modeling are now clearer.

\subsection{Effects of scattered light on the B$-$R and \jmk\  colors}

Since many of the 2MASS sources are highly polarized, we investigate
the effect of including AGN scattered light to our modeling.  We use
as the scatterer the average Milky Way dust, modeled by Drain (2003a)
as a mixture of carbonaceous grains and amorphous silicate grains,
with size distributions from the Weingartner \& Draine (2001) case A
model for $R_V$=3.1, and renormalized according to Draine (2003a).
The scattering cross section ($\sigma_{sca}(\lambda)$) for such a
mixture is obtained by multiplying the albedo (scattering cross
section/extinction cross section) by the extinction cross section per
H nucleon ($C_{ext}/H$ in cm$^2/H$), which depends on the wavelength
of incident light ($\lambda$) and are presented by Draine (2003b):

$$ \sigma_{sca} (\lambda) = albedo (\lambda) * C_{ext}/H (\lambda) $$

The amount of scattered light is then estimated at B,R,J,K effective
wavelengths by multiplying the scattering cross section
$\sigma_{sca}$, phase function $\phi_0(\theta)$ and scattering
dust+gas column \nh. The function $\phi_0(\theta)$ is the Henyey \&
Greenstein phase function from Drain (2003a) eq.(4):

$$\phi_0 =\frac{1}{4\pi}\frac{1-g^2}{(1+g^2-2gcos\theta)^{3/2}}$$
where $g=<cos\theta>$ is a function of $\lambda$ and its values are
also taken from Draine (2003b). For simplicity, we use a scattering
angle $\theta=90^o$. We do not assume any particular value for \nh ,
but indirectly obtain it when normalizing the amount of scattered AGN
light to the total (AGN + scattered + host galaxy) light at R band to
be consistent with the observed value.

In Figure~\ref{fig:two_color_polarization} we present the effect of
adding scattered intrinsic (unreddened) AGN light (E94 median SED) to
a reddened AGN nucleus (E94 median SED reddened by Milky Way dust), on
the B$-$R and \jmk\ colors.  Similar to the effect of host galaxy
emission, the addition of scattered AGN light changes primarily the
optical B$-$R colors, which become much bluer than the pure reddened
AGN colors (long-dash-short-dash line) once the direct AGN light is
absorbed away ($A_V$\gax $2-4$~mag. depending on the scattered light
contribution; see also Zakamska \etal\ 2006).  \jmk\ colors are
effected at higher ($P\gg 7$\%) scattered light contribution and
higher $A_V$ (see also Fig.~\ref{fig:colors_SED} which shows this from
the SED point of view).  Comparison of curves that include scattering
off dust and electrons shows that dust scattering, as expected, gives
bluer B$-$R and slightly redder \jmk\ colors than scattering off
electrons.

\subsection{The combined effects of scattered light and
host galaxy contributions.} 

In Figure~\ref{fig:two_color_host_polarization} we present the effects
of both host galaxy and polarization on the colors of a reddened AGN
(thick curves) and compare them with the colors of a pure reddened AGN
(long-dash-short-dash line).  The optical and IR colors both become
bluer relative to the pure reddened AGN light as scattered light and
host galaxy contribution is increased, although the effects on the
\jmk\ colors are weaker, especially for smaller scattered light and
host galaxy contributions. The optical/IR colors are also bluer when
compared to the colors of a reddened AGN+host galaxy (dotted curves).

\subsection{Colors and redshift}

\subsubsection{Redshift dependence of pure AGN color}

An AGN's observed optical/near-IR continuum colors change
significantly with redshift. In this section we calculate the J$-$K,
B$-$K and B$-$R colors of a pure median AGN SED from E94 (blue AGN
sample with host galaxy subtracted; the emission lines are {\it not}
explicitly included as discrete features\footnote{However please note
that the individual SEDs used to calculate the E94 median SED had the
emission lines included in their broad band bins.})  as a function of
redshift and present them in Fig.~\ref{fig:JmK}a--c (see Maddox \&
Hewett 2006 for simulations of redshift dependence of other colors
measured between passbands used by the SDSS and 2MASS). In all figures
the bottom (bluest) curve (thick line) is plotted for the unreddened
E94 median SED (see Table~8 for the approximate linear
fits of this dependence). Above this curve the colors of the reddened
AGN are plotted, where Milky Way dust extinction law from Savage \&
Mathis (1979) was used, with the 2200\AA\ feature removed, since it is
not observed in AGN.  The dependence of the 2MASS median SED on
redshift (dotted line) shows optical/near-IR colors that have values
in between the E94 SED reddened by $A_V$=1 and by $A_V$=2.

For the unreddened E94 SED, the J$-$K color becomes bluer with
redshift due to the optical blue bump moving into the near-IR spectral
region, which is one of the reasons for the predominance of
low-redshift QSOs in the 2MASS sample (Cutri \etal\ 2002, see also
Barkhouse \& Hall 2001).  In comparison, the behavior of the J$-$K
color for the 2MASS median (dotted curve) is different due to the
shape of the 2MASS SED which lacks the big blue bump (see
Fig.~\ref{fig:median}) Due to the recovery of the red J$-$K color at
z$\sim$2, objects with SEDs similar to the red 2MASS AGN SEDs at
$z>2$, should be picked up by deep near-IR surveys by virtue of their
red \jmk $> 2$ color.

In Fig.~\ref{fig:JmK}b and c, where the B$-$K and B$-$R dependence on
redshift is shown, the unreddened E94 median at low redshifts also
becomes bluer due to the blue bump moving towards longer wavelengths.
As $A_V$ increases from 1 to 10~mag., the J$-$K, B$-$K and B$-$R
colors become redder. As the redshift increases, the colors also
become redder, since the big blue bump becomes redder at shorter
wavelengths due to increasing extinction.  A strong increase in
Galactic dust extinction (used in our models without the 2200\AA\
feature) at $\lambda>$1500\AA (log$\nu$=15.3) explains the strong rise
to redder B$-$K and B$-$R colors at z$>$2.

\subsubsection{Redshift dependence of AGN + host galaxy + scattered
  light color} 

The redshift range of the red 2MASS AGN sample is small, but we have
shown that the colors change significantly even over a small range. We
therefore investigate the effects of redshift on our color-color
analysis.

Buzzoni (2005) presents fluxes and colors of the host galaxy templates
in the Johnson UBVRIJHK filters. Since the templates are not smooth,
we avoid interpolating between filters by calculating the colors of an
AGN+host galaxy redshifted to z=0.31 (which is also the approximate
highest redshift of the 2MASS sample), for which the fluxes in the U,
V, I, H bands at rest frame transform to the fluxes in B, R, J, K
bands in observed frame. We present this in
Fig.~\ref{fig:two_color_host_z_and_polarization_z}a and, for
comparison, we plot the observed colors of a pure reddened AGN at z=0.
In comparison with the z=0 models, the redshifted z=0.31 reddened
AGN+host galaxy models have less host galaxy (and more AGN)
contribution at the observed B and R bands and so extend to redder
B$-$R colors, closer to the original, pure reddened AGN colors. The
contribution of host galaxy at observed J and K bands is slightly
larger for AGN at z=0.31 than at z=0, hence the AGN+host galaxy curves
peak at slightly bluer \jmk\ color than the z=0 curves.

In Fig.~\ref{fig:two_color_host_z_and_polarization_z}b we compare the
colors of a reddened AGN + intrinsic AGN light scattered off dust
models from Section~5.3 at two different redshifts: z=0 (dotted lines)
and z=0.31 (solid lines). The observed B$-$R and \jmk\ colors at
z=0.31 become bluer (closer to the scattered intrinsic AGN light
colors) when compared to the z=0 models, as the dust scattering
efficiency increases with decreasing wavelength (a $\propto
\lambda^{-2}$ Reyleigh scattering dependence is assumed).

When both host galaxy emission and scattered AGN light are added to
the reddened AGN (Fig.~\ref{fig:agn_host_pol_z}) the z=0.31 models
reach redder B$-$R colors than the z=0 only if the host galaxy
contribution is high (e.g. compare (E15;5;1\%;z) and (E15;5;1\%)
curves). The z=0.31 models also peak at bluer \jmk\  colors, when
compared to the models where the contribution of host galaxy is low
(compare (Sd05;20;1\%;z) and (Sd05;20;1\%)).

\subsection{Estimates of the 2MASS AGN parameters from optical/IR colors}

In Table~7, columns (8)-(13) present model parameters of
the 2MASS AGN that were obtained from the B$-$R and \jmk\ color
modeling described in Sections~5.2 through 5.5.2. These are: the AGN
$A_V$ (column 8), host galaxy type and intrinsic AGN/host galaxy ratio
at R band (column 9), amount of scattered light relative to the
intrinsic and the reddened AGN (columns 10,~11). Since redshift
affects the colors significantly over our redshift range we use the
higher redshift models (Sections~5.5.1, 5.5.2) to interpret the
sources with z$\geq 0.3$ (these are marked in column 12). If more than
one model fit the optical/IR colors, we chose the reddening value
closest to the X-ray fitted \nh\ (column 6). If \nh\ was not available
we looked at the optical type (Table~1 and
Fig.~\ref{fig:spectra_SEDs_all}) and chose lower reddening values for
Type~1--1.5 and higher for Type~1.8--2.0.  Our modeling results were
compared with Marble \etal\ (2003), who studied the HST/WFPC2 I-band
images of the 2MASS AGN and present the AGN-to-total light ratio and
host galaxy type in which the AGN resides. Fifteen out of 16 objects
in common had an AGN-to-total light ratio and host galaxy Hubble type
consistent with Marble \etal\ (2003). We also compare our modeling
results with the host galaxy to total light ratios obtained from
stellar absorption features by Smith \etal\ (2003). Our sample
includes 11 sources in common, all having consistent ratios. 

Our modeling brakes down for one highly polarized ($P=11$\%), high~z
object 2222+1952, where our optical/IR modeling gives a 54\% observed
AGN, 34\% host galaxy and a 12\% scattered light contribution to total
light at R band, while Marble \etal\ (2003) quote 99\% AGN emission
with no host galaxy contribution at $\sim$8100\AA .  Our X-ray fitting
gives low values of \nh\ consistent with Marble \etal\ (2003) results,
but our modeling requires higher reddening to reproduce the red \jmk
=3.05~mag. color. In this object the red \jmk\ color is most likely
due to very hot dust lying close to the nucleus, and not due to the
reddening of the nucleus, predicted by the model.

\section{X-ray properties}

We will now concentrate on the X-ray properties of the red 2MASS AGN
and compare them with the optical/radio selected AGN from E94.  In
Figure~\ref{fig:Fx/FBvsFx/Fnn} we present the relations between the
ratio of intrinsic X-ray flux at 1~keV (corrected for Galactic and
intrinsic extinction) to the observed optical flux in B band
($F(1keV)/F_B$) and the ratios of intrinsic 1~keV flux to the observed
fluxes measured at: R and K bands ($F(1keV)/F_{R,K}$).  The
$F_{B,R,K}$ fluxes are corrected for Galactic dust absorption
only. Intrinsic $F_{B,R,K}$ fluxes are difficult to obtain since: 1)
the host galaxy contribution at optical/near-IR wavelengths is
non-negligible (see Section~5), 2) the reddening and host galaxy
contribution estimated from the B$-$R and \jmk\ colors (Section~5;
Table~7) are crude, and 3) estimates of dust reddening
from the measured X-ray absorption are not useful since optical dust
and X-ray gas column densities are frequently observed to disagree in
quasars by up to 3 orders of magnitude (Maiolino \etal\ 2001). The
2MASS sources in Figure~\ref{fig:Fx/FBvsFx/Fnn} cover the range of the
unreddened E94 sources and extend to 10$-100\times$ lower
$F_X/F_{B,R,K}$ ratios.

In Figure~\ref{fig:Fx/FBvsFx/Fnn}b ($F_X/F_B$ versus $F_X/F_K$
relation) one can roughly estimate the amount of dust reddening at
optical wavelengths in the 2MASS sample, relative to the unreddened
E94 sample. Since dust absorption is high at B and low at K band, the
$F_X/F_B$ changes strongly with reddening while the $F_X/F_K$ does
not. The 2MASS sample shows a shift/scatter towards larger $F_X/F_B$
values relative to the E94 sample, implying a range in reddening of
$E(B-V) \sim 0.86$ (\nh\ $=5\times10^{21}$~cm$^{-2}$ assuming a
standard Milky Way dust-to-gas ratio), as shown by the arrow. This is
a lower limit as some of the reddest sources have a non-negligible
contribution from the host galaxy (as shown in Section~5), which
decreases the effect of reddening on the optical/IR colors.

Since reddening by dust moves objects towards higher $F_X/F_{B,R,J,K}$
values in these figures, the 2MASS sources with $F_X/F_{B,R,J,K}$
ratios lower than E94 must be either intrinsically weaker or more
obscured in X-rays (a 10$\times$ lower 1~keV flux is obtained by
obscuring the intrinsic continuum with gas column of \nh\
=10$^{22}$~cm$^{-2}$). The sources with the lowest $F_X/F_{B,R,J,K}$
are mainly the lowest S/N ({\it A}) \chandra\ sources, where an
average \nh\ $=7.6\times10^{21}$~cm$^{-2}$ value was assumed (see
Section~3.1). Their properties will be discussed in the following
section.

\subsection{[O\,III]  vs. hard-X-ray relation}

The luminosity of the [O\,III]\,$\lambda$5007 emission line
(L$_{[O\,III]}$), originating from the narrow-line region (NLR), has
long been suggested to be an indicator of the intrinsic nuclear
luminosity of the AGN due to the similarity of the
[O\,III]-to-hard-X-ray flux ratio between Seyfert~1 and Seyfert~2
galaxies (Mulchaey \etal\ 1994, Alonso-Herrero, Ward \& Kotilainen
1997, Turner \etal\ 1997).  This, in turn, can be used as a test for a
Compton thick AGN (Ptak \etal\ 2003), in which the X-rays are so
heavily absorbed that the observed emission is dominated by scattered
light that appears unabsorbed, but unusually weak in comparison with
L$_{[O\,III]}$.  More recently, Maiolino \etal\ (1998) and Bassani
\etal\ (1999) suggest that an extinction-corrected L$_{[O\,III]}$ is
more appropriate, though the uncertainties in such extinction
corrections in AGN are generally large.  In Fig.~\ref{fig:O3vsFX} we
present the relation between the \chandra\ 2$-$10~keV X-ray flux and
the narrow emission line [O\,III]\,$\lambda$5007 flux.  Seyfert~1s and
Seyfert~2s from Mulchaey \etal\ (1994) sample have [O\,III] fluxes
uncorrected for extinction, and the intrinsic hard X-ray (2$-$10keV)
fluxes estimated from the EXOSAT and Ginga spectra by fitting a
power-law with Galactic and intrinsic absorption.  Most objects in the
Mulchaey \etal\ (1994) sample cluster around a line that represents
the $\log F([O\,III])/F(2-10keV)$ mean of $-$1.89$\pm$0.25 (1$\sigma$
uncertainty) for Seyfert~1s and Compton-thin Seyfert~2s. However
NGC~1068, a Seyfert~2, shows an exceptionally strong [O\,III] to
hard-X-ray flux ratio, that is interpreted as being due to obscuration
of the direct X-ray light by Compton thick (\nh\ $\ge
10^{24}$~cm$^{-2}$) material (Turner \etal\ 1997, Antonucci \& Miller
1985) so that scattered light dominates the X-ray emission.

We identify the 2MASS objects in Fig.~\ref{fig:O3vsFX} according to
the quality of their \chandra\ spectra.  We show two versions of the
figure: Fig.~\ref{fig:O3vsFX}a uses the observed [O\,III] flux
with no extinction correction applied and Fig.~\ref{fig:O3vsFX}b shows
the [O\,III] fluxes corrected for reddening estimated from the ratio
of the narrow H$\beta$ to H$\alpha$ lines, assuming an unreddened
(case B) value of 3. Emission from Fe\,II and [N\,II] lines were
subtracted for the measurements of H$\beta$ and H$\alpha$
respectively.  We circled objects in Fig.~\ref{fig:O3vsFX}b that have
undetermined reddening corrections because the extremely weak H$\beta$
defied measurement.  Such weak H$\beta$ is possibly due to high
extinction, so these objects may have higher intrinsic L([O\,III]).

All of the 2MASS objects with relatively well-determined \chandra\
{\it C} and {\it B} fluxes (with the exception of 0409+0758 and
2222+1952) follow the Mulchaey \etal\ (1994) relation to lower X-ray
fluxes, implying that the observed X-ray flux in these objects is
dominated by direct rather than scattered light and is thus a good
indicator of the AGN's intrinsic X-ray emission.  The remainder of the
sample, with low S/N ({\it A}) \chandra\ spectra, has lower
hard-X-ray-to-[O\,III] flux ratios by a factor of 10-100. Possible
explanations include:

\begin{enumerate}
\item{The X-ray fluxes are intrinsically weak relative to the AGN
  ionizing optical/UV flux (that produces [O\,III] emission). The
  [O\,III] vs. hard-X-ray luminosity correlation breaks down,
  because the hard X-ray flux is no longer a good indicator of AGN
  luminosity.}
\item{The X-ray absorption is higher than we assume (in {\it
  A} spectra we assumed  absorption of \nh = 7.6$\times 10^{21}$
  cm$^{-2}$, the median of the {\it C} sample).}
\item{The X-ray photon index is flatter than the normal, $\Gamma=2$
  photon index assumed for the {\it B} and {\it A} spectra, leading to
  an underestimate of the X-ray flux. The median photon index for the
  {\it C} sample, is $\Gamma_X=1.5$ (Wilkes \etal\ 2008, in
  preparation), however this change in photon index is too small to
  explain the 1--2 orders-of-magnitude lower X-ray fluxes.}
\end{enumerate}
The first two options are discussed in detail in the following section.

\subsection{Objects that do not follow the [O\,III] vs. X relation}

As discussed in the previous section, objects with 10-100$\times $
lower than typical (E94) hard-X-ray-to-[O\,III] flux ratios are either
intrinsically X-ray weak or have high obscuration or both. The first
group includes three AGN: {\it 0234+2438, 1258+2329, 2222+1959} with
the lowest $F(1keV)/F_K$ and $F(1keV)/F_B$ ratios in the sample. Their
optical and IR colors are blue (B$-$R$=0.68-1.18$~mag. and \jmk
$=2.04-2.16$~mag.) and, when modeled, imply an unreddened AGN
($A_V\le$ 1~mag.) with a weak ($<$6\%) host galaxy contribution at R
band (see Table~7).  Their optical spectra
(Fig.~\ref{fig:spectra_SEDs_all}) show strong Fe\,II emission and weak
(relative to H$\beta$) [O\,III] emission, resembling NLS1s and
BALQSOs, which are thought to have high $L/L_{Edd}$ ratios
(e.g. Boller, Brandt, Fink 1996, Pounds, Done \& Osborne 1995,
Kuraszkiewicz \etal\ 2000) that produce weak X-rays relative to the
Big Blue Bump (Witt, Czerny, \& \.Zycki 1997). The second group
includes: {\it 0348+1255, 0748+6947, 0955+1705, 1040+5934, 1307+2338,
1453+1353}, whose optical/IR colors are modeled with a highly obscured
AGN ($A_V=10-15$ i.e. \nh\ =$(1.6-2.3)\times 10^{22}$~cm$^{-2}$), and
the host galaxy contributing 96$-$100\% to the total observed light at
R band (Table~7).  These objects have more absorption than
assumed in the spectral fits, \nh = 7.6$\times 10^{21}$ cm$^{-2}$, and
thus their de-absorbed X-ray fluxes are underestimated causing them to
diverge from the hard-X-ray vs. [O\,III] correlation. In the next
section we will further confirm high obscuration in these sources by
studying their far-IR IRAS colors.

In summary, AGN that are highly obscured (\nh\ $\approx
10^{22}$~cm$^{-2}$) or have high $L/L_{Edd}$ ratios do not follow the
[O\,III] vs. hard-X-ray relation so the intrinsic hard-X-ray flux
cannot be estimated from their [O\,III] emission. In the first case,
the highly obscured X-rays are dominated by scattered light and hence
appear weak and unabsorbed.  In the second case, high $L/L_{Edd}$
produces a more luminous Big Blue Bump (see Witt, Czerny \& \.Zycki
1997), and so the X-rays seem relatively weak compared to the AGN's
ionizing optical/UV continuum that produces the [O\,III] emission.

\subsection{\nh\  estimates from the far-IR flux ratios}

Twelve of the objects in our red 2MASS AGN sample were observed by
IRAS. We plot their 12$\mu$m to 60$\mu$m flux ratio versus the
25$\mu$m to 60$\mu$m flux ratio dependence in
Fig.~\ref{fig:12_60to25_60}. As before, 2MASS AGN with \chandra\ flux
measurements of various S/N levels are indicated.  Again, we compare
the 2MASS sources with the optical/radio AGN sample of E94 (crosses)
and X-ray selected HEAO AGN (Kuraszkiewicz \etal\ 2003; open circles)
but this time add Seyfert~2 nuclei from Heisler, Lumsden \& Bailey
(1997), some of which have broad-line emission observed in polarized
light (stars) and some do not, possibly indicating higher obscuration
(open triangles). The diagonal line shows the change in far--IR colors
with differing amounts of dust reddening. As can be seen from the
figure, the 2MASS objects have \nh\ in the range of
$10^{22}-10^{23}$~cm$^{-2}$, consistent with \nh\ estimated from
X-rays, although most of the fluxes are upper limits and strong
conclusions cannot be drawn. The four low S/N ({\it A}) \chandra\
sources ({\it 0748+6947, 1040+5934, 1307+2338, 1453+1353}) show the
highest \nh\ $\sim 10^{23}$~cm$^{-2}$, confirming that these sources
may have more absorption than the assumed \nh = 7.6$\times 10^{21}$
cm$^{-2}$ (\chandra\ {\it C} fits median) that was also suggested by
the B$-$R and \jmk\ modeling.

\section{Emission line properties and comparisons with other samples}

We measure the rest frame equivalent widths, fluxes and full widths at
half maximum (FWHM) of the strongest optical emission lines
(H$\alpha$, H$\beta$, [O\,III]\,$\lambda$5007, Fe\,II, and
[O\,II]\,$\lambda$3728) in the red 2MASS AGN spectra presented in
Fig.~\ref{fig:spectra_SEDs_all}.  The emission line measurements were
made using the modeling software {\em
Sherpa}\footnote{\url{http://cxc.harvard.edu/sherpa/index.html}}\
(Freeman, Doe \& Siemiginowska 2001) originally developed for the
\chandra\ mission, but applicable to any grid of multidimensional
fitting.  Each spectrum was fitted by a reddened power-law continuum
(reddening curves of Cardelli, Clayton \& Mathis 1989 were used) to
regions of the spectrum away from strong emission lines and blended
iron emission. The blended optical (4400\AA\ $< \lambda_{rest}
<$7000\AA) iron emission was modeled using the Boroson \& Green (1992)
templates, broadened by convolving with Gaussian functions of width
between 900 and 10,000\kms, separated by steps of 250\kms. This
included several steps: 1) estimation of a crude flux normalization
for a 2000~km\,s$^{-1}$ template, 2) estimation of the width of the
iron template, and 3) a fit of both the amplitude and width of the
template. This was followed by another iteration of the continuum and
the iron emission modeling.  Finally the emission lines were fit with
a single Gaussian, unless 2 or 3 components (broad, narrow and
intermediate, in the case of H$\beta$ and H$\alpha$) were clearly
needed. Multiple-line components were usually required for higher S/N
spectra. The FWHM, peak amplitude of the Gaussian and the position of
the emission line were modeled.  All model parameters were determined
from a minimization of the $\chi^2$ statistic with the Gehrels
variance function (Gehrels 1986) and using the Powell optimization
method. This technique proved successful in the analysis of spectra
from the Large Bright Quasar sample (LBQS; Forster \etal\ 2001) and
the {\it FOS/HST} spectra (Kuraszkiewicz \etal\ 2002, 2004).  We
present the 2MASS AGN emission line measurements in
Table~9 and the fits are presented in a companion paper
Kuraszkiewicz \etal\ (2008, in preparation) and on our web site:
\url{http://hea-www.harvard.edu/\~{}joasia/2MASSAGN/emissionlinefits/table.html}.

We calculate the mean equivalent widths for the 2MASS red AGN and
compare them with those obtained from the SDSS composite quasar
spectrum (Vanden Berk \etal\ 2001) and the Large Bright Quasar Survey
(LBQS) sample (Forster \etal\ 2001) in Table~9. The mean
equivalent widths of H$\alpha$, H$\beta$, and Fe\,II of the 2MASS
sample agree with the means measured for the SDSS composite. The
Fe\,II mean is consistent with the LBQS mean, but mean
$W_{\lambda}(H\beta)$ is lower for the 2MASS sample than the LBQS
sample and is likely due to the presence of Type~1.8--2 objects that
lie preferably at lower H$\beta$ equivalent widths (see
Figure~\ref{fig:emission_lines_histo}a). The LBQS sample consists
mostly of broad lined QSO and so is biased against these edge-on
objects.

The mean equivalent widths of the forbidden [O\,III$]$ and [O\,II$]$
emission lines are much higher in the 2MASS sample than in the LBQS
sample and the SDSS composite (Table~9). The
distributions of the [O\,III$]$ equivalent widths and the comparison
with the LBQS sample is shown in
Figure~\ref{fig:emission_lines_histo}b. The K-S test showed a 99\%
probability that the distributions are different such that the 2MASS
sample is deficient in low EW([O\,III]) or overpopulated with high
EW([O\,III]) sources. Under the assumption that viewing angle is the
primary difference between the red 2MASS AGN and other AGN/QSOs (as
suggested by the SEDs), we can explain their higher [O\,II] and
[O\,III] equivalent widths as due to a lack of face-on objects (where
the continuum is unobscured and so brighter) and/or to a higher number
of inclined objects in the sample. This is the case since 85\% of the
sample consists of Type~1.2--2. In the unification model, type 1.0
sources are those thought to be face-on, while types 1.2, 1.5 etc.,
where the relative strength of the broad lines decreases relative to
the narrow lines, are progressively more edge-on.

Assuming that obscuration increases with inclination angle (Baker
1997), the moderate amount of obscuration \nh $\approx 10^{21}-
10^{22}$~cm$^{-2}$ found in our 2MASS sample is consistent with
moderate inclination angles for the obscuring torus or accretion disk,
such that we look through a wind or atmosphere above the main
structure. Alternatively these column densities are consistent with a
view through a host galaxy disk. The latter is supported by the
findings of Marble \etal\ (2003), who studied the HST/WFPC2 I-band
images of a different subset of 2MASS AGN, and concluded that these
AGN reside mostly in inclined host galaxies with inclination angles of
$i=50-75^{o}$.  AGN with such inclination angles will reshape the
EW([O\,III]) distribution by adding higher equivalent widths to the
distribution.

\section{The sample - a mixed bag of objects}

After analyzing the global properties of our red 2MASS AGN sample, we
attempt to understand its make-up by looking at groups of objects divided
according to optical type. We notice that the spectra become redder in
both B$-$R and \jmk\ colors as we progress from Type~1, through the
intermediate Type~1.2--1.9, to Type~2. This can be explained in terms
of orientation--dependent dust obscuration.

\subsection{The Type~1s}

Only one object ({\it 2344+1221}) in this class has an optical
spectrum that looks like a genuine Type~1 object with broad Balmer
emission lines and strong [O\,III] emission. It lies on the borderline
of our red \jmk\ selection, with \jmk =1.997. High S/N XMM-Newton
observations (Pounds, Wilkes, \& Page 2005) show an obscuring column,
\nh\ $\sim 10^{22}$~cm$^{-2}$ of moderately ionized gas or a lower
column, \nh\ $\sim 3\times 10^{21}$~cm$^{-2}$ of cold gas. Our
optical/IR color modeling of this object implies reddening of
$A_V=2-3$, favoring the second scenario. Six other objects in this
class show strong optical Fe\,II emission and extremely weak
(relatively to H$\beta$) or non--existent [O\,III] emission, resembling
the spectra of NLS1/BALQSOs (Osterbrock and Pogge 1985, Weymann,
Carswell \& Smith 1981). The red \jmk $> 2$ selection seems to pick a
high percentage (86\%) of these objects among Type~1s.  It has been
found that the IR SEDs of some NLS1s like Mrk~1239, Mrk~766, and
I~Zw~1 (Rodriguez-Ardila \& Mazzalay 2006, Rodriguez-Ardila, Contini
\& Viegas 2005, Rudy \etal\ 2000) show a significant 2.2$\mu $m bump,
that can be modeled with a blackbody function with T=1200~K, a
temperature that is close to the evaporation temperature of graphite
grains, indicating hot dust close to the nucleus. Hence the red \jmk
$> 2$ selection picks Type~1 sources with high $L/L_{Edd}$ {\it and}
large amounts of hot circumnuclear dust.  The optical/near-IR colors
of the red 2MASS Type~1s are mostly blue (B$-$R$\le 1.4$~mag. and
\jmk\ $<$2.2~mag.) and modeled as pure AGN ($A_V \leq 1$~mag.)  with
little/no ($< 17$\% at R band) host galaxy contribution (see
Table~7). An exception is {\it 1501+2329} with red optical
color (B$-$R$=1.9$~mag.) and \jmk =2.5~mag., modeled as a reddened AGN
($A_V=3$~mag.) with a 30\% host galaxy contribution. Absorbing column
densities of the red 2MASS Type~1s agree with those obtained from
X-ray fitting, where \nh $\le 5.6 \times 10^{21}$~cm$^{-2}$ is
consistent with typical values found in Type~1 AGN (Malizia \etal\
1997). The polarization values in our 2MASS Type~1s are generally low,
except in {\it 1501+2329}, where $P=3$\%, and in {\it 1516+1900} for
which $P=9.3$\%.

{\it 1516+1900} has a polarized light spectrum dominated by broad
Balmer lines and lacks narrow lines that are present in the total flux
spectrum (Smith \etal\ 2000). The total flux spectrum is blue in the
optical with broad permitted emission lines and is extremely steep and
red in the near-UV (see Fig.~\ref{fig:spectra_SEDs_all}). These
properties imply large amounts of dust lying near/within the NLR that
partially obscure the AGN, scatter the continuum and BLR light and
also redden the scattered light itself (as e.g. in the case of
IRAS+13349+2438; Hines \etal\ 2001).

\subsection{The Type~1.2 and 1.5s}

Most (10/17) of the optical spectra in this class are flat (B$-$R$\sim
1.4$~mag.) in $F_{\lambda}$ vs. $\lambda$, four are red
(B$-$R$>$1.5~mag.), and three are blue (B$-$R$\sim$1.2~mag.). {\it
  0234+2438}, the bluest source in this class, shows particularly
strong Fe\,{\small II}, weak [O\,{\small III}] emission and weak X-ray
emission, resembling the behavior of the Type~1s discussed above.
Most objects (13/17) in this class have \nh $<$few$\times
10^{21}$~cm$^{-2}$, consistent with the typical Type~1 values (Malizia
\etal\ 1997). Four: {\it 0420$-$2047, 1027+1219, 1659+1834} and {\it
  0955+1705}, have \nh $\sim$few$\times 10^{22}$~cm$^{-2}$, higher
than the typical Type~1 value. The first three are high S/N ({\it C})
\chandra\ sources for which \nh\ was measured and shown to be the
highest in the sample. {\it 0955+1705} is a low S/N ({\it A})
\chandra\ object with high $A_V=15$~mag. estimated from modeling of
the optical/IR colors.  The host galaxy contribution at R band,
obtained from optical/IR color modeling, is $<$50\% in about half
(8/17) of these intermediate type sources.

The group of Type~1.2--1.5 2MASS AGN includes the highest number
(5/17) of highly polarized ($5\%<P<$13\%) objects: ({\it 0420$-$2047,
0918+2117, 0938+0057, 1659+1834, 2222+1952}). This is consistent with
Smith \etal\ (2002) who find the highest polarization in intermediate
type objects.  Only three AGN from different optical types: {\it
1516+1900} (Type~1), {\it 1049+5837} (Type~1.8), and {\it 0108+2148}
(Type~1.9) show high polarization ($P$=9.27\%, $P\ge $8\% and
$P$=5.07\% respectively).

\subsection{The Type~1.8 and 1.9s}

All spectra in $F_{\lambda}$ vs. $\lambda$ are either flat (6/9) or
red (3/9).  Two objects, {\it 0108+2148} (Type~1.9) and {\it
1049+5837} (Type~1.8), in this class have high polarization ($P\ge
$5\%). {\it 1049+5837} shows variations in the degree and position
angle of polarization with wavelength implying two scattering
components: one with $P\ge $8\% from material originating from the
polar scattering lobes and one dominating at $\lambda$\lax 4500\AA\
with $P\ge $20\% lying along a less reddened line of sight (Schmidt
\etal\ 2007; complex absorbing and scattering properties are also
visible in X-rays, Wilkes \etal\ 2008). The B$-$R and \jmk\ colors of
Type~1.8--1.9s, as expected, are redder than those of the Type~1--1.5
discussed above. Modeling of the optical/IR colors finds six
moderately absorbed ($A_V=3-6$~mag. i.e. \nh\ $=
(4.8-9.7)\times10^{21}$~cm$^{-2}$) and three highly absorbed
($A_V=10$~mag.; \nh $=1.6\times10^{22}$~cm$^{-2}$) sources with
substantial ($>$60\%) host galaxy contribution at R band. An exception
is 2024$-$5723 with a $\sim$30\% host galaxy contribution.

\subsection{The Type~2s}

This group of 11 objects includes the two reddest \jmk\ sources in our
red 2MASS AGN sample: {\it 0348+1255} and {\it 1307+2338} (\jmk =3.294
and 3.314 respectively). Column densities obtained from X--ray
spectral fitting and optical/IR color modeling are: \nh=
$(0.7-3.5)\times 10^{22}$~cm$^{-2}$, which are at the low end of the
\nh\ distribution for Seyfert~2s (Risaliti, Maiolino \& Salvati 1999).
Since typical, highly absorbed (i.e.  edge-on \nh \gax
$10^{23}$~cm$^{-2}$) Type~2s have AGN completely obscured both in
optical and near-IR, their colors are consistent with pure host galaxy
colors i.e. B$-$R$=1-1.8$ and \jmk $=0.8-1$. However, the low \nh ,
red 2MASS Type~2s are obscured at optical wavelengths, but not in the
near-IR, so the \jmk\ colors are dominated by the reddened AGN light
and will be picked up by the red \jmk $>2~$mag. selection. The host
galaxy contribution in the red 2MASS Type~2s from optical/IR color
modeling is strong and responsible for 76$-$100\% of the total
observed flux at R band. This is also confirmed by strong galactic
absorption lines, a 4000~\AA\ dip in the optical spectra of {\it
1453+1353} and {\it 2225+1958}, and extended images on the B and R
plates. Dilution by the large contribution of the host galaxy is
likely to explain the low observed polarization of Type~2 red 2MASS
AGN (Smith \etal\ 2002; 2003).

Malizia \etal\ (1997) suggested that, since low--absorption (\nh\
$\sim 10^{22}$~cm$^{-2}$) Type~2 AGN show little/no change in \nh\
over time, they must have an absorber lying further away from the
nucleus, possibly in the host galaxy inclined to our line of
sight. This may be the case in half of the red 2MASS Type~2 AGN, which
have typical Type~2 optical spectra with strong [O\,{\small III}]
emission ({\it 0050+2933, 1021+6311, 1300+1632, 1307+2338,
2225+1958}). The other half ({\it 0157+1712, 1453+1353, 1755+6751})
have weak or non--existent ({\it 0348+1255, 1507-1225}) [O\,{\small
III}] emission, implying either absorption closer to the nucleus (to
dampen the ionizing photons before they reach the NLR), or perhaps a
host galaxy absorber {\it and} an AGN with high $L/L_{Edd}$, similar
to the high $L/L_{Edd}$ NLS1s and BALQSOs which have weak [O\,{\small
III}] emission.

\section{Conclusions}

We have analyzed a sample of 44 AGN selected from the 2MASS survey
based on their red \jmk $>2$ color and later observed by \chandra.
The sample includes a mixed bag of objects: 7 Type~1s, whose spectra
(except for one) resemble NLS1/BALQSOs (strong Fe\,{\small II} and
weak [O\,{\small III}] emission), 11 Type~2s with relatively low (for
Type~2s) \nh $<$ few $\times 10^{22}$~cm$^{-2}$, that allows the
reddened AGN colors to dominate in the near-IR, and 26 
intermediate--type sources (1.2--1.9), where 8 show high optical
polarization.

The red \jmk\ selection picks AGN in which circumnuclear and/or host
galaxy obscuration result in unusually red near-IR colors. It also
picks high $L/L_{Edd}$ Type~1 sources with hot circumnuclear dust
emission. Sources which are obscured are most likely viewed at an
intermediate angle, and offer an opportunity to study the
contributions of weaker components that are normally outshone by the
AGN light. The sample shows the following properties:

\begin{enumerate}

\item{The 2MASS AGN median SED is redder (by $\sim$1~mag. in B$-$R) in
the optical/UV than the blue optically/radio selected AGN from Elvis
\etal\ (1994), and redder than the hard--X--ray selected AGN from
Kuraszkiewicz \etal\ (2003).}

\item{Seven of the 44 sample objects show pure AGN optical and IR
colors (with $A_V=0-1$~mag.; SED and emission line properties indicate
high $L/L_{Edd}$). The remaining sources show redder colors, which are
modeled as a mixture of a reddened AGN ($A_V=0-22$ i.e. \nh $\leq
4.2\times 10^{22}$~cm$^{-2}$), host galaxy emission (approaching 100\%
at R band in 1/3 of the sample, with intrinsic AGN/host galaxy ratio =
few--40 at R band), and, in a few sources, AGN scattered light
emission.}

\item{Sources with high polarization, when modeled, turn out to have a
$<$2\% scattered AGN light contribution at R relative to the intrinsic
AGN light. The scattered light becomes significant as the direct AGN
light is absorbed in our line of sight.}

\item{The column densities obtained from X-ray spectral fitting are
between $10^{21}$~cm$^{-2}$ and $7\times 10^{22}$~cm$^{-2}$. These
values are consistent with those found from modeling of the
optical/near-IR colors (\nh\ $\leq 4.2\times 10^{22}$~cm$^{-2}$) and the
far-IR IRAS colors where the 12$\mu$m/60$\mu$m and 25$\mu$m/60$\mu$m
flux ratios give crude estimates of \nh\ $< 10^{23}$~cm$^{-2}$.  The
consistency of \nh\ values obtained from optical colors and X-ray
spectral fitting is due to our detailed modeling of the optical
colors, which accounts for the affects of reddening, host galaxy
emission and scattered AGN light emission.  Without such detailed
analysis \nh\ values may disagree by up to 3 orders of magnitude
(Maiolino \etal\ 2001).}

\item{The deficiency of low equivalent widths in the [O\,{\small III}]
distribution, relatively to the LBQS and SDSS QSO samples, implies a
predominance of inclined objects with intermediate viewing angles in
the red 2MASS sample.}

\item{The analysis of SED and emission line properties implies that
the weakness of X-ray emission (low F(1keV)/F$_K$ and
F(2$-$10keV)/F([O\,{\small III}]) ratios) shown by a large number of
the red 2MASS AGN (Wilkes \etal\ 2002) relative to the blue
optically/radio selected QSOs (Elvis \etal\ 1994) is due to either
higher intrinsic obscuration or high $L/L_{Edd}$ ratios, or both.}

\item{We find that objects with weak (relative to the big blue bump)
X-ray emission (high obscuration or/and $L/L_{Edd}$) depart from the
F(2$-$10keV) versus [O\,{\small III}] relation found for Seyfert~1s
and 2s (Mulchaey \etal\ 1994) which limits its usefulness for
estimating the AGN's intrinsic X-ray emission.}

\end{enumerate}

\acknowledgements 

We wish to thank the referee, Gordon Richards, for comments that
helped improve this paper.  BJW and JK gratefully acknowledge the
financial support of NASA Chandra grants: GO1-2112A, GO3-4138A and
NASA XMM-Newton grants: NNG04GD27G, NNG05GM24G, which supported
various aspects of this work.  We also gratefully acknowledge the
financial support of grants: NAS8-39073, GO-09161.05-A (HST). PSS
acknowledges support from NASA/JPL contract 1256424. This publication
makes use of data products from the Two Micron All Sky Survey, which
is a joint project of the University of Massachusetts and the Infrared
Processing and Analysis Center/California Institute of Technology,
funded by the National Aeronautics and Space Administration and the
National Science Foundation. This research has also made use of the
NASA/IPAC Extragalactic Database (NED). SuperCOSMOS Sky Survey
material is based on photographic data originating from the UK,
Palomar and ESO Schmidt telescopes and is provided by the Wide-Field
Astronomy Unit, Institute for Astronomy, University of Edinburgh,
which is funded by the UK Particle Physics and Astronomy Research
Council.


\clearpage
\begin{figure}
\plotone{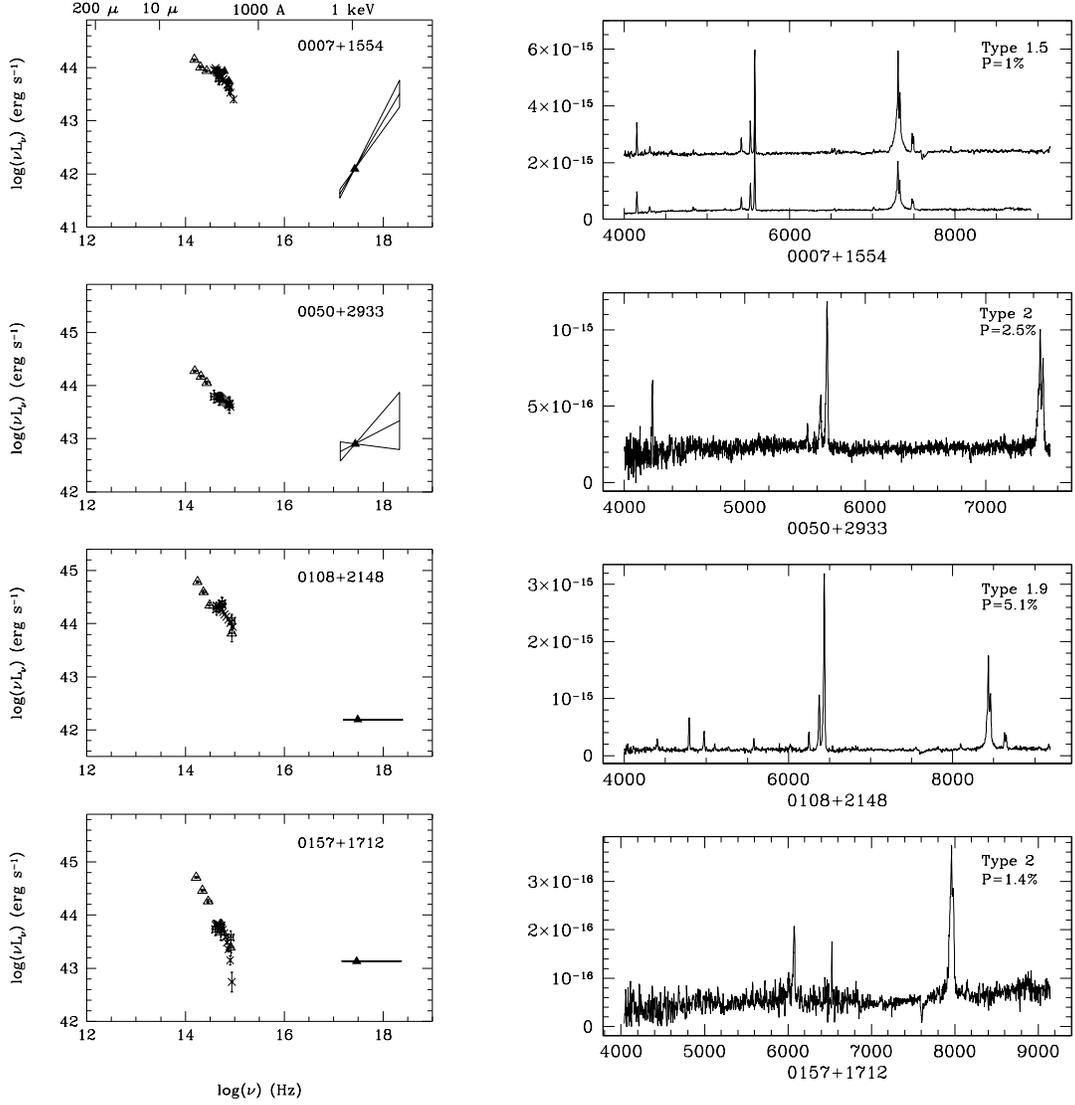}
\caption{{\it Left column} - Restframe Far--IR to X-ray spectral
energy distributions (SEDs) for AGN in our sample. The host galaxies
have not been subtracted. IRAS, 2MASS (J, H, K$_S$) and USNO--A2.0 (B,
R) photometry is indicated by open triangles, SDSS and Chandra data by
filled triangles, SuperCOSMOS photometry and HST spectrophotometry by stars,
optical spectrophotometry by crosses, ROSAT data by open
pentagons. {\it Right column} - Optical spectra on a $F_{\lambda}$
(erg~s$^{-1}$~cm$^{-2}$~\AA$^{-1}$) vs $\lambda$ (\AA) scale. If two
spectra are present for one object then spectrum ``a'' from
Table~4 is greyshifted and plotted above spectrum
``b''.}
\label{fig:spectra_SEDs_all}
\end{figure}

\begin{figure}[t]
\plotone{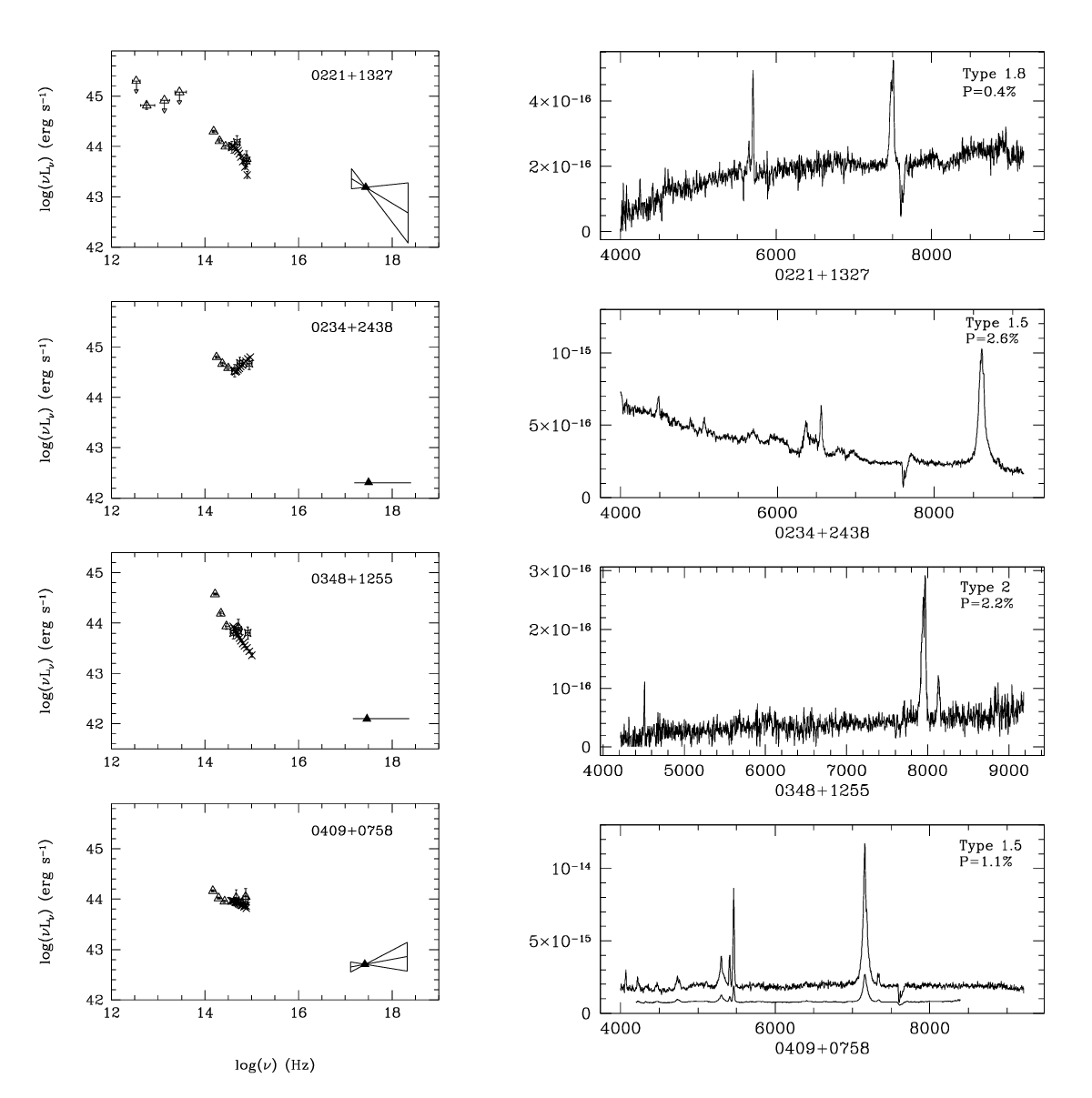}
\setcounter{figure}{0}
\caption{--continued}
\end{figure}

\begin{figure}[t]
\plotone{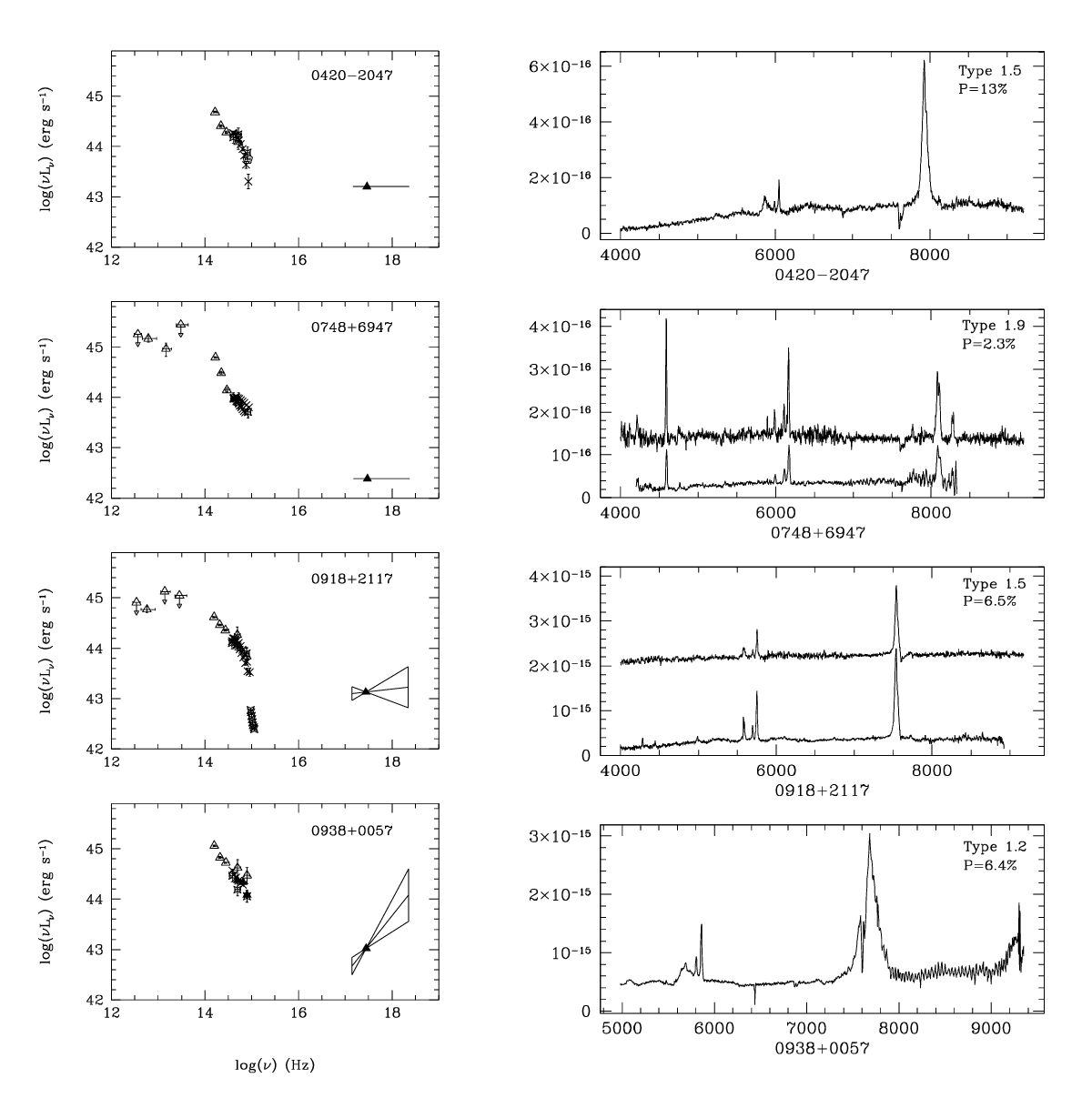}
\setcounter{figure}{0}
\caption{--continued}
\end{figure}

\begin{figure}[t]
\plotone{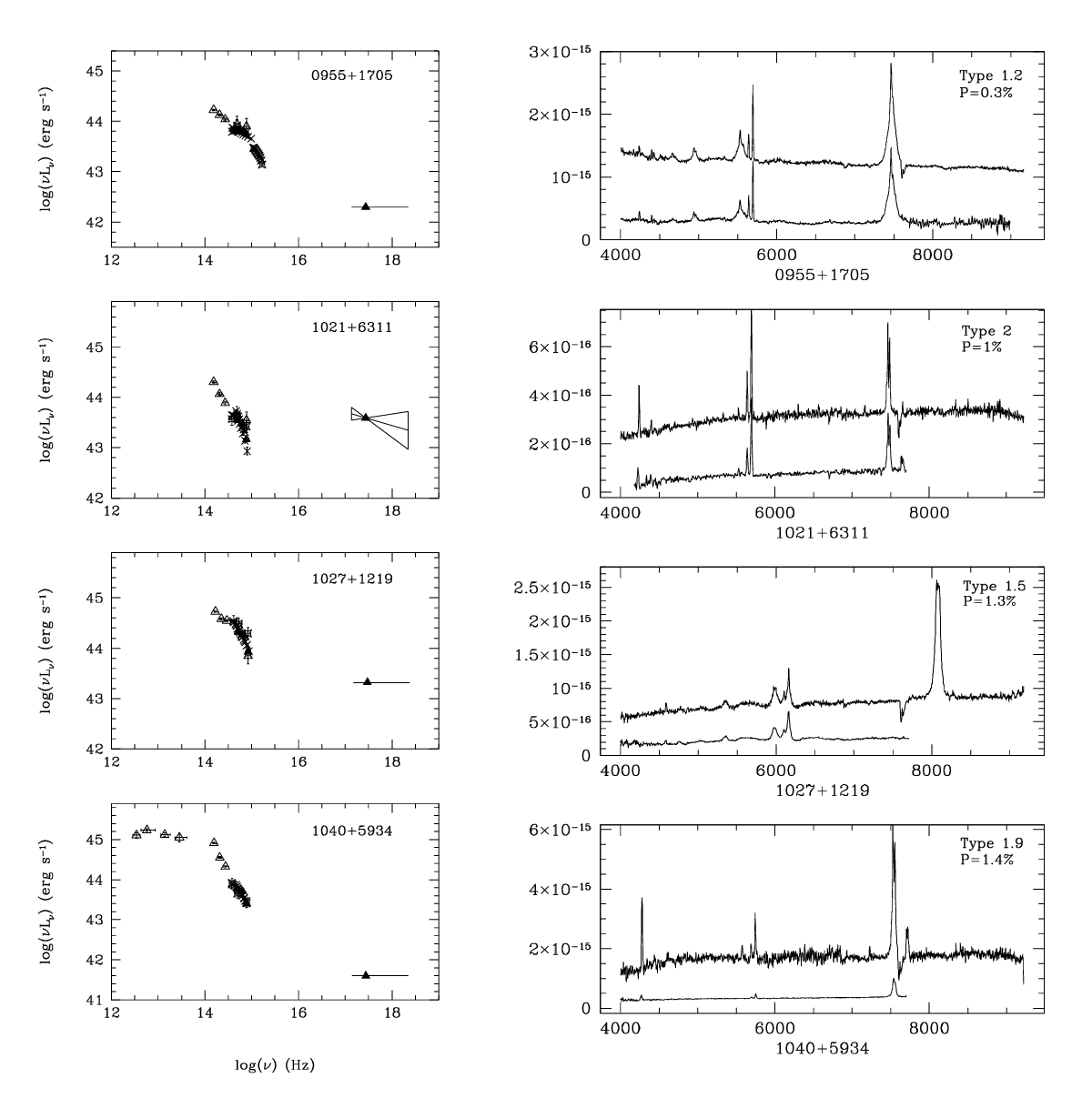}
\setcounter{figure}{0}
\caption{--continued}
\end{figure}

\begin{figure}[t]
\plotone{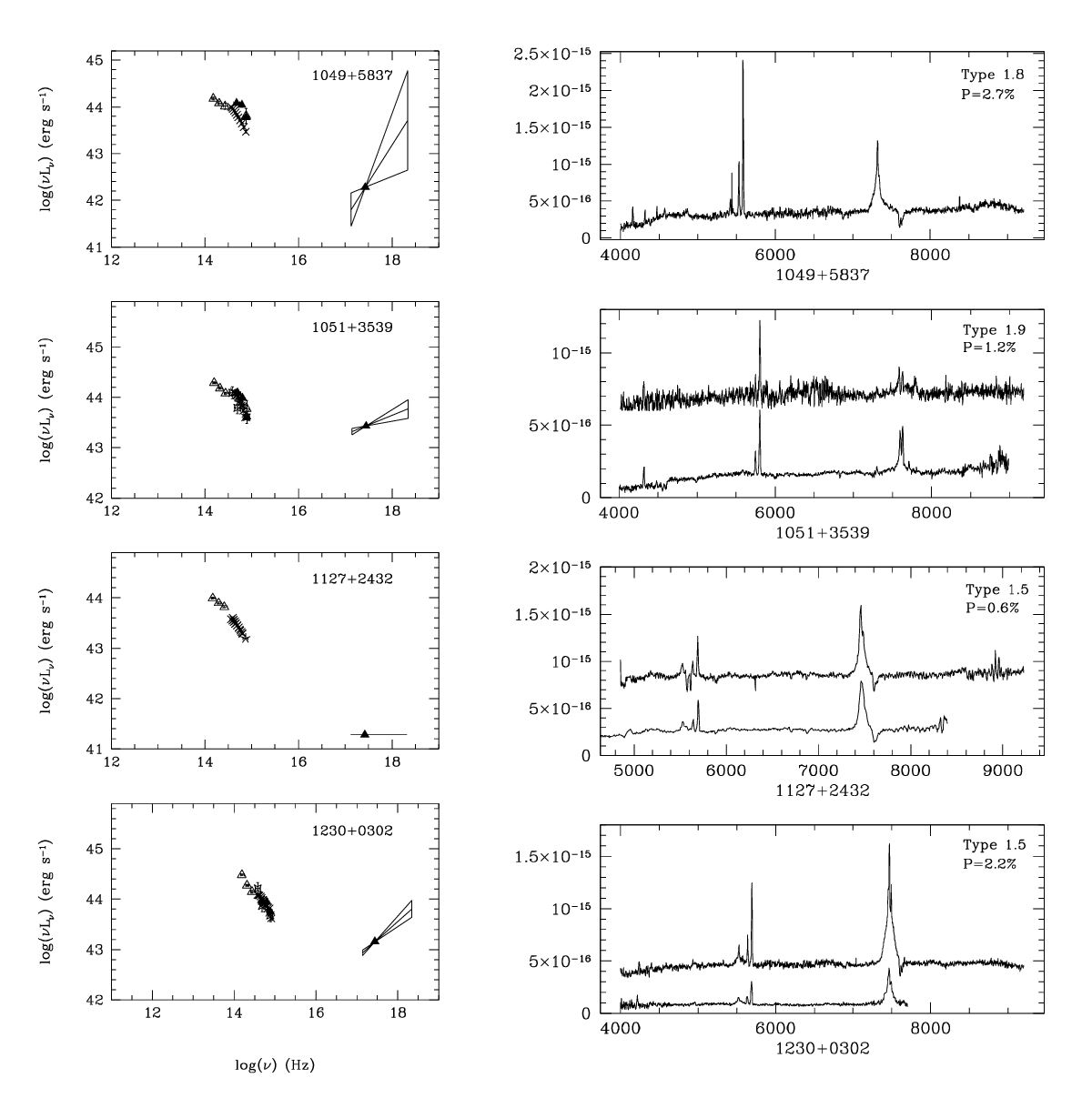}
\setcounter{figure}{0}
\caption{--continued}
\end{figure}

\begin{figure}[t]
\plotone{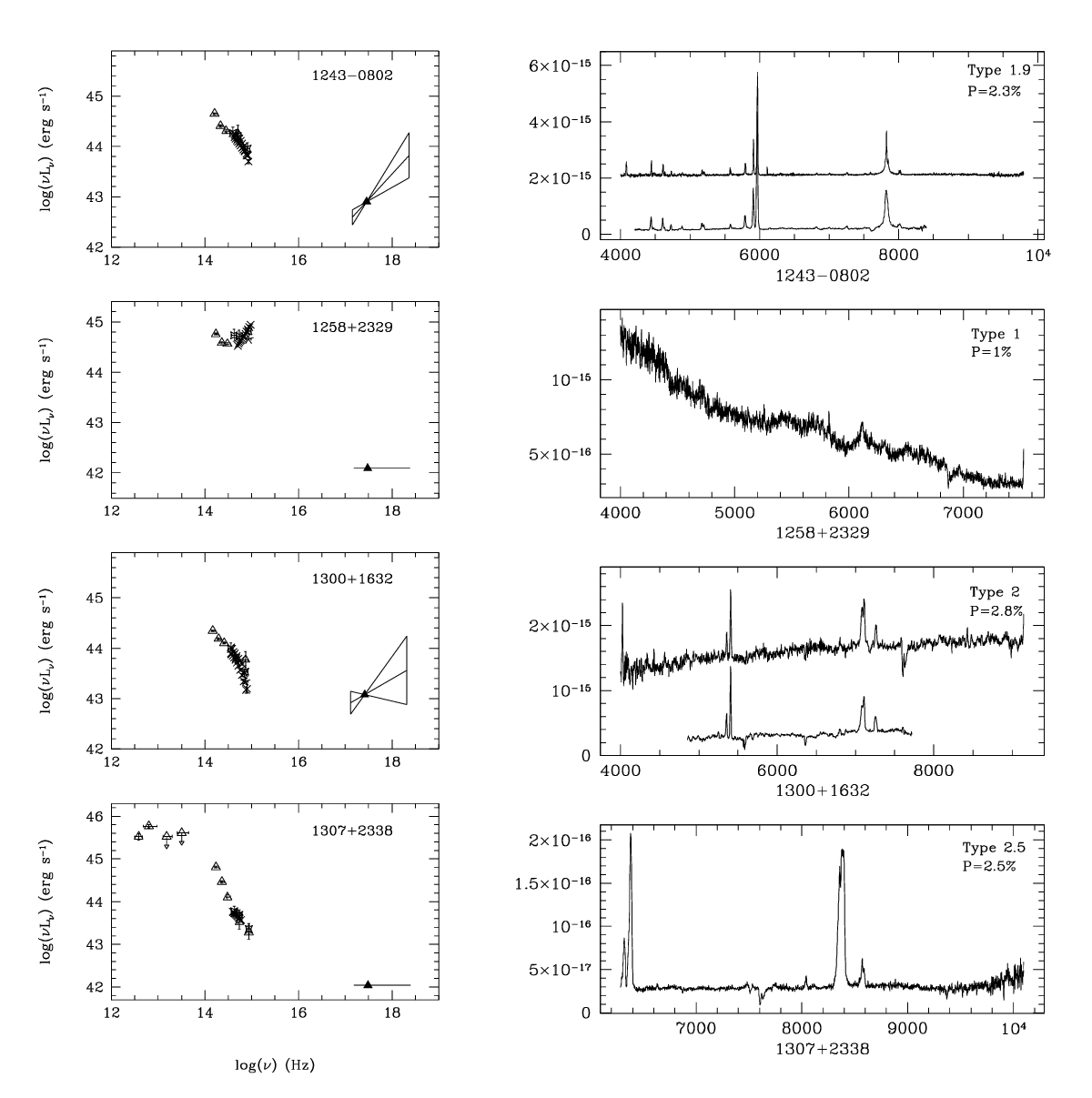}
\setcounter{figure}{0}
\caption{--continued}
\end{figure}

\begin{figure}[t]
\plotone{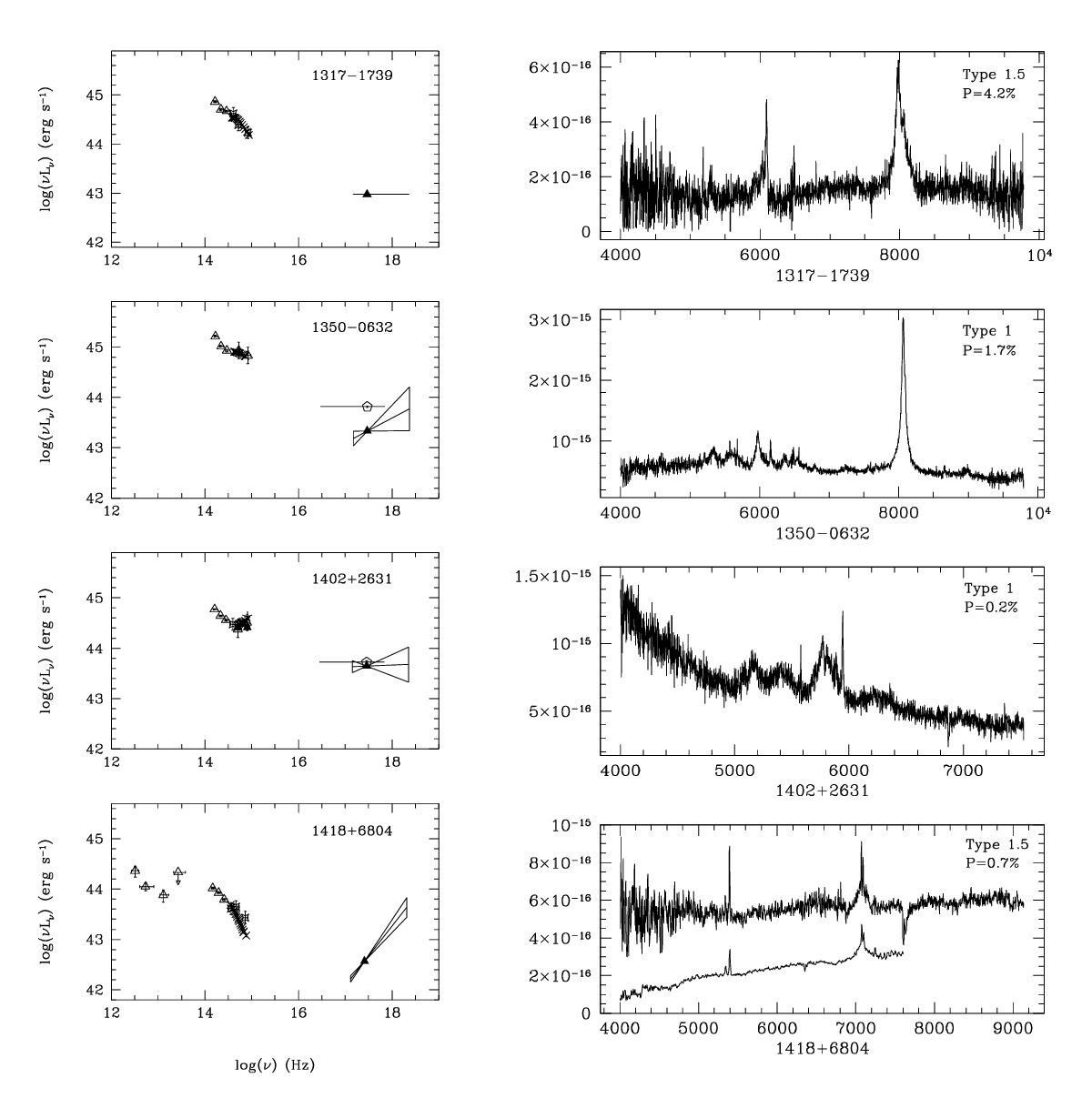}
\setcounter{figure}{0}
\caption{--continued}
\end{figure}

\begin{figure}[t]
\plotone{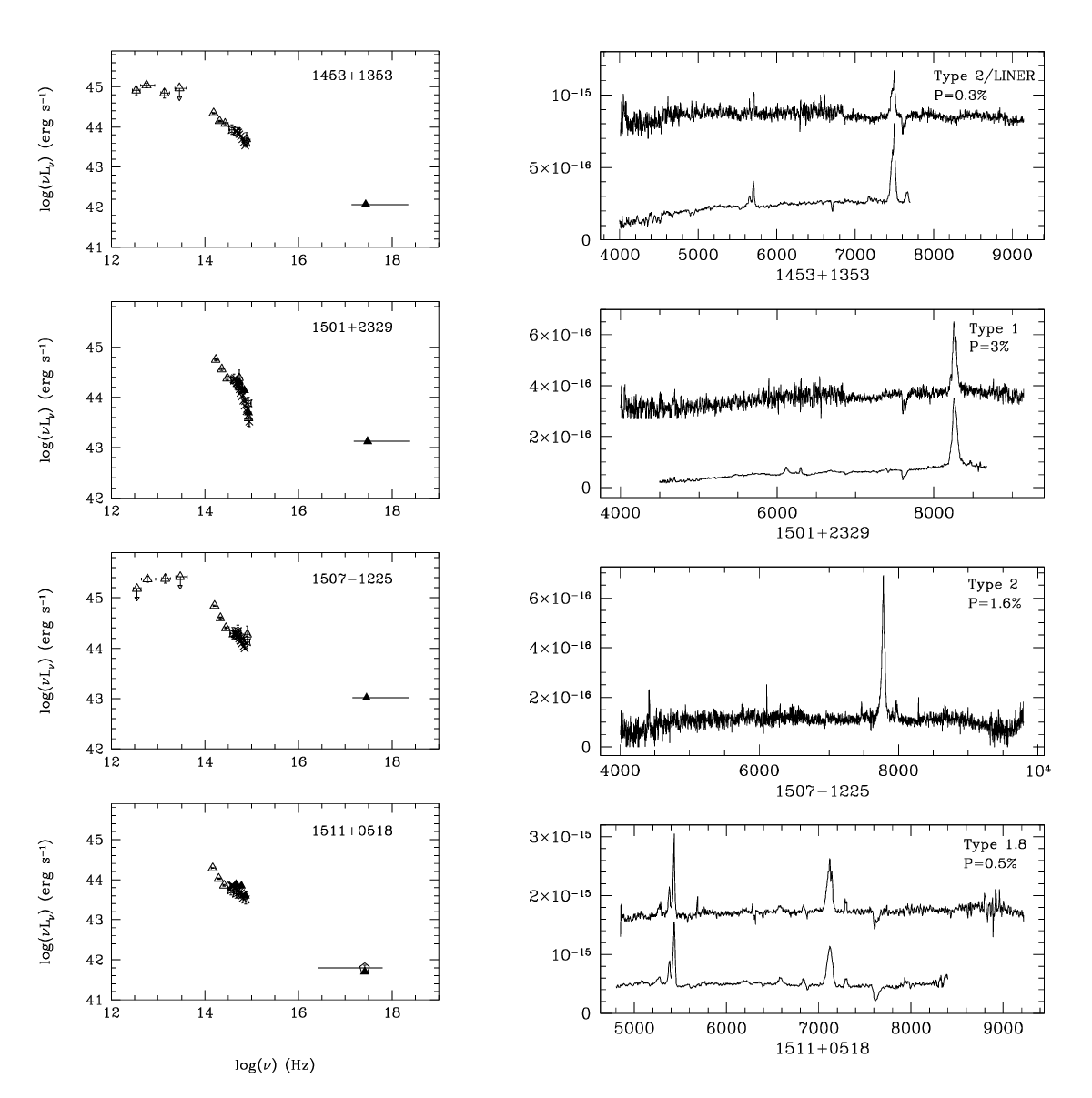}
\setcounter{figure}{0}
\caption{--continued}
\end{figure}

\begin{figure}[t]
\plotone{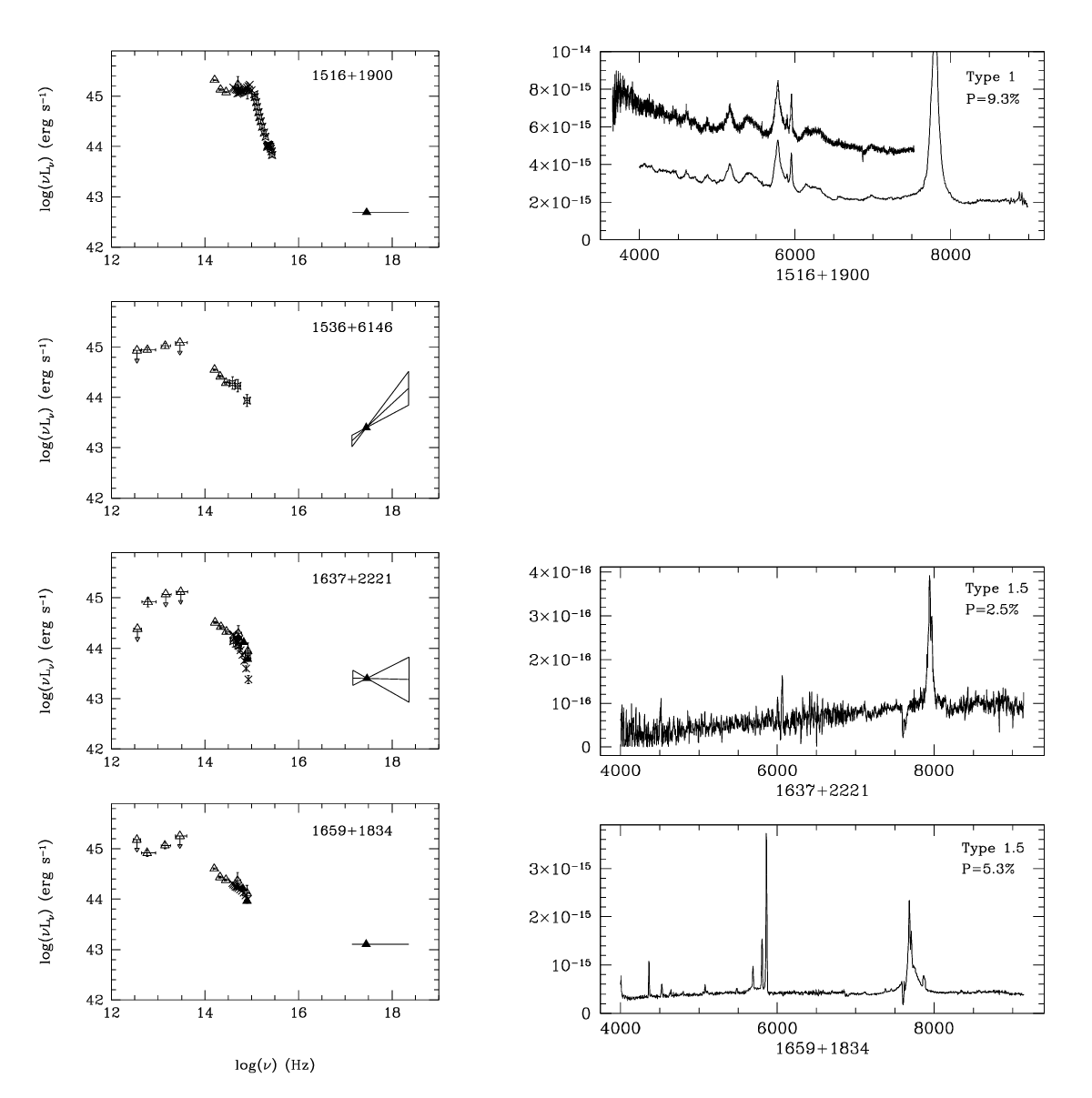}
\setcounter{figure}{0}
\caption{--continued}
\end{figure}

\begin{figure}[t]
\plotone{f1j.eps}
\setcounter{figure}{0}
\caption{--continued}
\end{figure}

\begin{figure}[t]
\plotone{f1k.eps}
\setcounter{figure}{0}
\caption{--continued}
\end{figure}

\clearpage
\begin{figure}[t]
\plotone{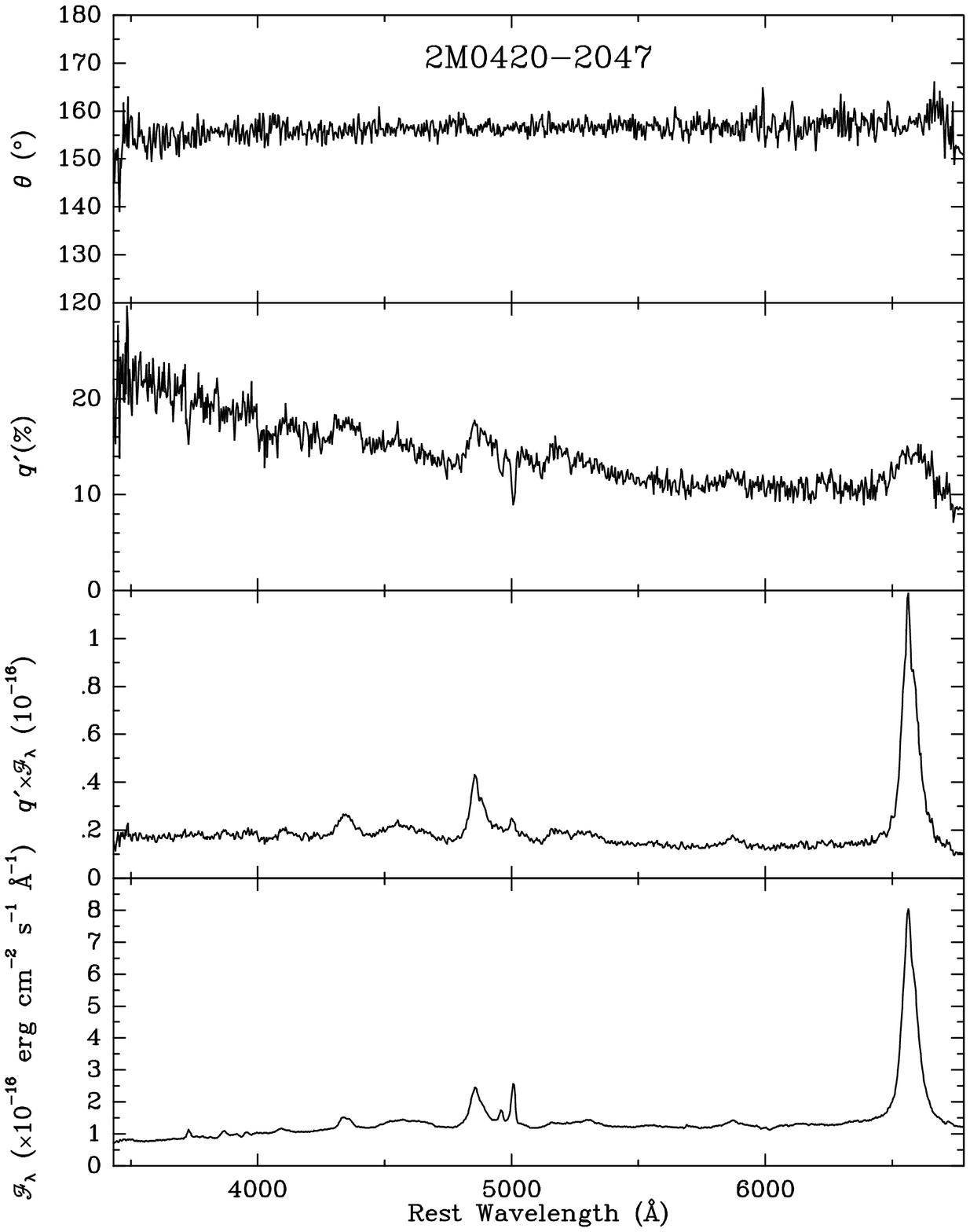}
\caption{Optical Spectropolarimetry for 2MASSJ04203206$-$2047592,
  2MASSJ09384445+0057156, 2MASSJ13170436$-$1739126,
  2MASSJ13503735$-$0632153, and 2MASSJ15362773+6146417. The
  panels show from top to bottom, the position angle of polarization
  ($\theta$), the percentage polarization ($q'$), the polarized flux
  density ($q' \times F_{\lambda}$), and total flux density
  ($F_{\lambda}$).}
\label{fig:specpol}
\end{figure}

\begin{figure}[t]
\plotone{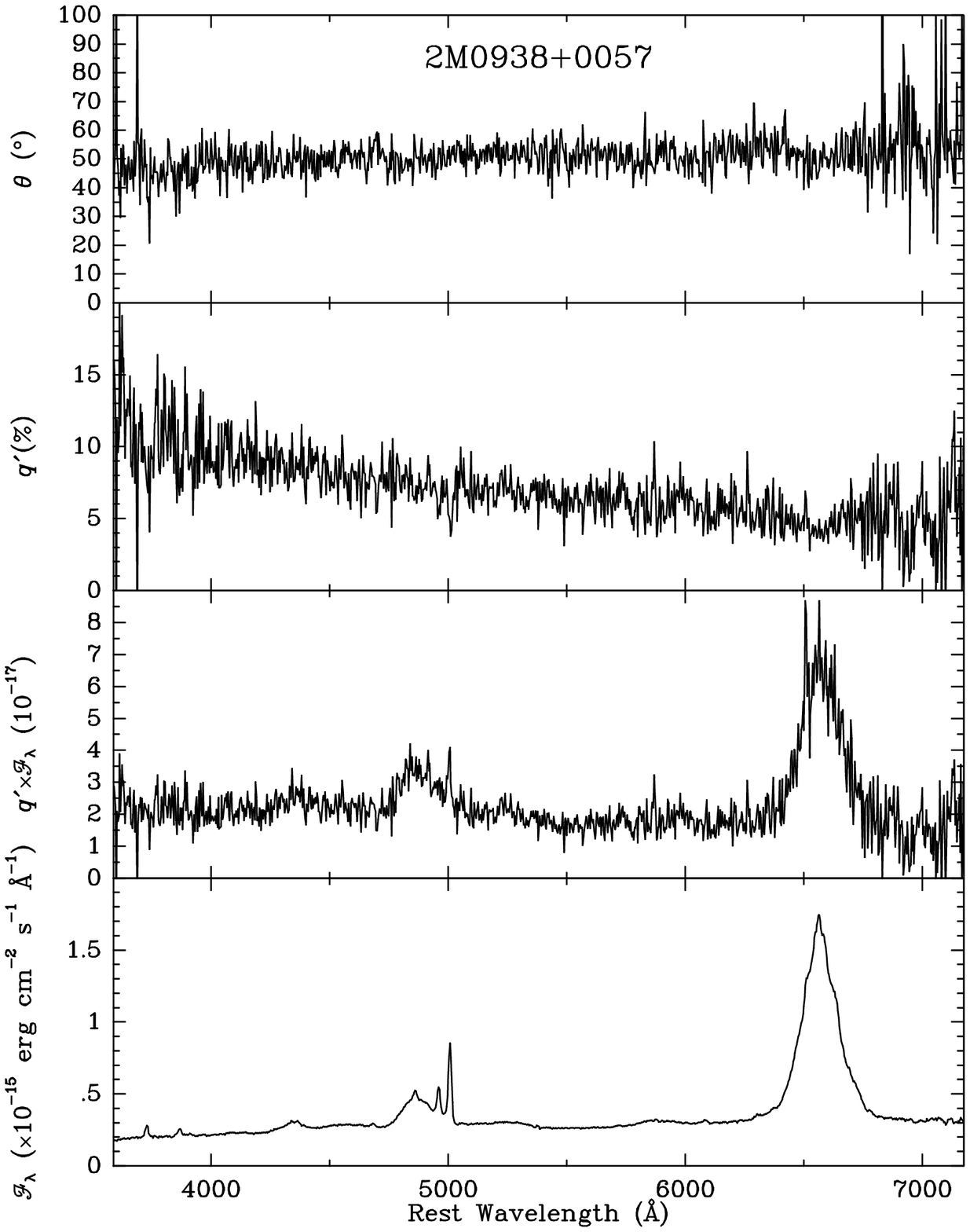}
\setcounter{figure}{1}
\caption{--continued}
\end{figure}

\begin{figure}[t]
\plotone{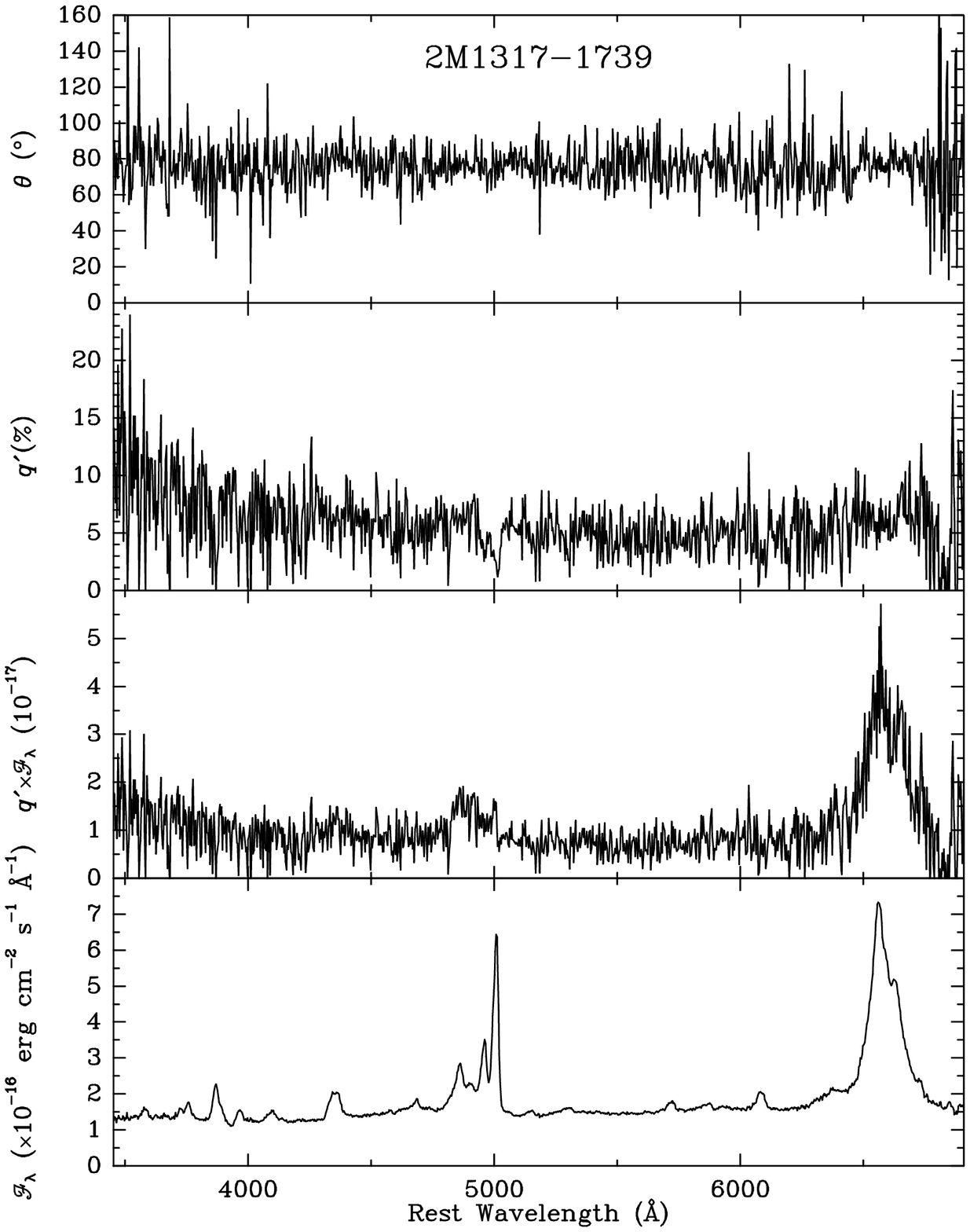}
\setcounter{figure}{1}
\caption{--continued}
\end{figure}

\begin{figure}[t]
\plotone{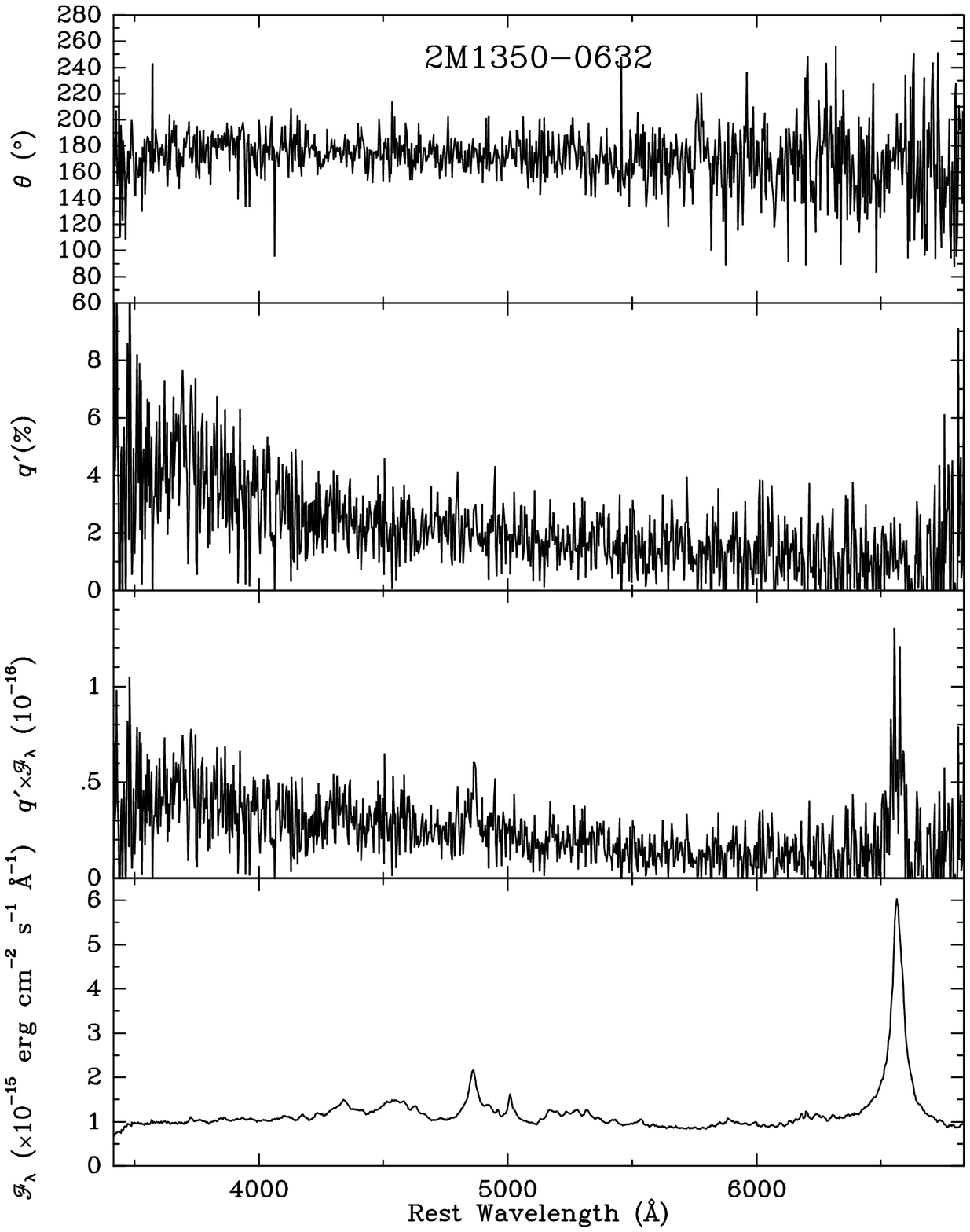}
\setcounter{figure}{1}
\caption{--continued}
\end{figure}

\begin{figure}[t]
\plotone{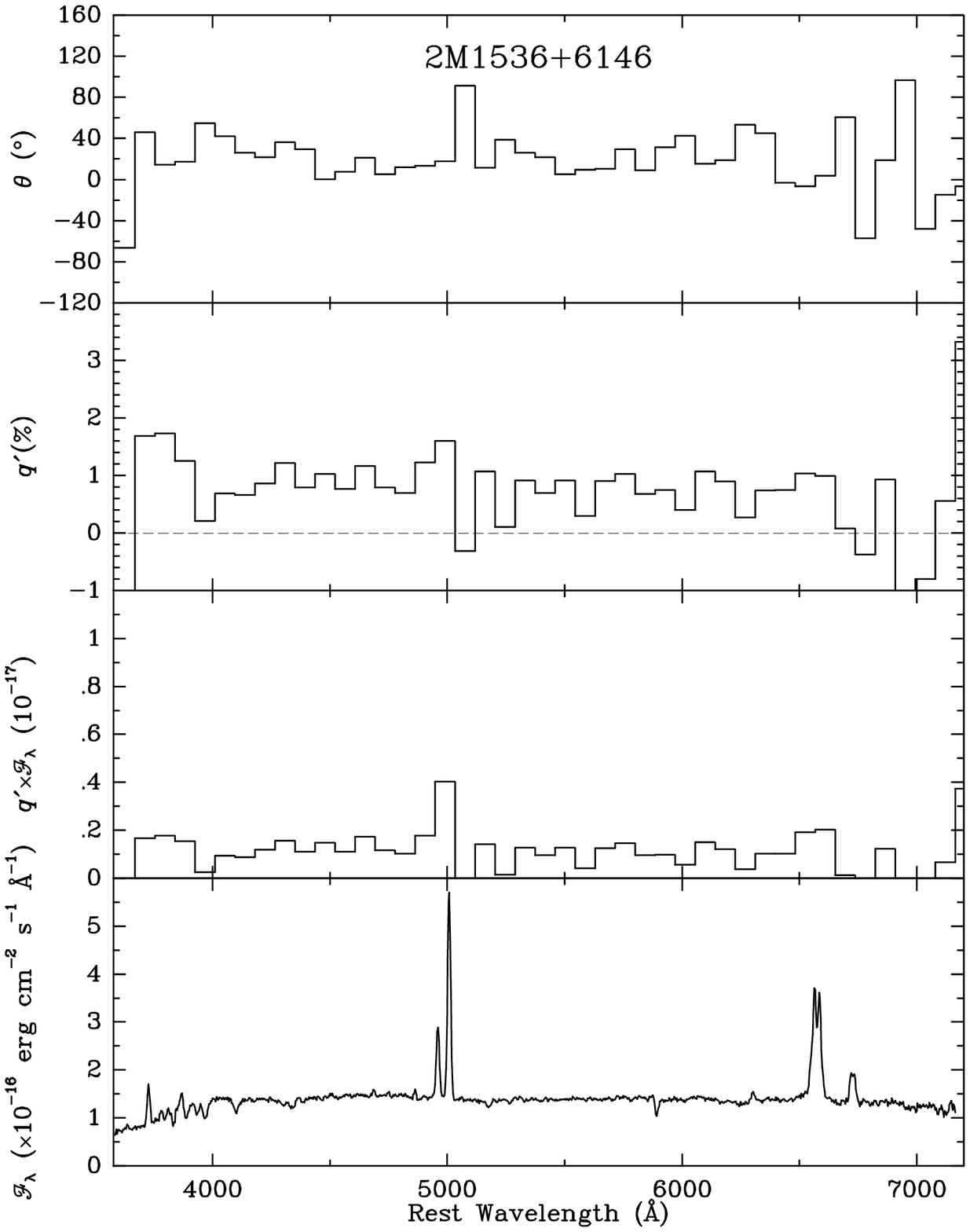}
\setcounter{figure}{1}
\caption{--continued}
\end{figure}

\clearpage
\begin{figure}
\plottwo{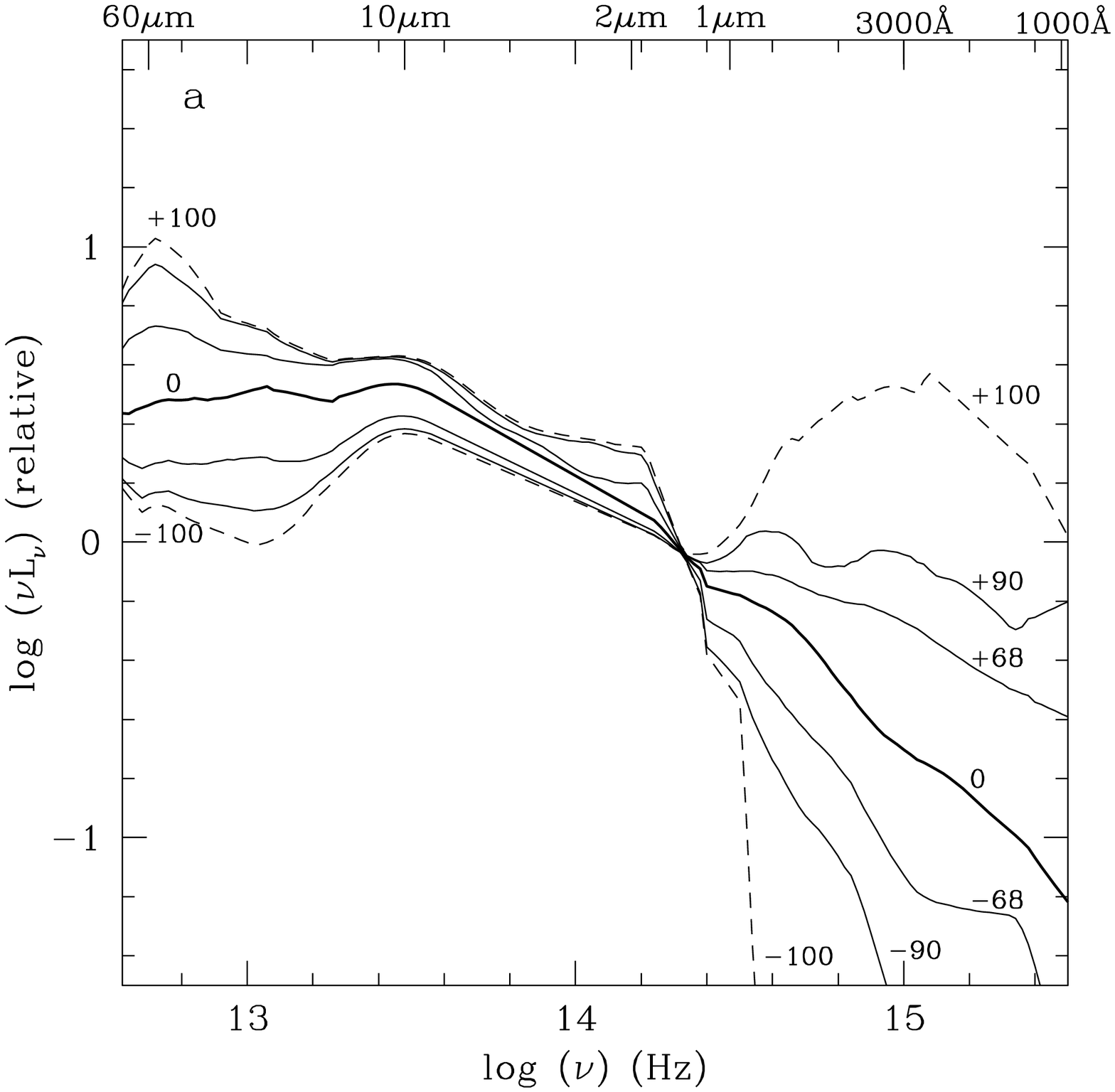}{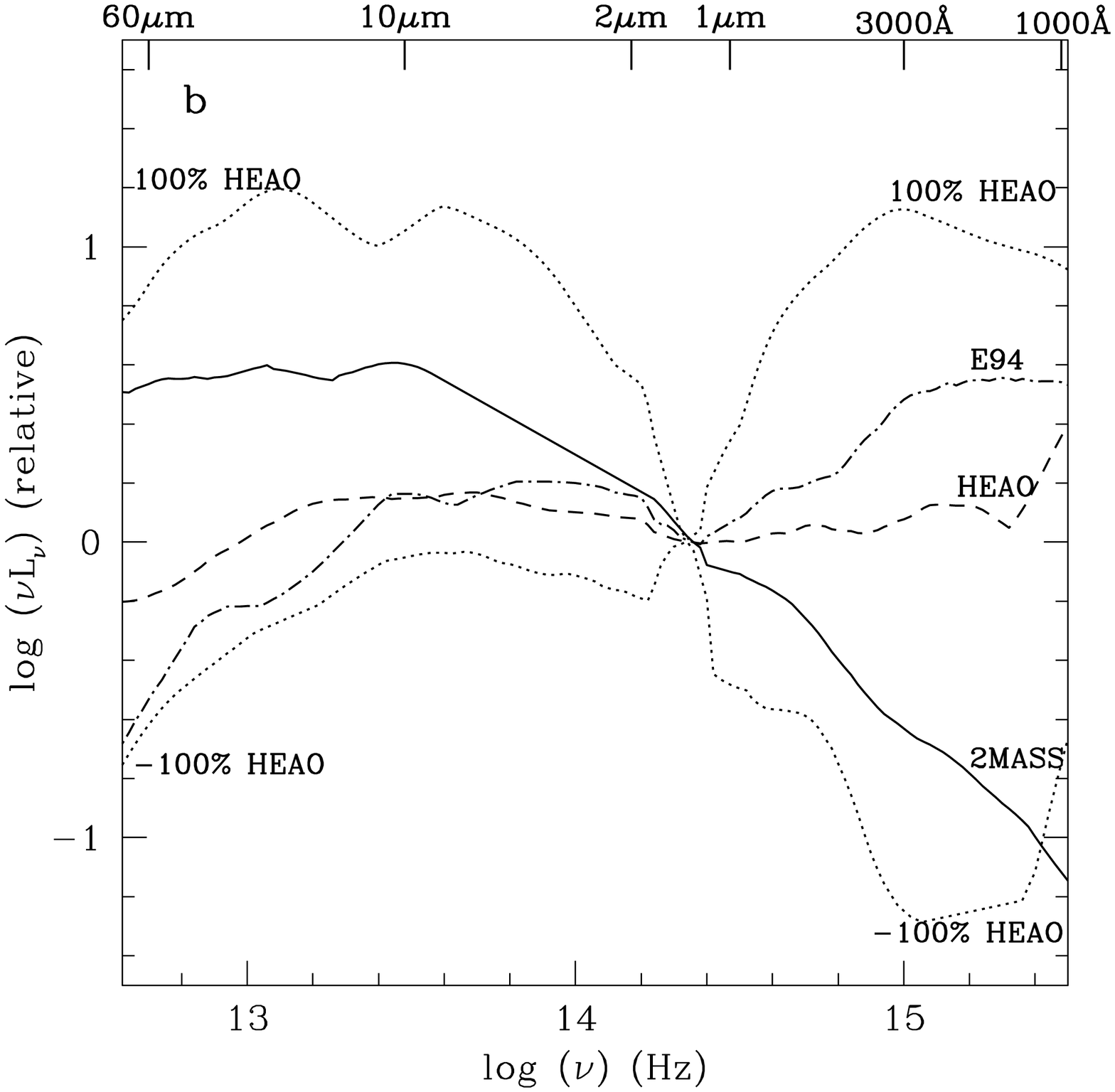}
\caption{a) The median energy distributions for the red 2MASS AGN
  sample normalized at 1.5~$\mu$m along with the 68, 90, and 100
  (dashed line) Kaplan-Meier percentile envelopes. b) Comparison
  between medians normalized at 1.5~$\mu$m of the red 2MASS AGN sample
  (solid line), hard-X-ray selected HEAO AGN sample from Kuraszkiewicz
  \etal\ (2003) (dashed line) and optical/radio selected Einstein QSO
  sample from Elvis \etal\ (1994) (dot-dash line). The dotted lines
  show the Kaplan-Meier 100 percentile envelopes for the HEAO sample.}
\label{fig:median}
\end{figure}

\clearpage
\begin{figure}
\plotone{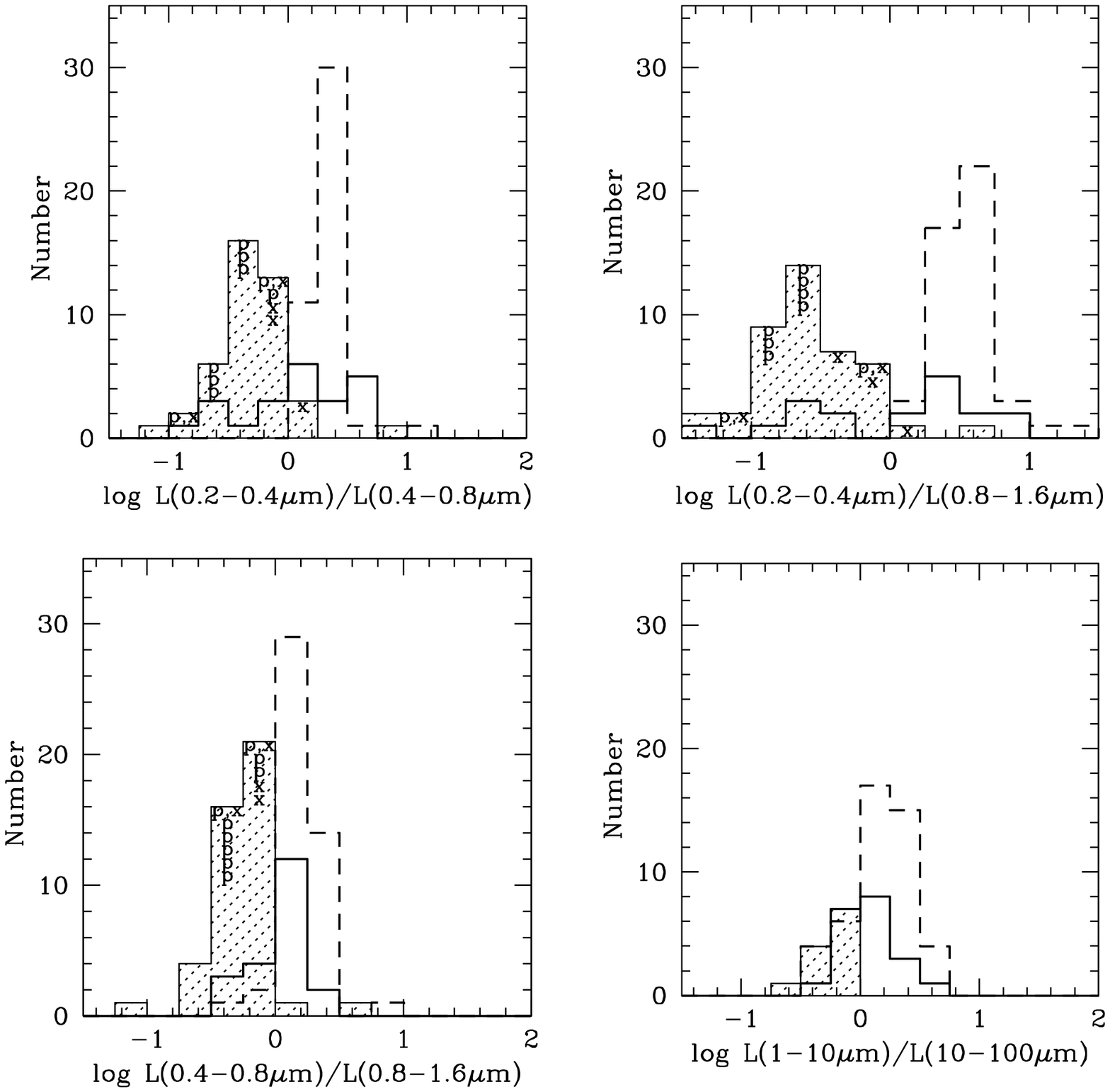}
\caption{Histograms of the optical, UV and IR luminosity ratios. The
shaded area: current 2MASS AGN sample; thick solid line: the
hard-X-ray selected HEAO AGN sample of Kuraszkiewicz \etal\ (2003);
and the dashed line: the optical/radio selected Einstein QSO sample
from Elvis et al. (1994). Objects with HST spectra are marked by ``x''
and objects with high polarization ($P>3$\%) with ``p''.}
\label{fig:histogram_L}
\end{figure}

\clearpage
\begin{figure}
\plotone{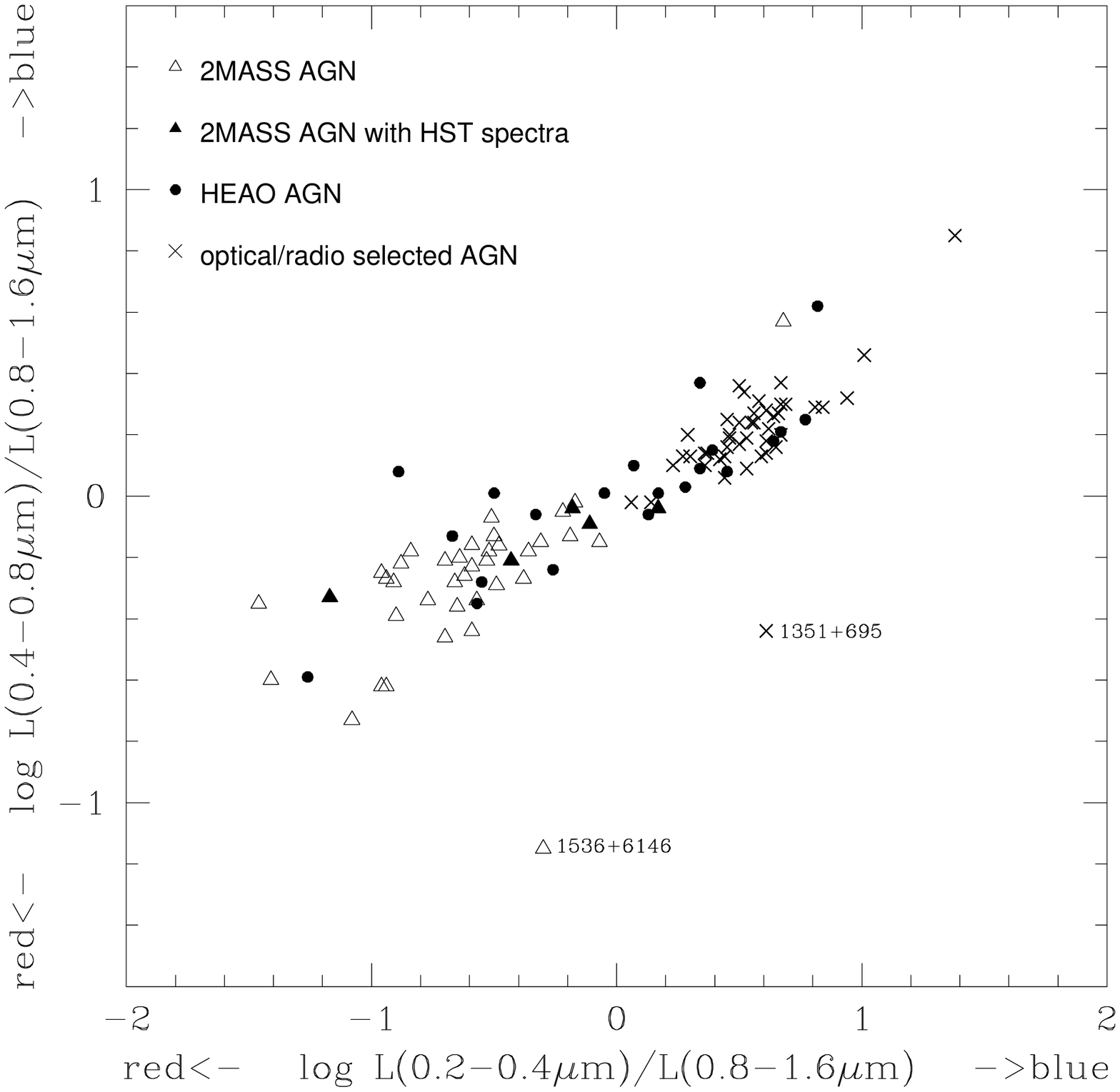}
\caption{The L(0.4-0.8$\mu$m)/L(0.8-1.6$\mu$m) versus
L(0.2-0.4$\mu$m)/L(0.8-1.6$\mu$m) color-color diagram.  2MASS AGN are
indicated by triangles (with those having HST spectra being filled
triangles), hard-X-ray selected HEAO AGN (Kuraszkiewicz \etal\ 2003) by
filled circles, and optical/radio selected AGN (Elvis et
al. 1994) with crosses.}
\label{fig:opttwo_color}
\end{figure}

\clearpage
\begin{figure}
\plotone{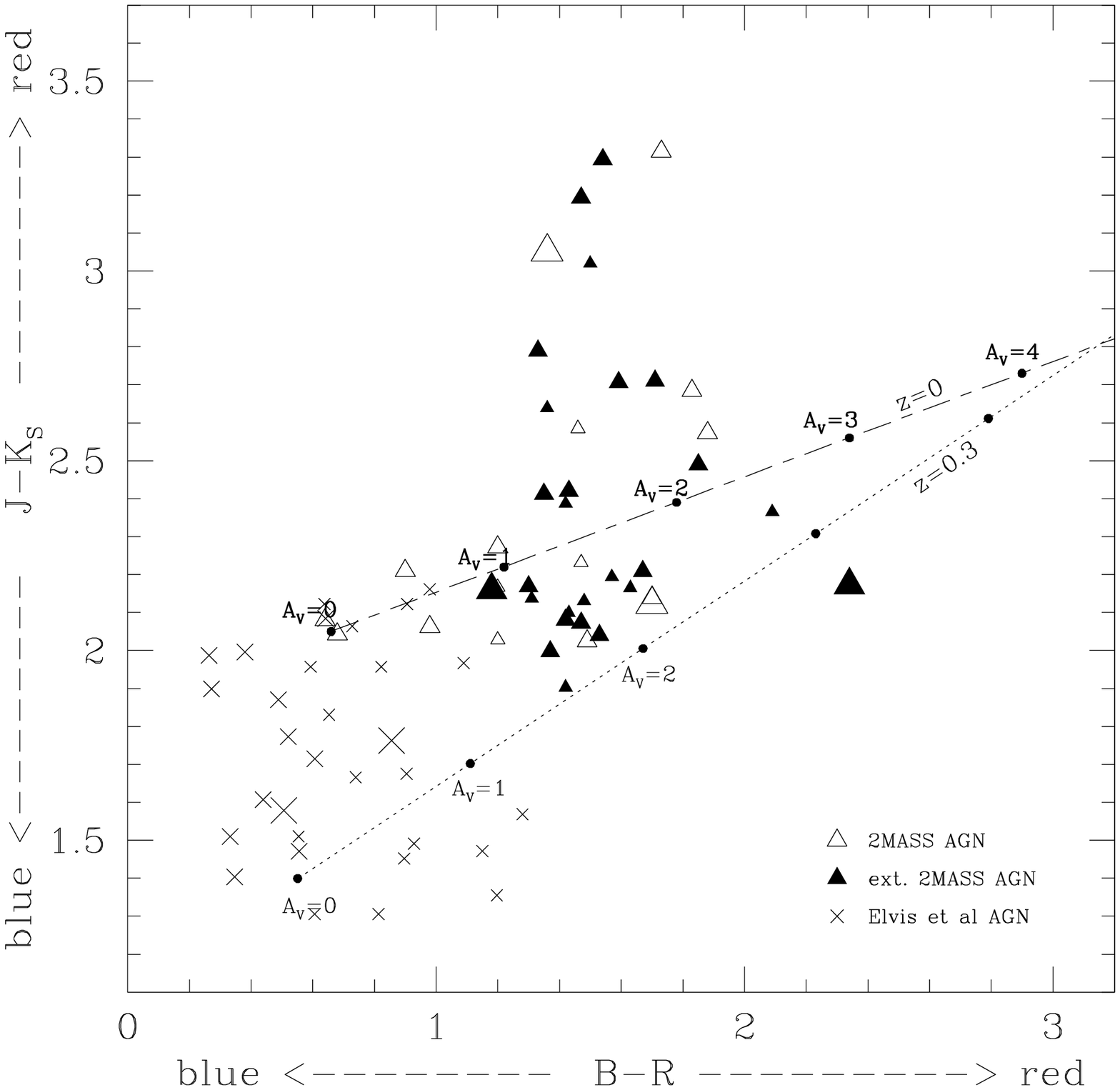}
\caption{Observed \jmk\ color versus observed B$-$R color
diagram. Triangles denote 2MASS objects, where filled triangles are
optically extended sources on B and R plates, and open triangles are
point sources. Crosses denote objects in the Elvis \etal\ (1994)
sample (colors not corrected for redshift or host galaxy; J$-$K
converted to \jmk\ using Bessell 2005). The size of the triangles and
crosses is proportional to redshift (smallest symbols are for
$z\le0.15$, medium symbols for $0.15<z\le0.3$, and largest symbols for
$0.3<z<0.4$).  The long-dash-short-dash line shows colors of the
median blue optical/radio selected AGN SED (Elvis \etal\ 1994; z=0,
corrected for host galaxy) at z=0 reddened by dust with $A_V$ changing
from 0 to 4~mags.  The dotted line shows colors of this median at
z=0.3 ($\simeq$ highest redshift in our red 2MASS sample).}
\label{fig:two_color}
\end{figure}

\clearpage
\begin{figure}
\epsscale{0.9}
\plotone{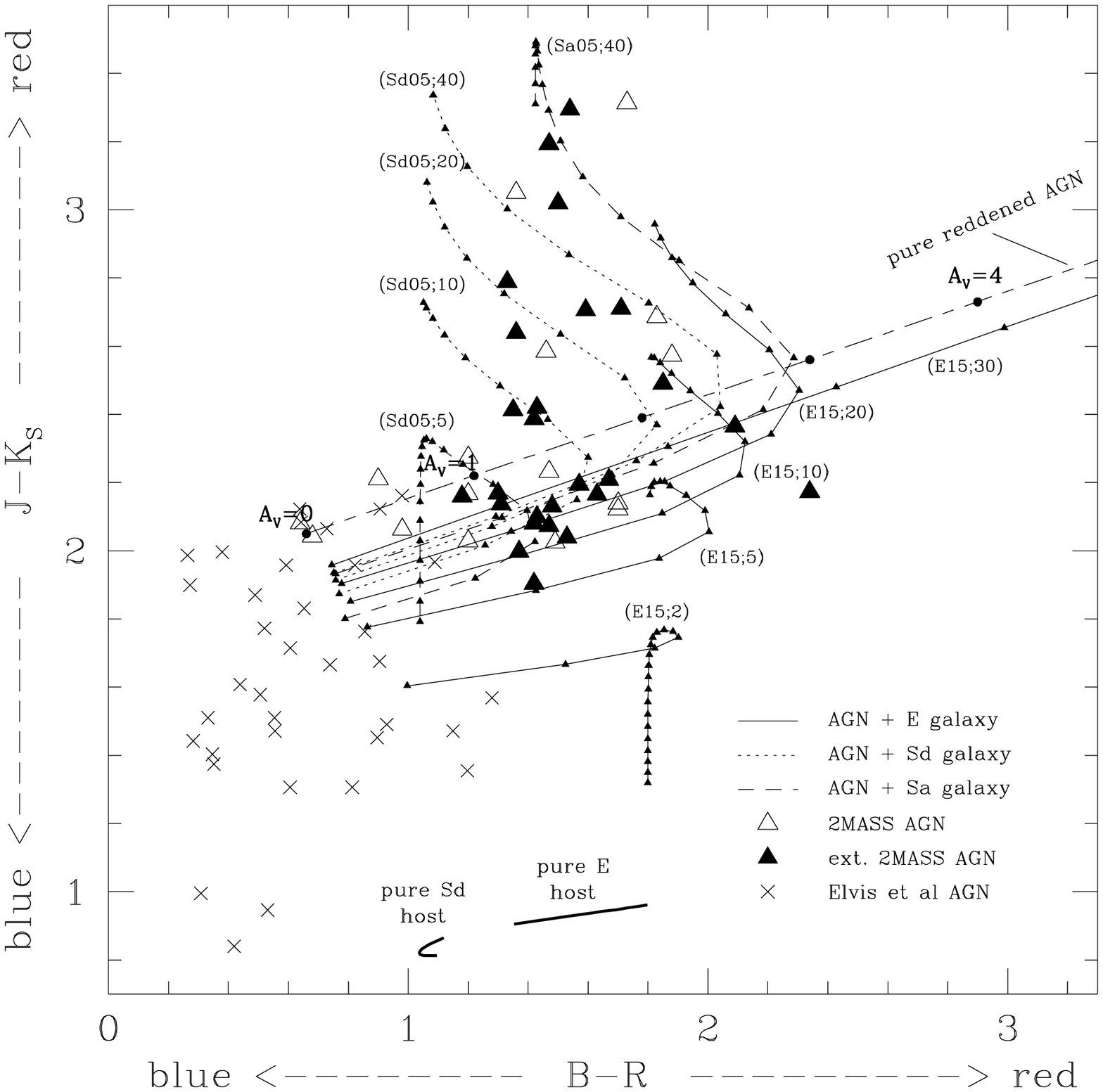}
\caption{Colors of a reddened AGN combined with contributions from a
host galaxy (z=0). Solid lines represent AGN with an elliptical host
with a 15--Gyr stellar population (reddest in B$-$R color), dotted
lines an AGN with a Sd host with a 5--Gyr stellar population (bluest
in B$-$R), and dashed line and AGN with a Sa host with a 5--Gyr
stellar population (intermediate B$-$R color). Each curve starts at
the colors of an unreddened ($A_V$=0) AGN + host galaxy (bluest
optical/IR colors) and extends to the reddened $A_V$=10~mag AGN + host
galaxy in steps of 1~mag.  denoted by small triangles on the curves.
Three curves: (E15;2), (Sd05;5) and (Sa05;40) have been extended to
$A_V=20$.  Numbers in parenthesis give the host galaxy type and age of
stars followed by the intrinsic, unreddened AGN/host galaxy flux ratio
at R band (e.g. (E15;20) is a E host with 15--Gyr stars and AGN/host
flux ratio = 20). Thick solid curves indicate pure host galaxy (E and
Sd) colors changing with star age (from 1 to 15--Gyr). Triangles and
crosses represent the observed colors of red 2MASS and Elvis \etal\
(1994) AGN samples respectively.}
\label{fig:two_color_host}
\end{figure}

\clearpage
\begin{figure}
\epsscale{0.9}
\plotone{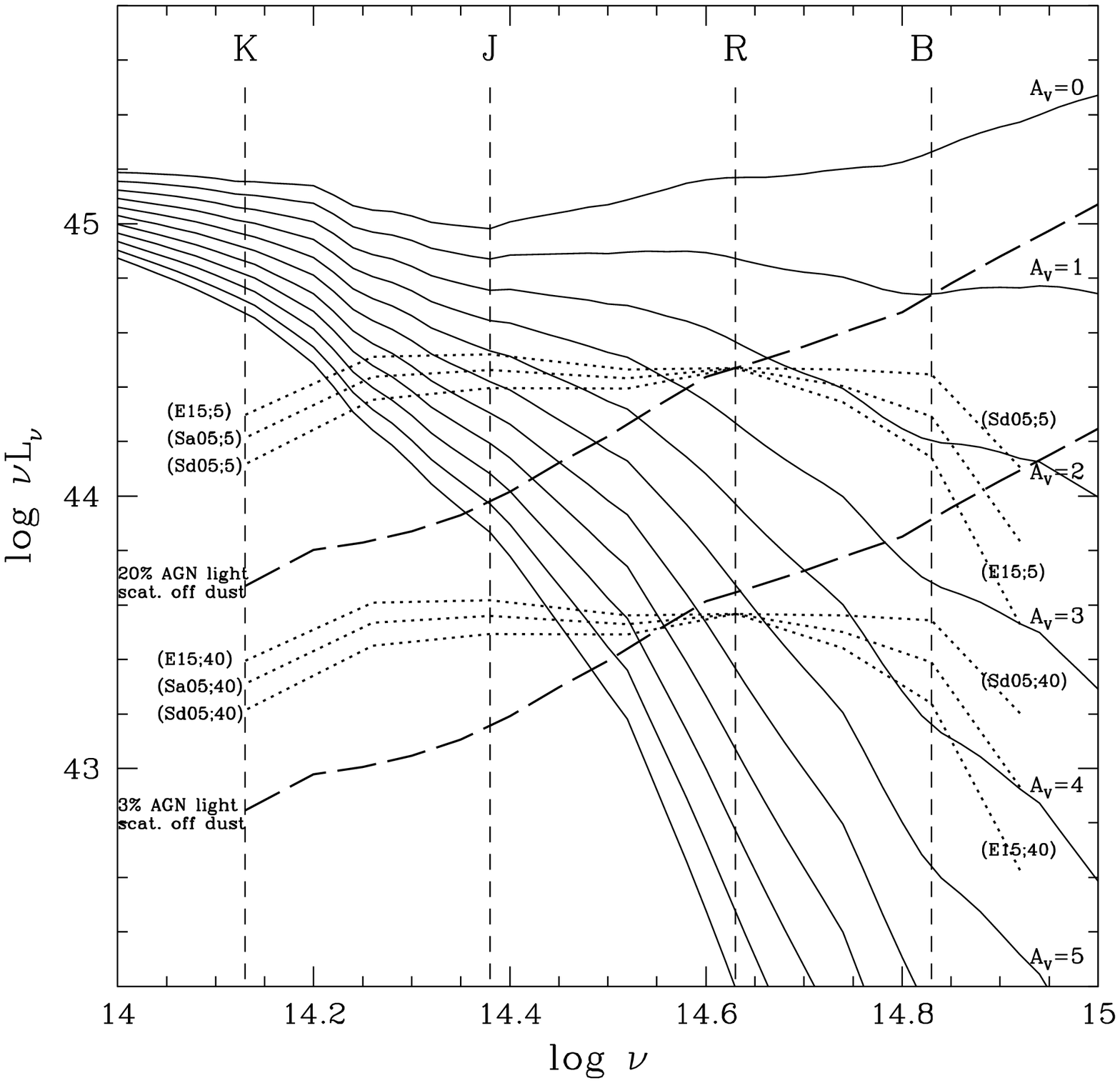}
\caption{SEDs of a pure AGN (Elvis \etal\ 1994 median SED) reddened by
  $A_V$ ranging from 0 to 10~mags (solid lines). Dotted lines show
  SEDs of three host galaxy templates from Buzzoni \etal\ (2005;
  elliptical galaxy with a 15~Gyr stellar population, and Sd and Sa
  galaxies with a 5~Gyr stellar population) normalized in the R band
  to be 5 times and 40 times (top three and bottom three dotted curves
  respectively) weaker than the AGN. Long-dashed lines show SED of
  nuclear AGN light scattered off dust normalized to 3\% (lower dashed
  line) and 20\% (upper dashed line) in the R band. The frequency of
  the J,K,R and B band effective wavelengths are indicated by
  short-dashed lines. SEDs are at z=0.}
\label{fig:colors_SED}
\end{figure}

\clearpage
\begin{figure}
\epsscale{0.85}
\plotone{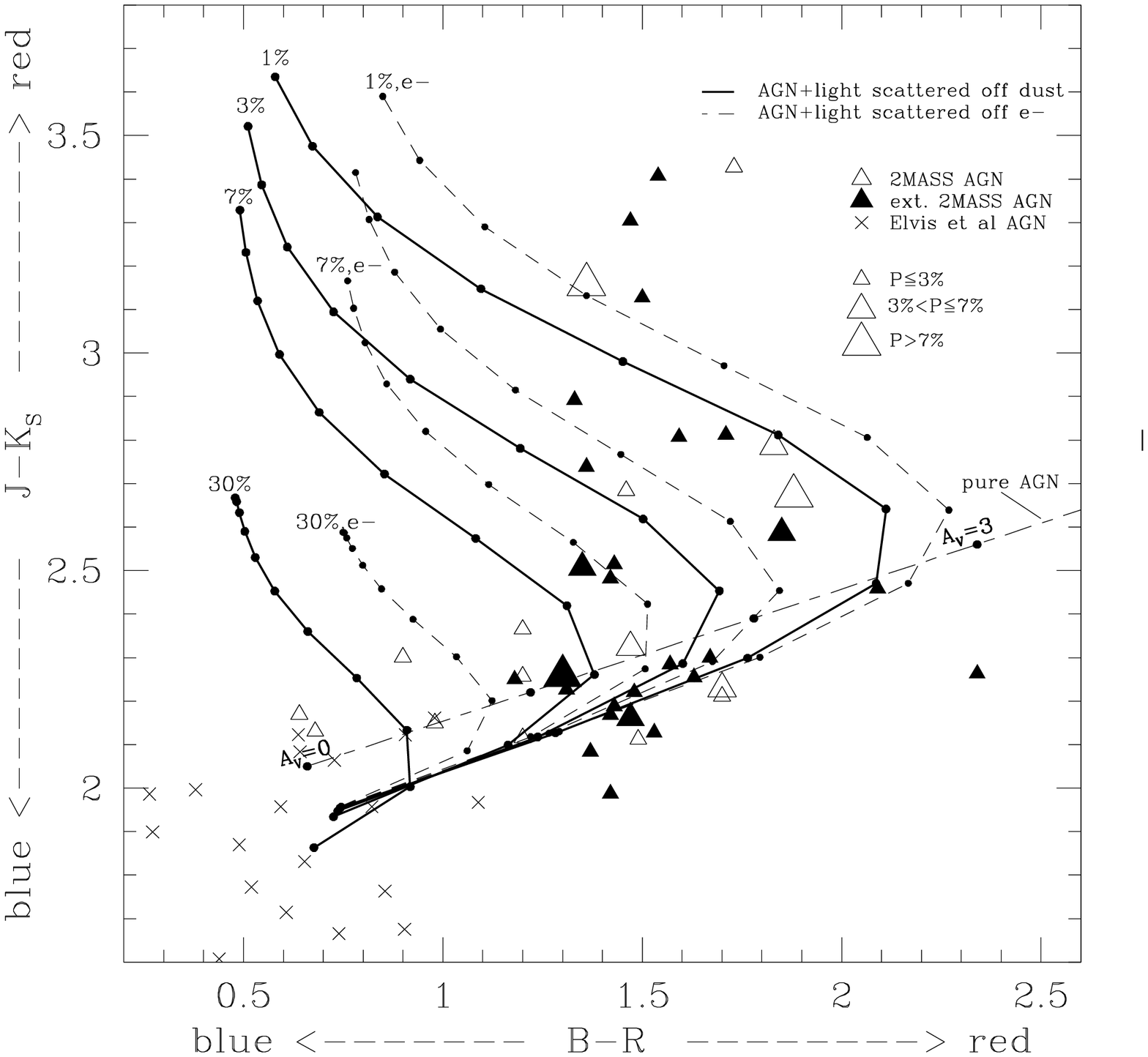}
\caption{The effect of adding scattered AGN light to a reddened AGN
  continuum on the optical/near-IR colors (z=0). The
  long-dash-short-dash line shows the colors of a pure reddened AGN
  (small dots are at $A_V$=0,1,2,3~mag.).  Thick solid curves show the
  effect of adding to this reddened AGN continuum (Elvis \etal\ 1994
  median AGN SED reddened by dust with $A_V$ ranging from 0 to
  10~mag. -- loci represented by small circles) an unreddened AGN
  continuum (Elvis \etal\ 1994 median SED) scattered off dust at
  scattering angle $\theta=90^{o}$.  Dashed lines show the same but
  with scattering off electrons at $\theta=90^{o}$ (i.e. scattering is
  independent of wavelength). Scattered intrinsic AGN light at levels
  1\%, 3\%, 7\%, and 30\% in R band were added, which translates to
  1\%, 3\%, 6.5\%, and 23\% scattered light relative to the total
  observed (AGN+scattered) flux at R band, if the observed AGN is not
  reddened, or to a higher scattered light contribution if the
  observed AGN is reddened. Each 2MASS source is represented by a
  triangle, with size proportional to the amount of polarization
  measured at R band.  Extended sources are denoted by filled
  triangles. Triangles and crosses represent the observed colors of
  red 2MASS and Elvis \etal\ (1994) AGN samples respectively.}
\label{fig:two_color_polarization}
\end{figure}

\clearpage
\begin{figure}
\plotone{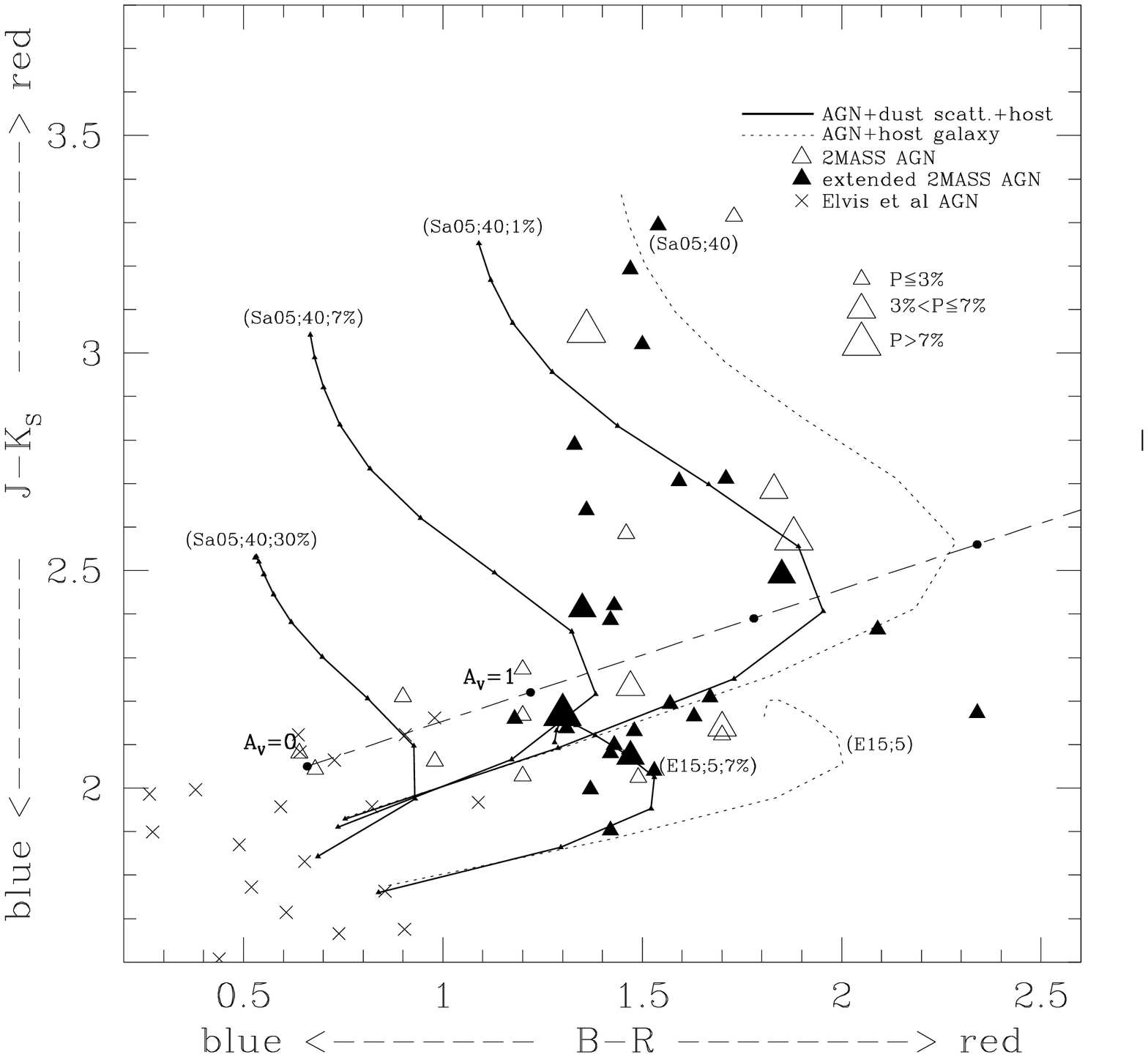}
\caption{The effect of adding host galaxy and scattered AGN light to a
  reddened AGN continuum on the optical/near-IR colors (z=0). Solid
  curves represent the colors for a reddened AGN ($A_V$ changing from
  0 to 10~mag. -- small triangles at loci) + host galaxy (Sa spiral
  with 5--Gyr stars and elliptical with 15--Gyr stars: Sa05 and E15
  respectively) + unreddened AGN light scattered off dust
  (1\%,7\%,30\% of unreddend Elvis \etal\ 1994 median AGN SED added at
  R band). Dotted lines show AGN + host galaxy colors as in
  Fig.~\ref{fig:two_color_host}.  The long-dash-short-dash line
  represents a pure reddened AGN where dots are at
  $A_V$=0,1,2,3~mag. loci.}
\label{fig:two_color_host_polarization}
\end{figure}

\clearpage
\begin{figure}
\plotone{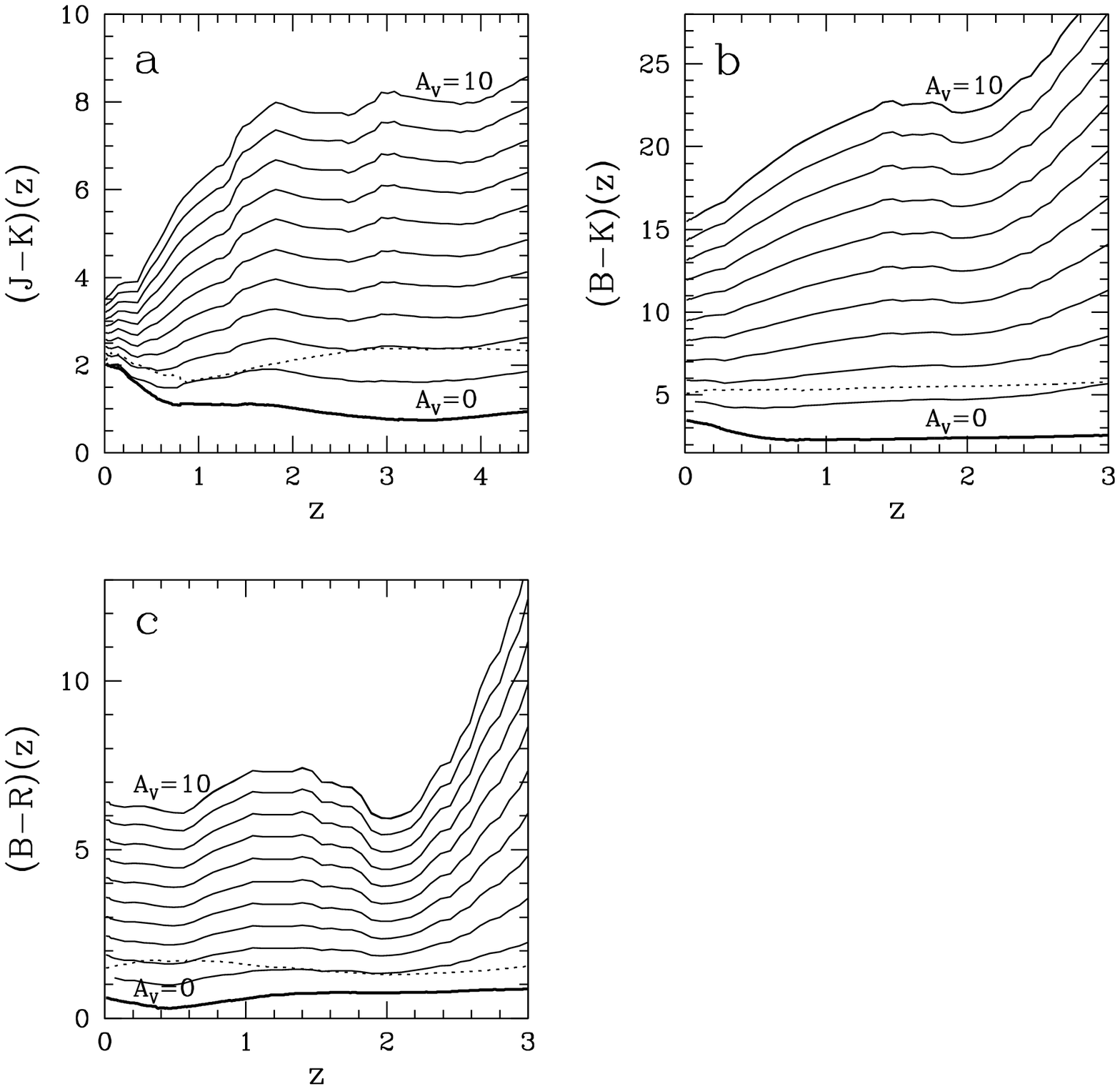}
\caption{J$-$K, B$-$K, and B$-$R observed color dependence on
redshift. The bottom solid and bold curve in a), b) and c) shows the redshift
dependence for the unreddened Elvis \etal\ (1994) median. Solid curves
above this curve show the redshift dependence of the reddened Elvis et
al. (1994) median where $A_V$ changes from 1 to 10~mag. The
dotted line in all figures shows the color redshift dependence when a
2MASS median is used ($A_V=0$).} 
\label{fig:JmK}
\end{figure}

\clearpage
\begin{figure}
\plottwo{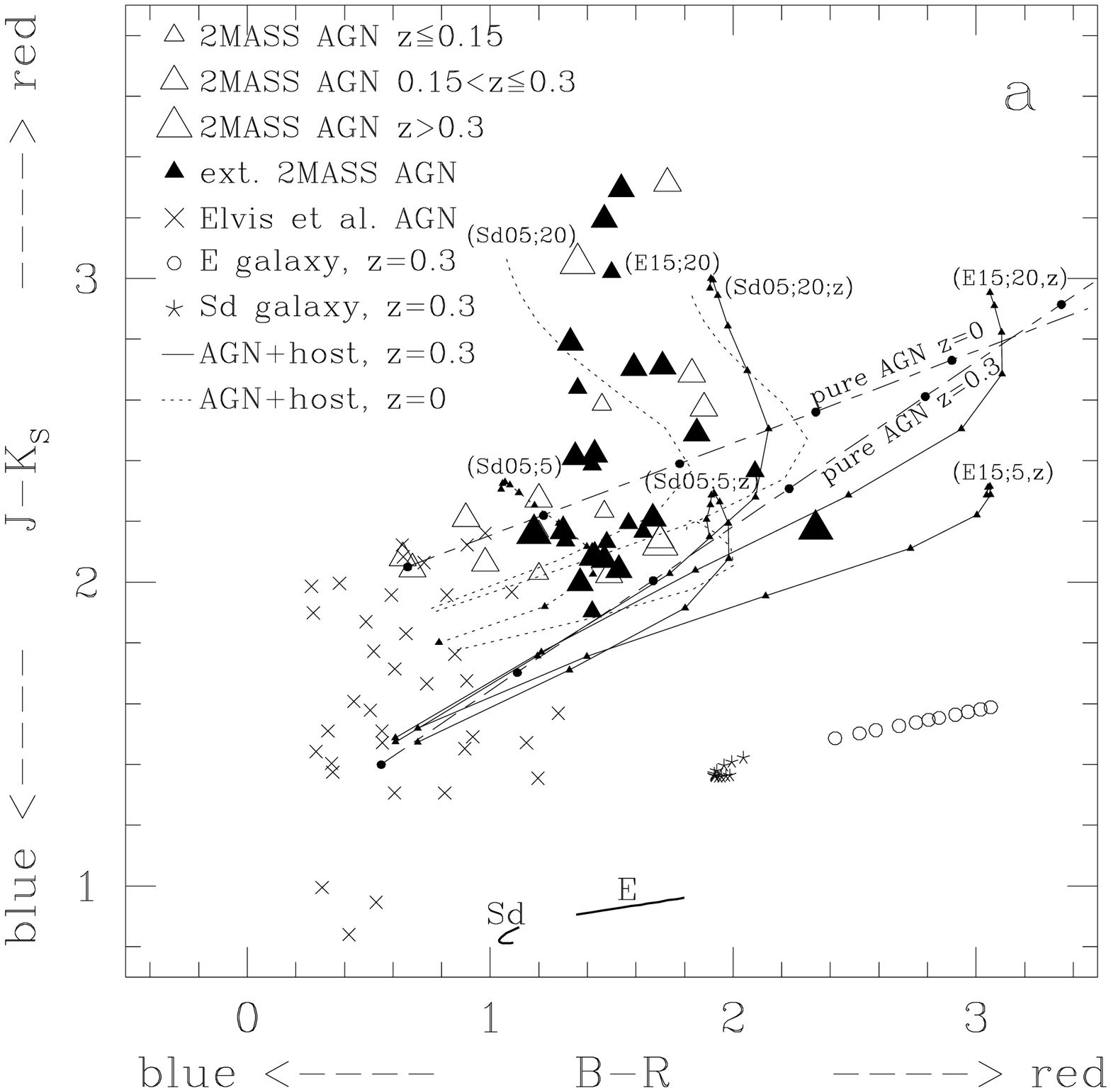}{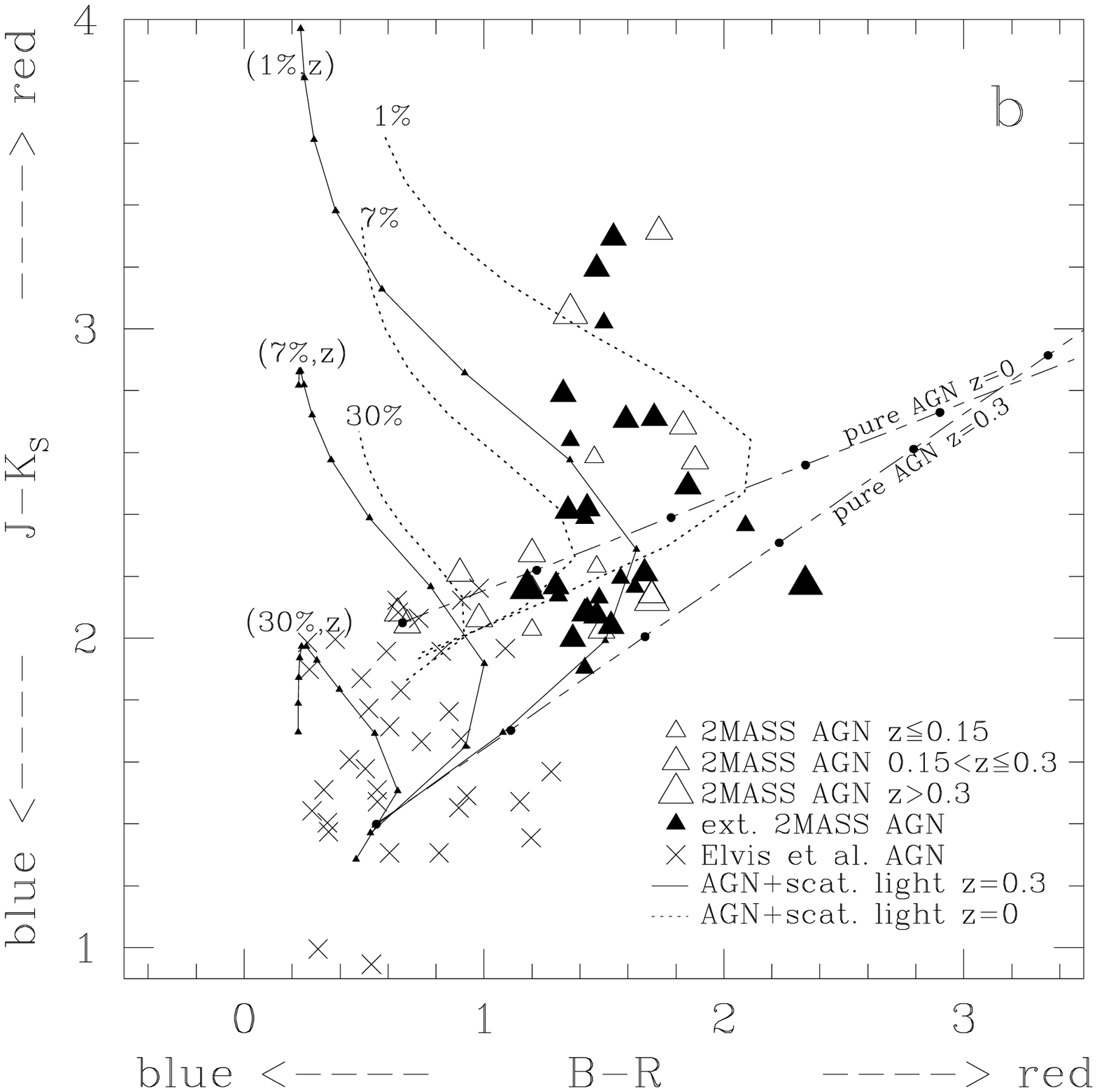}
\caption{a) Redshift dependence of the reddened AGN+host galaxy
   colors. Four reddened AGN + host galaxy curves were chosen from
   Fig.~\ref{fig:two_color_host}: ((E15;5), (E15;20), (Sd05;5) and
   (Sd05;20) (represented here by dotted lines) and redshifted to
   z=0.31 (solid lines). (Sd05;5;z) denotes colors of a reddened AGN
   ($A_V=0-10$~mag.) + Sd (5--Gyr stars) host galaxy at z=0.31, where
   the {\it restframe} R band intrinsic AGN to host galaxy ratio is 5.
   At the bottom of the figure, thick solid lines marked with ``Sd''
   and ``E'' show colors of a pure Sd and elliptical host galaxy at
   z=0. These colors redshifted to z=0.31 are marked as stars and
   circles respectively.  b) Redshift dependence of the reddened
   AGN+scattered light colors. Dotted lines are the reddened AGN +
   dust-scattered intrinsic AGN light curves from
   Fig.~\ref{fig:two_color_polarization}. Solid lines are the same
   curves (i.e. same 1\%,7\%,30\% normalization of scattered light at
   {\it restframe} R and $A_V$ changing from 0 to 10~mag) redshifted
   to z=0.31. In both figures the size of triangles is proportional to
   the redshift of the 2MASS AGN. Filled triangles are extended
   sources, open triangles are point sources.}
\label{fig:two_color_host_z_and_polarization_z}
\end{figure}

\clearpage
\begin{figure}
\plotone{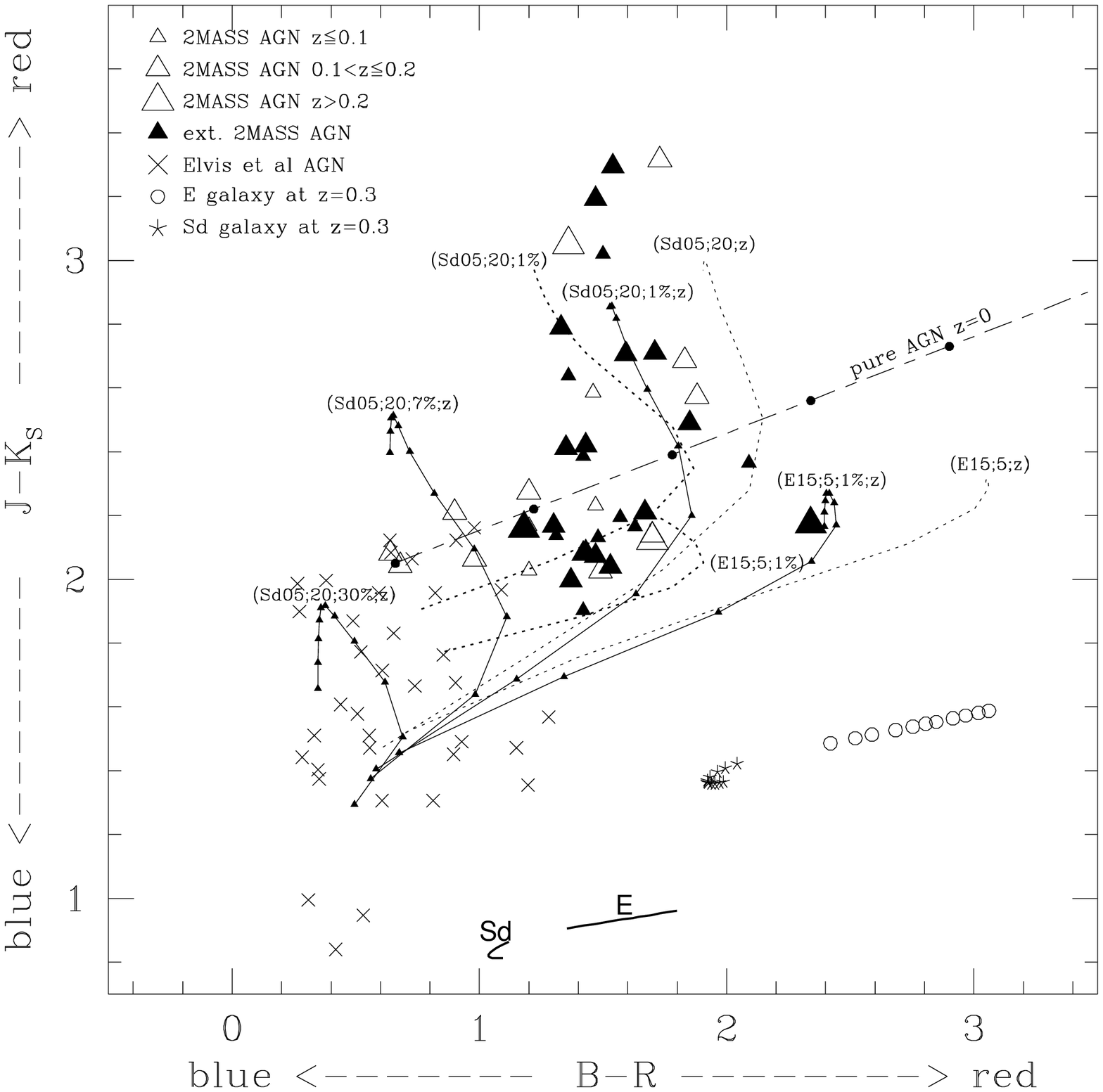}
\caption{Redshift dependence of the B$-$R and \jmk\ colors of a
  reddened AGN contaminated by both host galaxy emission and scattered
  AGN light. Two reddened AGN + host galaxy curves were chosen from
  Fig.~\ref{fig:two_color_host}: (E15;5) and (Sd05;20). 1\%,7\% and
  30\% of scattered intrinsic AGN light at R band was added to the
  (Sd05;20) curve and 1\% to (E15;5) curve, and then redshifted to
  z=0.31 yielding (Sd05;20;1\%;z), (Sd05;20;7\%;z), (Sd05;20;30\%;z)
  and (E15;5;1\%;z) curves (solid line). For comparison we also plot
  the redshifted AGN+host galaxy curves (E15;5;z) and (Sd05;20;z) from
  Fig.~\ref{fig:two_color_host_z_and_polarization_z}a (dotted
  line). Triangles and crosses represent the observed colors of red
  2MASS and Elvis \etal\ (1994) AGN samples respectively. The size of
  the triangles is proportional to the redshift of the 2MASS AGN.}
\label{fig:agn_host_pol_z}
\end{figure}

\clearpage
\begin{figure}
\plotone{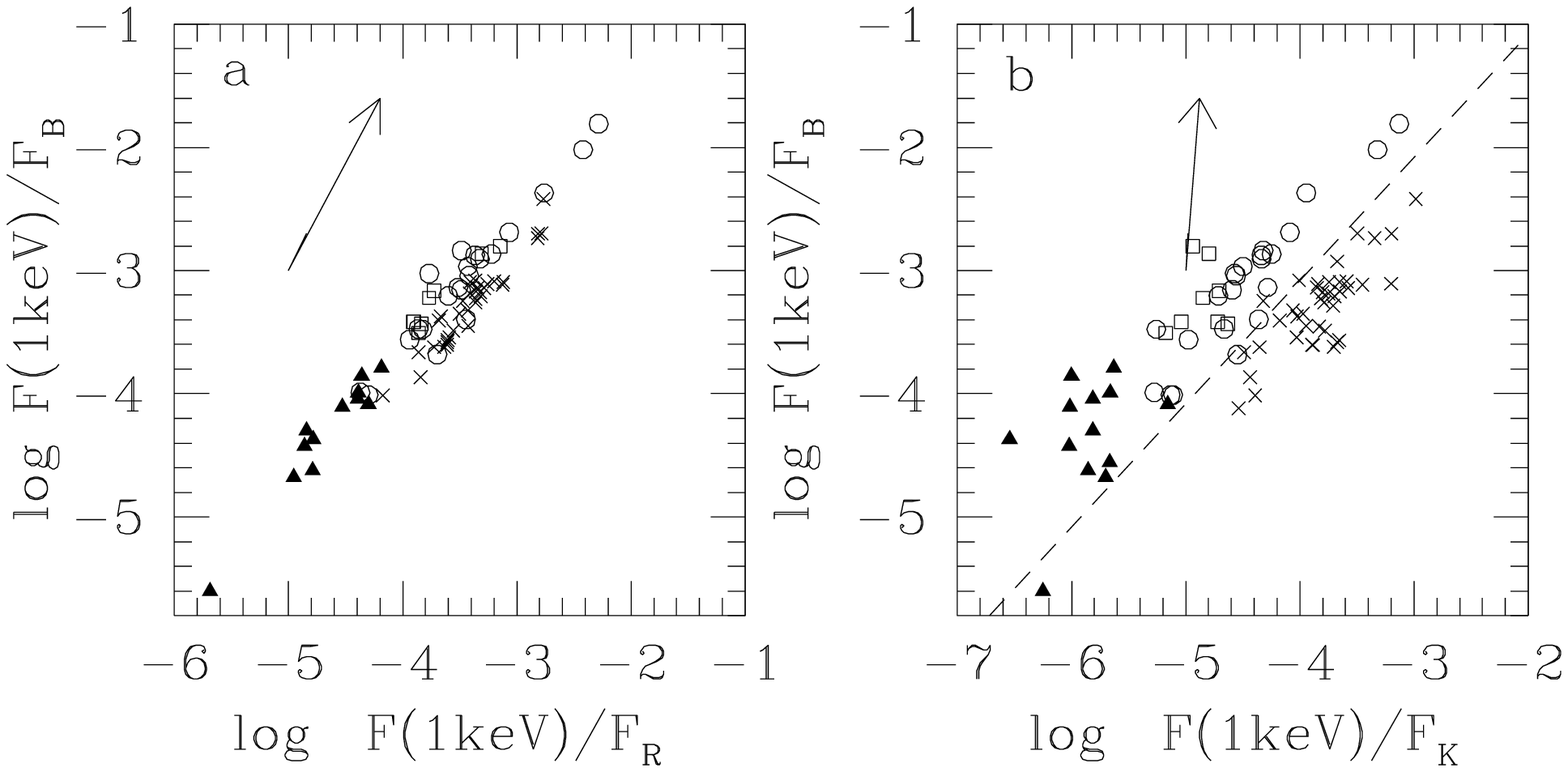}
\caption{Relation between the intrinsic 1~keV X-ray flux to observed B
 flux ratio and the intrinsic 1~keV X-ray flux to observed R and K
 flux ratios. The X-ray flux is corrected for Galactic and intrinsic
 extinction while the B,R and K fluxes are corrected only for Galactic
 extinction. Crosses represent optical/radio selected AGN from Elvis
 \etal\ (1994).  The 2MASS sources are delineated by the S/N: high S/N
 \chandra\ spectra ({\it C} fits) by circles, medium S/N \chandra\
 spectra ({\it B} fits) by squares, and low S/N \chandra\ spectra
 ({\it A} fits) by filled triangles.  The arrows indicate how the
 ratios change when intrinsic dust with $A_V$=3.1~mag. (\nh\
 $=5\times10^{21}$~cm$^{-2}$ assuming a Milky Way dust to gas ratio)
 reddens the optical/IR fluxes. The dashed line in b) shows the
 B$-$K$_S$=4.3 color cut used to select the red 2MASS AGN. 1258+2329
 (object with lowest F(1keV)/F$_B$) lies below this color cut, since
 SuperCOSMOS, not the USNO--A2 B magnitude (initially used for color
 selection), is used here to calculate the F(1keV)/F$_B$ ratio.}
\label{fig:Fx/FBvsFx/Fnn}
\end{figure}

\clearpage
\begin{figure}
\plottwo{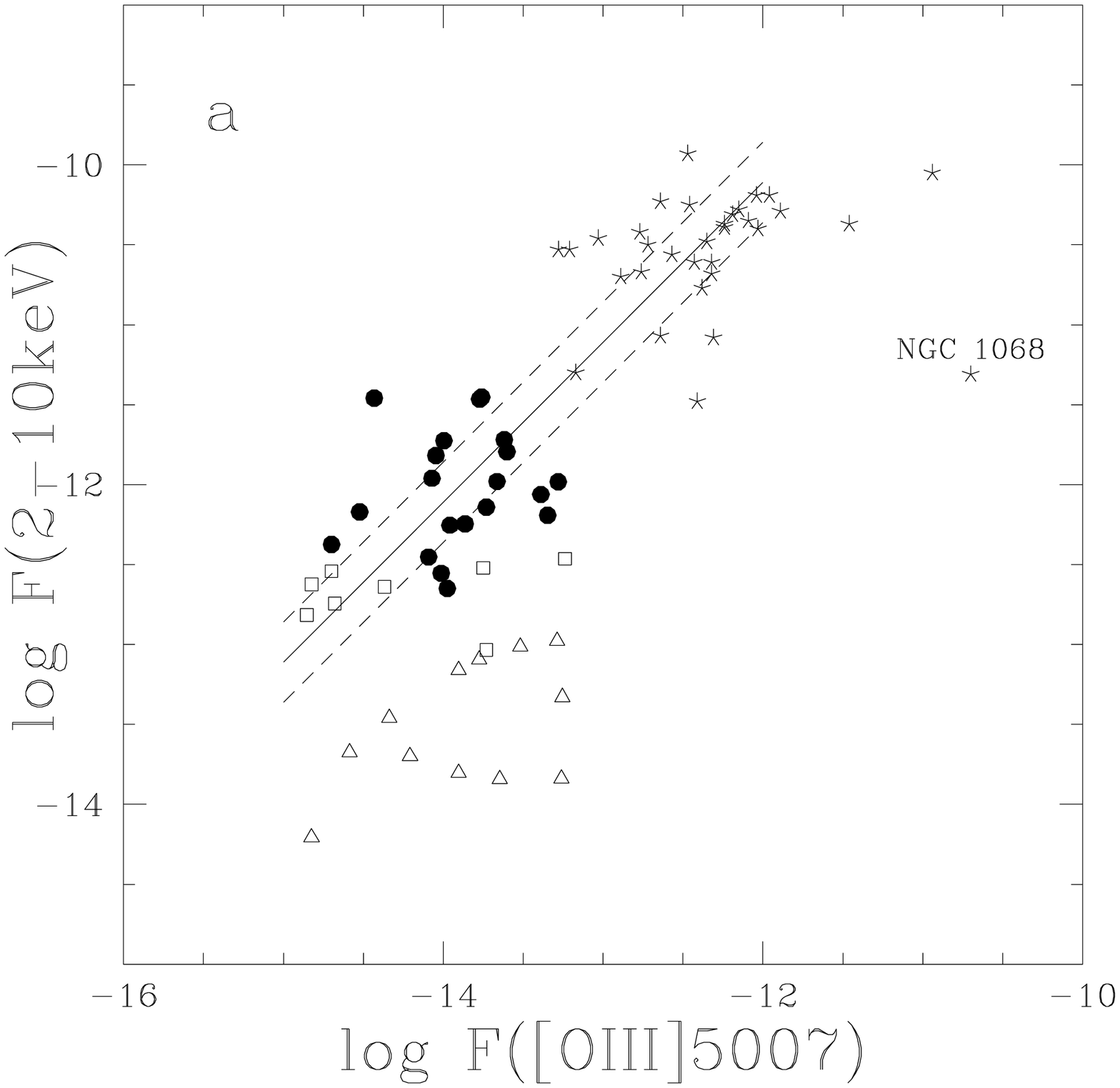}{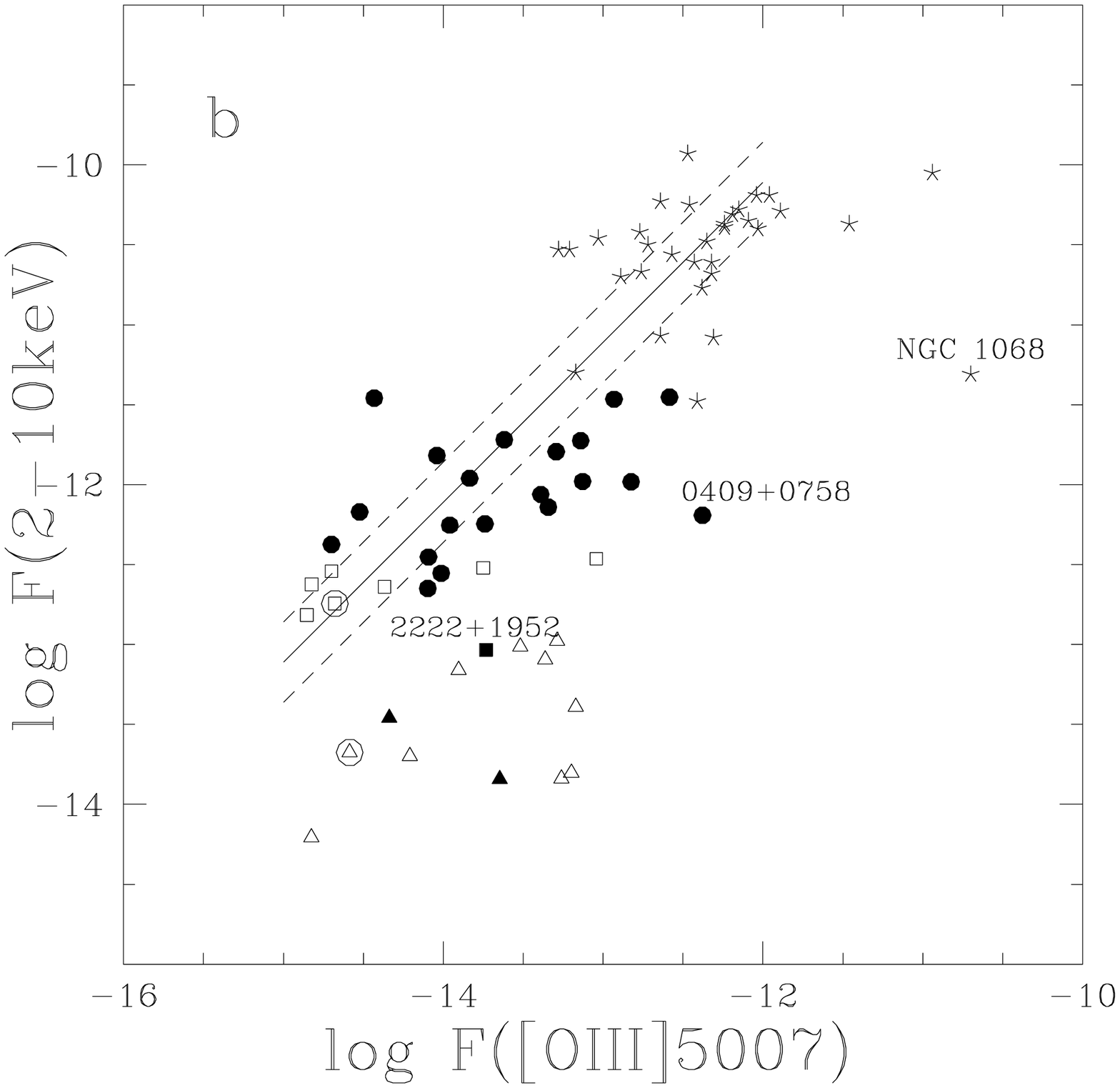}
\caption{Relation between the narrow emission line
[O\,III]\,$\lambda$5007 flux and the hard-X-ray flux at 2-10keV
(corrected for intrinsic absorption): a) with no reddening correction
to [O\,III]; b) applying a reddening correction based on the
H$\beta$/H$\alpha$ narrow-line ratio to the 2MASS AGN. In both plots
stars represent Seyfert 1s and 2s from Mulchaey \etal\ (1994).  The
2MASS objects are delineated by their Chandra S/N classifications,
with filled circles, open squares and open triangles representing our
2MASS objects with {\it C} (highest S/N), {\it B} (medium S/N) and
{\it A} (low S/N) spectral types respectively. Filled triangles and
squares represent the reddest \jmk\ $>3$ objects. Two objects
with undetermined reddening corrections (due to lack of H$\beta$
emission) are surrounded by an open circle. Solid and dashed
lines represent the mean $\log F[O\,III]/F(2-10keV) = -1.89\pm0.25$
found for Seyfert 1s and 2s in Mulchaey \etal\ (1994).}
\label{fig:O3vsFX}
\end{figure}

\clearpage
\begin{figure}
\plotone{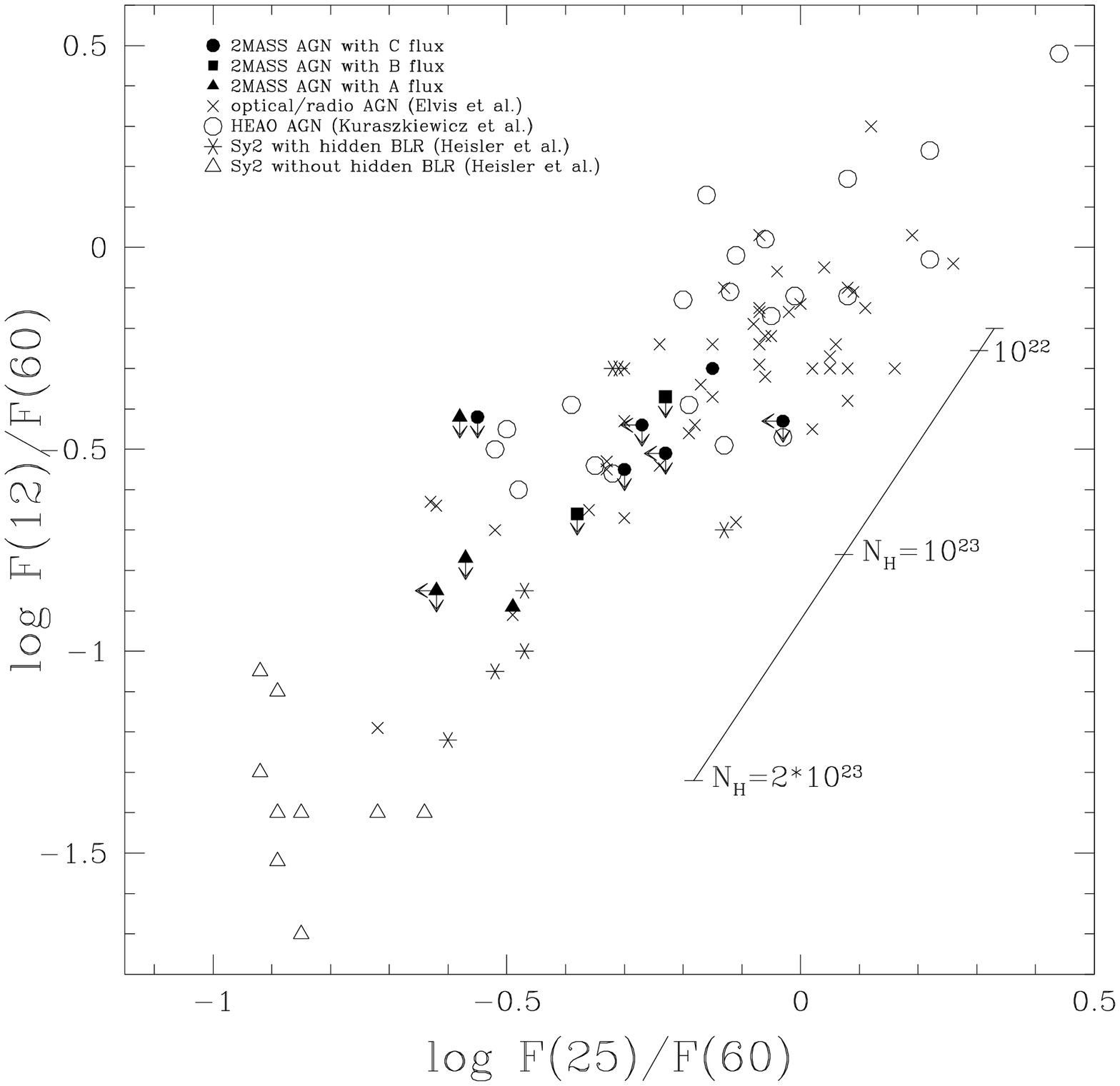}
\caption{The 12/60 versus 25/60 flux ratios. Filled circles are 2MASS
  AGN with high S/N \chandra\ spectra ({\it C} flux), filled squares
  2MASS AGN with medium S/N \chandra\ spectra ({\it B} flux), and
  filled triangles 2MASS AGN with low S/N \chandra\ spectra ({\it A}
  flux). X-ray selected Seyferts from the HEAO sample (Kuraszkiewicz
  \etal\ 2003) are represented by open circles, blue QSOs from Elvis
  \etal\ (1994) by crosses, Seyfert~2s with a hidden broad line region
  by stars, Seyfert~2s without a hidden broad line region (higher \nh
  ) by open triangles (Heisler, Lumsden, \& Bailey 1997).}
\label{fig:12_60to25_60}
\end{figure}

\clearpage
\begin{figure} 
\plotone{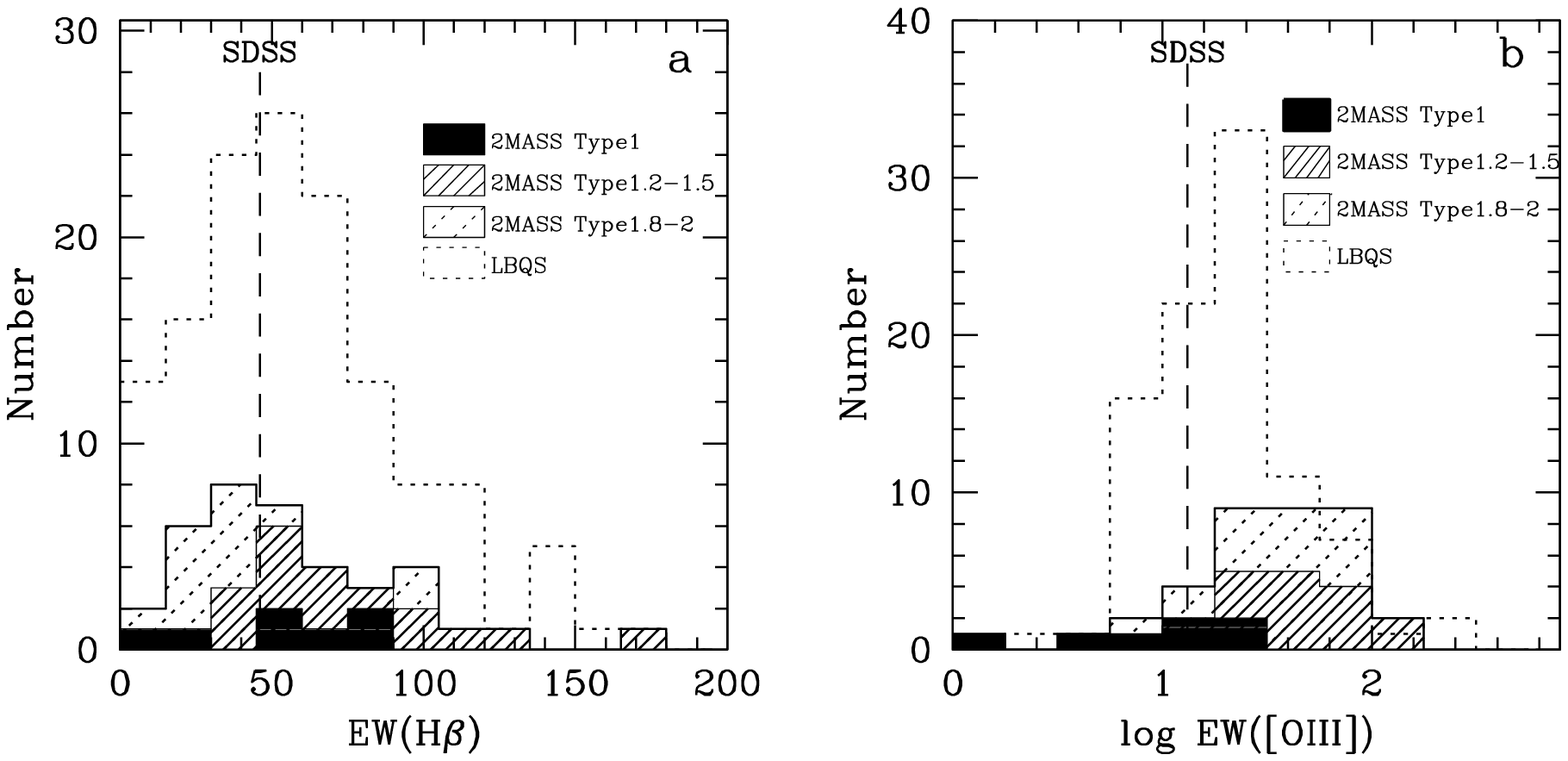}
\caption{Comparison of the distributions of H$\beta$, and [O\,III]
  emission line equivalent widths between 2MASS AGN (shaded areas) and
  LBQS sample (dotted line; Forster \etal\ 2001). Dashed lines show
  equivalent widths from the SDSS mean composite spectrum (Vanden Berk
  \etal\ 2001).}
\label{fig:emission_lines_histo}
\end{figure}



\clearpage
\begin{figure}
\plotfiddle{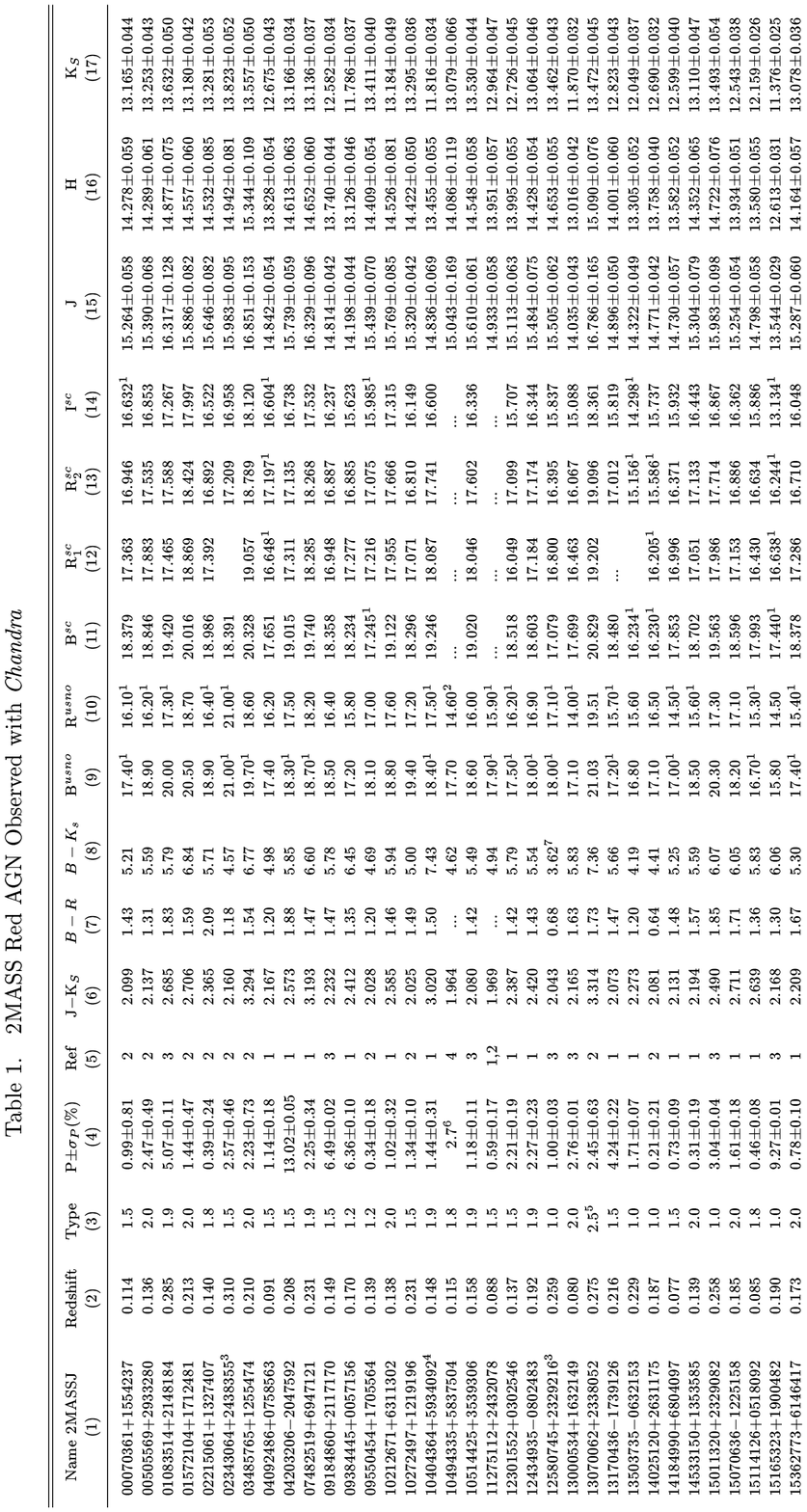}{6in}{0.}{90}{90}{-240}{-250} 
\label{tab:objects} 
\end{figure}

\clearpage
\begin{figure}
\plotfiddle{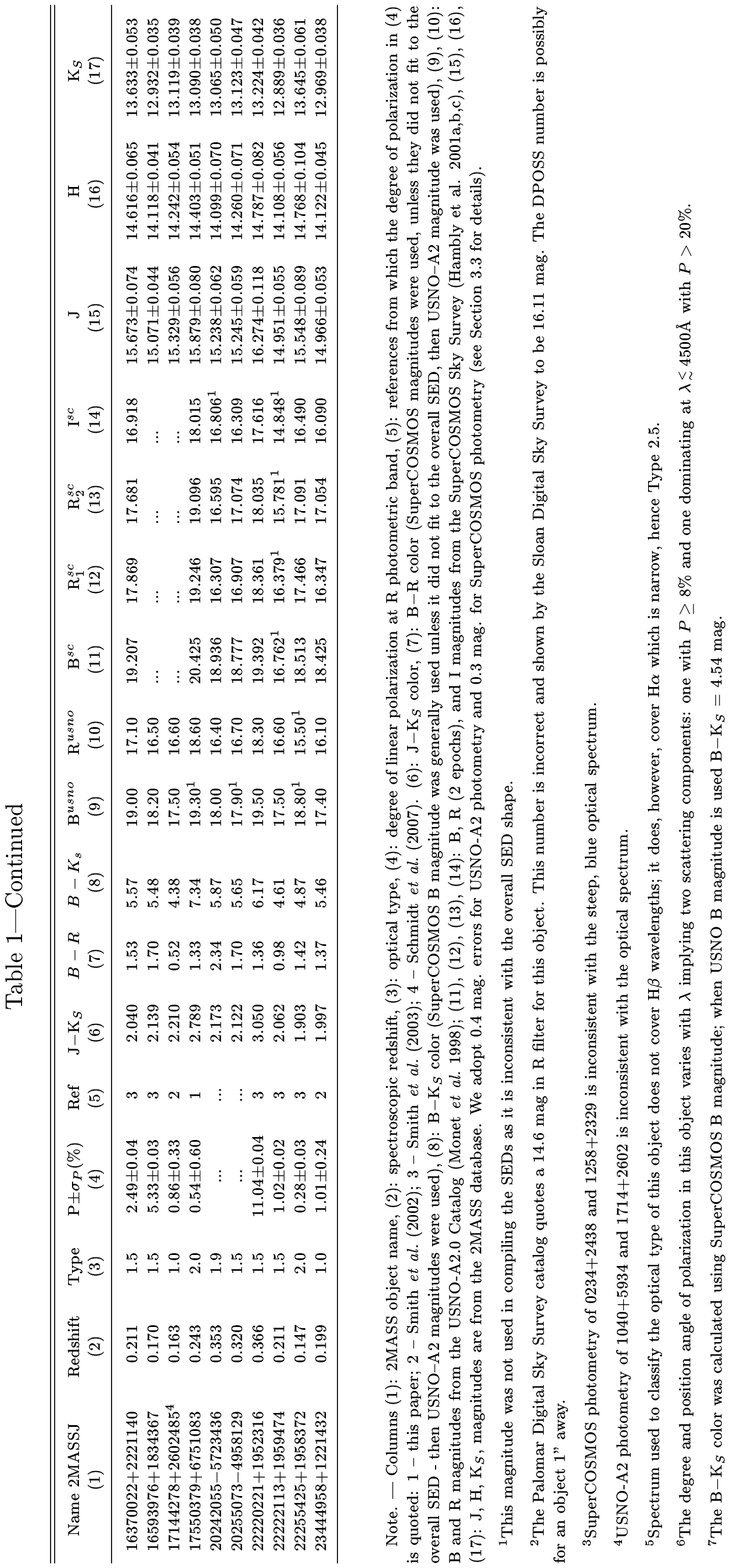}{6in}{0.}{90}{90}{-240}{-250} 
\end{figure}

\clearpage
\begin{deluxetable}{llll}
\tablenum{2}
\tablecaption{HST Spectroscopy}
\tablewidth{0pt}
\tablehead{
\colhead{Name} &
\colhead{Date} &
\colhead{Instrument} &
\colhead{Grating} 
}
\startdata
0918+2117& 2002/04/27&HST/STIS &PRISM \\
0955+1705& 2002/04/26&HST/STIS &PRISM \\
1516+1900& 2002/02/11&HST/STIS &G140L,G230L \\
1714+2602& 2002/01/25&HST/STIS &PRISM \\
2222+1959& 2002/05/03&HST/STIS &PRISM \\
\enddata
\end{deluxetable}


\begin{table}
\label{tab:hstspec} 
\end{table}

\begin{deluxetable}{lllllllll}
\tabletypesize{\scriptsize}
\tablewidth{0pt}
\tablenum{3}
\tablecaption{SDSS photometry}
\tablehead{
\colhead{Name} &
\colhead{u}&
\colhead{g}&
\colhead{r}&
\colhead{i}&
\colhead{B}&
\colhead{V}&
\colhead{R}&
\colhead{I}
}
\startdata
0007+1554& 18.695$\pm$ 0.035& 17.594$\pm$ 0.008& 16.840$\pm$ 0.005& 16.072$\pm$ 0.005& 17.891 & 17.172 & 16.610 & 15.787  \\
0938+0057$^2$& 18.927$\pm$ 0.021& 17.584$\pm$ 0.005& 16.664$\pm$ 0.004& 15.702$\pm$ 0.003& 17.922 & 17.076 & 16.440 & 15.477  \\
1021+6311& 20.879$\pm$ 0.117& 19.123$\pm$ 0.015& 18.113$\pm$ 0.010& 17.530$\pm$ 0.009& 19.532 & 18.568 & 18.076 & 17.386  \\ 
1027+1219$^2$& 20.082$\pm$ 0.044& 18.525$\pm$ 0.008& 17.441$\pm$ 0.005& 16.833$\pm$ 0.005& 18.900 & 17.931 & 17.430 & 16.723  \\
1040+5934& 20.450$\pm$ 0.165& 18.664$\pm$ 0.011& 17.712$\pm$ 0.008& 16.876$\pm$ 0.006& 19.078 & 18.139 & 17.551 & 16.679  \\
1049+5837$^1$& 18.568$\pm$ 0.061& 17.192$\pm$ 0.008& 16.344$\pm$ 0.005& 15.604$\pm$ 0.004& 17.536 & 16.721 & 16.170 & 15.367  \\
1051+3539& 20.105$\pm$ 0.147& 18.291$\pm$ 0.011& 17.125$\pm$ 0.007& 16.469$\pm$ 0.006& 18.709 & 17.655 & 17.135 & 16.393  \\ 
1230+0302$^2$& 18.815$\pm$ 0.028& 17.819$\pm$ 0.007& 17.103$\pm$ 0.005& 16.400$\pm$ 0.005& 18.098 & 17.417 & 16.880 & 16.103  \\ 
1402+2631& 17.146$\pm$ 0.009& 16.971$\pm$ 0.004& 16.609$\pm$ 0.004& 16.087$\pm$ 0.004& 17.111 & 16.753 & 16.284 & 15.639  \\
1501+2329& 21.042$\pm$ 0.188& 19.220$\pm$ 0.015& 17.809$\pm$ 0.008& 17.189$\pm$ 0.007& 19.640 & 18.456 & 17.951 & 17.234  \\
1511+0518& 18.721$\pm$ 0.028& 17.190$\pm$ 0.004& 16.301$\pm$ 0.004& 15.804$\pm$ 0.004& 17.560 & 16.698 & 16.239 & 15.611  \\
1637+2221& 20.125$\pm$ 0.086& 18.710$\pm$ 0.013& 17.583$\pm$ 0.008& 16.988$\pm$ 0.009& 19.061 & 18.094 & 17.598 & 16.899  \\
1659+1834$^2$& 19.024$\pm$ 0.028& 17.992$\pm$ 0.006& 17.069$\pm$ 0.005& 16.455$\pm$ 0.004& 18.277 & 17.482 & 16.979 & 16.267  \\
1714+2602& 17.060$\pm$ 0.010& 17.027$\pm$ 0.005& 16.927$\pm$ 0.005& 16.564$\pm$ 0.005& 17.143 & 16.945 & 16.537 & 16.006  \\
\enddata
\tablecomments{u,g,r,i, magnitudes are from SDSS. These are converted
  to B, V, R, I magnitudes using transformations for quasars from
  Jester et al. (2005).} 
\tablenotetext{1}{SDSS Photometry does not fit the overall SED.}
\tablenotetext{2}{These objects appear in the SDSS Quasar Catalog 
  by Schneider \etal\ (2007).}
\end{deluxetable}


\begin{table}
\label{tab:SDSSmags} 
\end{table}

\clearpage
\begin{deluxetable}{lllllc}
\tabletypesize{\scriptsize}
\tablenum{4}
\tablecaption{Optical Spectroscopy}
\tablewidth{0pt}
\tablehead{
\colhead{Name} &
\colhead{Date} &
\colhead{Telescope} &
\colhead{Slit('')}&
\colhead{Factor$^a$} \\
}
\startdata
0007+1554a &  1998/07/17 & Palomar200/NORRIS &  2  & 1.8 \\      
0007+1554b &  2001/10/18 & KPNO/BOK          &  4.5& 1.7 \\    
0050+2933  &  1998/01/24 & FAST/Tillinghast  &  3  & 1.2 \\ 
0108+2148  &  1998/11/15 & Palomar200/NORRIS &  2  & 1.2 \\      
0157+1712  &  1998/09/20 & Palomar200/NORRIS &  2  & 1.0 \\      
0221+1327  &  1998/11/15 & Palomar200/NORRIS &  2  & 2.0 \\     
0234+2438  &  1998/09/20 & Palomar200/NORRIS &  2  & 0.65\\   
0348+1255  &  1998/11/14 & Palomar200/NORRIS &  2  & 2.0 \\        
0409+0758a &  1998/07/17 & Palomar200/NORRIS &  2  & 0.4 \\        
0409+0758b &  2005/12/30 & KPNO/BOK          &  3  & 1.0 \\
0420$-$2047&  2000/12/29 & Palomar200/NORRIS &  2  & 2.2 \\        
0748+6947a &  2000/01/11 & Palomar200/NORRIS &  2  & 2.5 \\        
0748+6947b & 2005/04/14,15 & KPNO/BOK        &  2  & 2.5 \\
0918+2117a &  1999/01/10 & Palomar200/NORRIS &  2  & 1.3 \\      
0918+2117b &  2001/10/18 & KPNO/BOK          &  4.5& 2.0 \\      
0938+0057  &  2000/12/02 & KPNO/BOK          &  2.5& 1.0 \\        
0955+1705a &  1997/12/24 & Palomar200/NORRIS &  2  & 0.7 \\      
0955+1705b &  2002/01/13 & KPNO/BOK          &  4.5& 0.6 \\      
1021+6311a &  2000/12/27 & Palomar200/NORRIS &  2  & 1.0 \\        
1021+6311b &  2005/04/13 & KPNO/BOK          &  1  & 2.0 \\ 
1027+1219a &  1998/11/14 & Palomar200/NORRIS &  2  & 0.65\\        
1027+1219b & 2005/04/14,15 & KPNO/BOK        &  2  & 1.0 \\ 
1040+5934a &  2000/12/28 & Palomar200/NORRIS &  2  & 0.18\\        
1040+5934b &  2005/04/14 & KPNO/BOK          &  2  & 1.3 \\ 
1049+5837  &  2001/02/16 & Palomar200/NORRIS &  2  & 3.8 \\        
1051+3539a &  1999/01/10 & Palomar200/NORRIS &  2  & 3.3 \\      
1051+3539b &  2002/01/13 & KPNO/BOK          &  4.5& 4.2 \\      
1127+2432a &  2001/04/11 & KPNO/BOK          &  2.5& 2.4 \\        
1127+2432b &  2005/12/30 & KPNO/BOK          &  3  & 1.0 \\ 
1230+0302a &  2001/02/16 & Palomar200/NORRIS &  2  & 3.7 \\        
1230+0302b & 2005/04/13  & KPNO/BOK          &  2  & 4.0 \\ 
1243$-$0802a& 2000/05/01 & SSO/2.3m          &  2  & 2.5 \\      
1243$-$0802b& 2005/12/30 & KPNO/BOK          &  3  & 1.0 \\ 
1258+2329  &  1998/03/01 & FAST/Tillinghast  &  3  & 1.0 \\ 
1300+1632a &  1998/07/19 & Palomar200/NORRIS &  2  & 4.0 \\     
1300+1632b &  2002/01/13 & KPNO/BOK 	     &  4.5& 4.0 \\    
1307+2338  &  2001/09/06 & Keck II	     &  2  & 1.2 \\    
1317$-$1739&  2001/04/11 & SSO/2.3m          &  2  & 3.2 \\       
1350$-$0632&  2000/01/06 & SSO/2.3m          &  4.5& 2.7 \\      
1402+2631  &  1998/03/03 & FAST/Tillinghast  &  3  & 1.2 \\ 
1418+6804a &  2000/07/03 & Palomar200/NORRIS &  2  & 1.4 \\        
1418+6804b & 2005/04/13,15 & KPNO/BOK        &  1  & 1.3 \\
1453+1353a &  1998/09/21 & Palomar200/NORRIS &  2  & 3.2 \\        
1453+1353b & 2005/04/13,15 & KPNO/BOK        &  1  & 1.0 \\
1501+2329a &  1998/09/21 & Palomar200/NORRIS &  2  & 0.7 \\        
1501+2329b &  2001/03/30 & MMT               &  2  & 2.2 \\        
1507$-$1225a& 2000/01/05 & SSO/2.3m          &  2  & 2.2 \\      
1507$-$1225b& 2005/04/13 & KPNO/BOK          &  2  & 4.6 \\
1511+0518a &  2001/04/12 & KPNO/BOK          &  2.5& 0.6 \\        
1511+0518b &  2005/12/30 & KPNO/BOK          &  3  & 1.0 \\
1516+1900a &  2002/01/16 & KPNO/BOK          &  4.5& 1.0 \\        
1516+1900b &  1998/03/03 & FAST/Tillinghast  &  3  & 1.0 \\  
1536+6146$^b$  &  2001/06/12 & KPNO/BOK       &  2.5& ...  \\
1637+2221  &  1998/09/21 & Palomar200/NORRIS &  2  & 3.0 \\        
1659+1834  &  1998/09/20 & Palomar200/NORRIS &  2  & 1.0 \\        
1714+2602a &  1998/07/17 & Palomar200/NORRIS &  2  & 1.0 \\     
1714+2602b &  2001/10/18 & KPNO/BOK          &  4.5& 0.7 \\      
1755+6751a &  2000/07/02 & Palomar200/NORRIS &  2  & 0.9 \\        
1755+6751b &  2005/04/13,14 & KPNO/BOK       &  2  & 1.0 \\
2024$-$5723&  2000/01/02 & SSO/2.3m          &  2  & 1.4 \\    
2025$-$4958&  2000/01/01 & SSO/2.3m          &  2  & 1.0 \\      
2222+1952  &  1998/07/17 & Palomar200/NORRIS &  2  & 0.7 \\        
2222+1959a &  1998/07/17 & Palomar200/NORRIS &  2  & 1.0 \\      
2222+1959b &  2001/10/18 & KPNO/BOK          &  4.5& 0.63\\      
2225+1958  &  1998/07/17 & Palomar200/NORRIS &  2  & 0.5\\        
2344+1221  &  1998/09/20 & Palomar200/NORRIS &  2  & 1.0\\  
\enddata
\tablenotetext{a}{Factor by which the spectrum was multiplied
  (grayshifted) to match the optical photometry.}
\tablenotetext{b}{Spectrum not used in SEDs and emission line measurements.} 

\end{deluxetable}


\begin{table}
\label{tab:optspec} 
\end{table}

\clearpage
\begin{deluxetable}{llrr}
\tabletypesize{\scriptsize}
\tablenum{5}
\tablecaption{New Polarimetry of 2MASS AGN\tablenotemark{1}}
\tablewidth{0pt}
\tablehead{\colhead{Object} &
\colhead{Date} &
\colhead{$P\tablenotemark{2}\pm\sigma_P$} &
\colhead{$\theta\tablenotemark{2}\pm\sigma_\theta$ }
\\
& \colhead{(yyyymmdd)}
& \colhead{(\%)}
& \colhead{(deg)}
}

\startdata
 0409+0758   & 2003/09/22 & $1.14\pm0.18$ & $135.7\pm4.4$ \\
 0420$-$2047 & 2003/09/22 & $10.94\pm0.31$ & $158.3\pm0.8$ \\
             & 2003/10/29 & $13.02\pm0.05$ & $156.7\pm0.1$ \\
 0748+6947   & 2004/02/22 & $1.34\pm0.48$ & $142.4\pm11.4$ \\
             & 2005/04/14,15 & $2.25\pm0.34$ & $129.3\pm4.3$ \\
 0938+0057   & 2004/02/21 & $7.12\pm0.24${\tablenotemark{3}} & $49.3\pm1.0${\tablenotemark{3}} \\
             & 2004/04/25 & $6.36\pm0.10$  & $50.9\pm0.5$ \\
 1021+6311   & 2004/02/21 & $1.02\pm0.32$  & $2.1\pm8.9$ \\
             & 2005/04/13 & $2.22\pm0.55$  & $8.6\pm7.1$ \\
 1027+1219   & 2005/04/14,15 & $1.34\pm0.10$  & $147.5\pm2.1$ \\
 1040+5934   & 2004/02/21 & $1.43\pm0.31$  & $66.0\pm6.1$ \\
             & 2005/04/14 & $1.77\pm0.29$  & $91.4\pm4.7$ \\
 1127+2432   & 2004/02/21 & $0.59\pm0.17$  & $60.3\pm8.2$ \\
 1230+0302   & 2004/02/21 & $2.21\pm0.19$  & $17.8\pm2.4$ \\
             & 2005/04/13 & $3.15\pm1.25$  & $10.3\pm11.4$ \\
 1243$-$0802 & 2004/02/22 & $2.27\pm0.23$  & $67.6\pm2.8$ \\
 1317$-$1739 & 2004/02/22 & $4.24\pm0.22$  & $73.7\pm1.5$ \\
             & 2004/04/25 & $4.91\pm0.15$  & $75.9\pm0.9$ \\
 1350$-$0632 & 2004/05/23 & $1.71\pm0.07$  & $172.7\pm1.1$ \\
 1418+6804   & 2004/05/23 & $1.28\pm0.18$  & $63.2\pm4.1$ \\
             & 2005/04/13,15 & $0.73\pm0.09$  & $65.9\pm3.3$ \\
 1453+1353   & 2005/04/13,15 & $0.31\pm0.09$  & $63.2\pm8.6$ \\
 1507$-$1225 & 2004/05/23 & $1.62\pm0.18$  & $84.7\pm3.2$ \\
             & 2005/04/13 & $2.35\pm1.20$  & $123.6\pm14.6$ \\
 1511+0518   & 2004/05/23 & $0.46\pm0.08$  & $4.0\pm5.2$ \\
 1536+6146   & 2004/04/26 & $0.78\pm0.10$  & $17.8\pm3.7$ \\
 1755+6751   & 2003/09/22 & $0.54\pm0.60$ & \nodata \\
             & 2005/04/13,14 & $1.56\pm0.35$ & $24.8\pm6.4$\\
\enddata
\tablenotetext{1}{Measurements are in the $R\/$ photometric band.}
\tablenotetext{2}{$P$ is linear polarization in the R band; $\theta$
 is the position angle of the polarization.}
\tablenotetext{3}{Measured in the $V\/$ photometric band.}
\end{deluxetable}

\begin{table}
\label{tab:pol} 
\end{table}

\begin{deluxetable}{lrrrr}
\tablenum{6}
\tablecaption{IRAS photometry}
\tablewidth{0pt}
\tablehead{
\colhead{Name}&
\colhead{12$\mu$m}&
\colhead{25$\mu$m}&
\colhead{60$\mu$m} &
\colhead{100$\mu$m}\\
\colhead{}&
\colhead{Jy}&
\colhead{Jy}&
\colhead{Jy} &
\colhead{Jy}}
\startdata
0221+1327&  $<$ 0.119     & $<$ 0.172      & 0.327$\pm$0.052& $<$ 1.674\\
0748+6947&  $<$ 0.098     & 0.068$\pm$0.020& 0.261$\pm$0.039& $<$ 0.531\\
0918+2117&  $<$ 0.097     & $<$ 0.245      & 0.260$\pm$0.039& $<$ 0.597\\
1040+5934& 0.100$\pm$0.023 & 0.248$\pm$0.025& 0.770$\pm$0.054& 0.954$\pm$0.143\\
1307+2338&  $<$ 0.101     & $<$ 0.173      & 0.721$\pm$0.058& 0.686$\pm$0.123\\
1418+6804&  $<$ 0.073     & 0.054$\pm$0.015& 0.191$\pm$0.032& 0.651$\pm$0.163\\
1453+1353&  $<$ 0.094     & 0.149$\pm$0.037& 0.556$\pm$0.061& 0.702$\pm$0.168\\
1507$-$1225&$<$ 0.146     & 0.283$\pm$0.051& 0.666$\pm$0.087& $<$ 0.717\\
1536+6146&  $<$ 0.079     & 0.143$\pm$0.014& 0.286$\pm$0.029& $<$ 0.459\\
1637+2221&  $<$ 0.056     & $<$ 0.105      & 0.179$\pm$0.036& $<$ 0.086\\
1659+1834&  $<$ 0.119     & 0.163$\pm$0.024& 0.277$\pm$0.041& $<$ 0.853\\
2024$-$5723&0.097$\pm$0.023& 0.193$\pm$0.029& 0.273$\pm$0.049& $<$ 0.737\\
\enddata
\end{deluxetable}


\begin{table}
\label{tab:IRAS} 
\end{table}

\clearpage
\begin{deluxetable}{lllclcrcccccr}

\tabletypesize{\scriptsize}
\tablewidth{0pt}
\tablenum{7}
\tablecaption{Summary of the Data Sources for the SEDs and Parameters
  Used in Optical/IR Color Modeling}
\tablehead{
\colhead{Name} &
\colhead{IR} &
\colhead{Optical} &
\colhead{UV} &
\colhead{X-ray} &
\colhead{${\rm N_H}$}  &
\colhead{$A_V(N_H)$} &
\colhead{$A_V^{mod}$} &
\colhead{host galaxy} &
\colhead{$P_{mod}$} &
\colhead{$P_{obs}$} &
\colhead{high} &
\colhead{host} \\
\colhead{} &
\colhead{} &
\colhead{} &
\colhead{} &
\colhead{} &
\colhead{$10^{22}$cm$^{-2}$} &
\colhead{mag} &
\colhead{mag} &
\colhead{} &
\colhead{\%} &
\colhead{\%} &
\colhead{z?} &
\colhead{\%} \\
\colhead{(1)} &
\colhead{(2)} &
\colhead{(3)} &
\colhead{(4)} &
\colhead{(5)} &
\colhead{(6)} &
\colhead{(7)} &
\colhead{(8)} &
\colhead{(9)} &
\colhead{(10)} &
\colhead{(11)} &
\colhead{(12)} & 
\colhead{(13)} 
}
\startdata
0007+1554   &    2&5,6  & & C   & 1.22$\pm$0.15&  7.60$\pm$0.92    &3   &(Sd05,5)    &... &...&...&61   \\
0050+2933   &    2&5,6  & & C   & 1.12$\pm$0.66&  6.96$\pm$4.07    &4	&(Sd05,5)    &... &...&...&76   \\
0108+2148   &    2&5,6  & & A   & ...          &  ... \ \ \ \      &5   &(Sd05,20)   &0.5 &6  & y &58   \\
0157+1712   &    2&5,6  & & B   & 2.83$\pm$1.40& 17.60$\pm$8.70    &7   &(Sa05,20)   &... &...&...&86   \\
0221+1327   & 1, 2&5,6  & & C   & 1.66$\pm$0.58& 10.29$\pm$3.60    &4-5 &(E15,10)    &... &...&...&62-76\\
0234+2438   &    2&5,6  & & A   & ...          &  ... \ \ \ \      & 1  &pure AGN    &... &...&...&0    \\
0348+1255   &    2&5,6  & & A   & ...          &  ... \ \ \ \      &11  &(Sa05,40)   &... &...&...&98   \\
0409+0758   &    2&5,6  & & C   & 0.33$\pm$0.13&  2.04$\pm$0.79    &1   &pure AGN    &... &...&...&0    \\
0420$-$2047 &    2&3,5,6& & B   & 4.01$\pm$1.51& 24.91$\pm$9.38    &5	&(E15,20)    &1.2 &13 &...&53   \\
0748+6947   & 1, 2&5,6  & & A   & ...          &  ... \ \ \ \      &10  &(Sa05,40)   &... &...&...&96   \\
0918+2117   & 1, 2&5,6  &7& C   & 0.22$\pm$0.21&  1.39$\pm$1.28$^b$&3   &(Sa05,20)   &1.2 &6.5&...&27   \\
0938+0057   &    2&5,6  & & C   & 0.76$\pm$0.62&  4.73$\pm$3.82    &5   &(Sa05,40)   &0.6 &7  &...&57   \\
0955+1705   &    2&5,6  &7& A   & ...          &  ... \ \ \ \      &1   &(Sd05,10)   &... &...&...&17   \\
1021+6311   &    2&5,6  & & C   & 2.37$\pm$0.55& 14.72$\pm$3.39    &22  &(Sa05,20)   &... &...&...&100  \\
1027+1219   &    2&5,6  & & B   & 4.34$\pm$2.57&26.98$\pm$15.95    &15  &(Sa05,5)    &... &...&...&100  \\
1040+5934   & 1, 2&5,6  & & A   & ...          &  ... \ \ \ \      &10  &(Sa05,20)   &... &...&...&98   \\
1049+5837$^a$&   2&6    & & C   & 4.22$\pm$4.18&26.21$\pm$25.95    &... &...         &... &...&...&...  \\
1051+3539   &    2&5,6  & & C   & 0.56$\pm$0.14&  3.48$\pm$0.89    &3   &(Sd05,5)    &... &...&...&61   \\
1127+2432$^a$&   2&4,6  & & A   & ...         &  ... \ \ \ \       &... &...         &... &...&...&...  \\
1230+0302   &    2&5,6  & & C   & 0.33$\pm$0.10&  2.08$\pm$0.61    &1-2 &pure AGN    &... &...&...&0    \\
1243$-$0802 &    2&5,6  & & C   & 0.78$\pm$0.43&  4.84$\pm$2.67    &4   &(Sd05,10)   &... &...&...&61   \\
1258+2329   &    2&4,5,6& & A   & ...          &  ... \ \ \ \      &0   &pure AGN    &... &...&...& 0   \\
1300+1632   &    2&4,5,6& & C   & 1.79$\pm$1.08& 11.09$\pm$6.68    &10  &(E15,5)     &... &...&...&96   \\
1307+2338   & 1, 2&5,6  & & A   & ...          &  ... \ \ \ \      &11  &(Sa05,$>$40)&... &...&...&98   \\
1317$-$1739 &    2&5,6  & & A   & ...          &  ... \ \ \ \      &2   &(Sa05,40)   &1.5 &5  &...&13   \\
1350$-$0632 &    2&5,6  & & C, 8& 0.02$\pm$0.16&  0.14$\pm$1.00    & 1  &pure AGN    &... &...&...&0    \\
1402+2631   &    2&4,5,6& & C, 8& 0.01$\pm$0.11&  0.09$\pm$0.71    &0   &pure AGN    &... &...&...&0    \\
1418+6804   & 1, 2&5,6  & & C   & 0.73$\pm$0.22&  4.54$\pm$1.34    &4   &(Sd05,10)   &... &...&...&61   \\
1453+1353   & 1, 2&5,6  & & A   & ...          &  ... \ \ \ \      &14  &(Sd05,10)   &... &...&...&100  \\
1501+2329   &    2&5,6  & & B   & 0.56$\pm$0.31&  3.46$\pm$1.94    &3   &(Sd05,20)   &... &...&...&29   \\
1507$-$1225 & 1, 2&5,6  & & B   & 2.67$\pm$1.11& 16.61$\pm$6.88    &20  &(E05,20)    &... &...&...&100  \\
1511+0518   &    2&4,5,6& & B, 8& 0.03$\pm$0.00&  0.21$\pm$0.00    &6   &(Sd05,20)   &... &...&...&76   \\
1516+1900   &    2&5,6  &7& A   & ...          &  ...  \ \ \ \     &3   &(Sa05,20)   &2   &10 &...&26   \\  
1536+6146   & 1, 2&4,5,6& & C   & 2.27$\pm$0.61& 14.12$\pm$3.82    &18  &(E15,10)    &... &...&...&100  \\
1637+2221   & 1, 2&5,6  & & C   & 0.83$\pm$0.32&  5.16$\pm$1.98    &2-3 &(Sa05,5)    &... &...&...&44-61\\
1659+1834   & 1, 2&6    & & A   & ...          &  ...  \ \ \ \     &3   &(E15,10)    &1.2 &5  &...&42   \\
1714+2602   &    2&6    &7& C   & 0.14$\pm$0.17&  0.89$\pm$1.03    &0-1 &pure AGN    &... &...&...&0    \\
1755+6751   &    2&5,6  & & B   & 2.87$\pm$1.25& 17.81$\pm$7.74    &22  &(Sa05,20)   &... &...&...&100  \\
2024$-$5723 & 1, 2&4,5,6& & C, 8& 0.33$\pm$0.13&  2.06$\pm$0.83    &3   &(E15,20)    &... &...& y &29   \\
2025$-$4958 &    2&5,6  & & C   & 0.71$\pm$0.24&  4.42$\pm$1.47    &4-5 &(Sd05,4)    &... &...& y &80-89\\
2222+1952   &    2&5,6  & & B   & 0.05$\pm$0.22&  0.29$\pm$1.36    &10  &(Sd05,50)   &0.7 &12 & y &34   \\
2222+1959   &    2&4,5,6&7& A   & ...          &  ...  \ \ \ \     &0-1 &(E15,30)    &... &...&...&6    \\
2225+1958   &    2&5,6  & & C   & 3.07$\pm$2.06&19.04$\pm$12.78    &17  &(Sa05,5)    &... &...&...&100  \\
2344+1221   &    2&5,6  & & C   & 0.40$\pm$0.18&  2.50$\pm$1.09    &1   &(E15,10)    &... &...&...&17   \\
\enddata				      

\tablecomments{Column (1) - object name, columns (2)-(5) show data
  sources used for the SEDs:  
1 -- The IRAS Faint Source Catalog Moshir et al. (1990), 
2 -- 2MASS database photometry,
3 -- Maddox et al. (1990), 
4 -- B magnitude from NED,
5 -- SuperCOSMOS photometry,
6 -- USNO-A2 photometry,
7 -- HST/STIS archive,
8 -- WGA Catalog;  
column (5) also  information on whether the \chandra\ spectrum 
had high (A), medium (B), or low (C) signal-to-noise. Column (6) gives
  column density obtained from \chandra\ spectral  fitting, column (7)
  reddening obtained from column (6) using the following
  conversion: \nh  
  $=A_V*1.61\times10^{21}$~~cm$^{-2}$, which assumes the Galactic
  gas--to--dust ratio; columns (8)-(13) give AGN parameters obtained from
  the optical/IR color modeling described in Section~5 where: (8): AGN
  reddening, (9): type of host galaxy and intrinsic
  AGN/host galaxy ratio (e.g. (Sd05,5) denotes a Sd host galaxy with
  5~Gyr stars, and intrinsic AGN/host galaxy ratio at R band = 5),
  (10): polarization fraction at R band relative to
  the intrinsic AGN, (11): observed polarization fraction at R band
  relative to the reddened AGN, (12): shows whether the
  AGN is at high (z$\ge$0.3) redshift which needed to be
  taken into account during modeling (see Sections~5.5.1 and
  5.5.2), (13): observed host galaxy contribution at R band.}
\tablenotetext{a}{This object does not have good B and R photometry to
 model the optical/IR colors.}
\tablenotetext{b}{High S/N XMM-Newton spectra found {\it 0918+2117} 
  to be variable in X-rays and to have absorption, \nh\
  $\sim4\times 10^{21}$~cm$^{-2}$ in the higher flux state (Pounds \&
  Wilkes 2007), which is consistent with our optical/IR color modeling.} 
\end{deluxetable}


\begin{table}
\label{tab:SED} 
\end{table}

\clearpage
\begin{deluxetable}{ccr}
\tablenum{8}
\tablecaption{Dependence of the Observed Optical and Near-IR Colors on
  Redshift}  
\tablewidth{0pt}
\tablehead{
\colhead{Observed Color} &
\colhead{Redshift Dependence} &
\colhead{Redshift Range} \\
}
\startdata

$(J-K)=$ &$(2.013\pm 0.005) - (0.585\pm 0.062)\times z$ &
 $0<z < 0.17 $  \\   
$(J-K)=$ &$(2.248\pm 0.056) - (1.947\pm 0.269)\times z$ &
 $0.18\le z < 0.24$ \\
$(J-K)=$ &$(2.132\pm 0.021) - (1.575\pm 0.047)\times z$ &
$0.25 \le z < 0.64$ \\
$(J-K)=$ &$(1.143\pm 0.001) - (0.035\pm 0.001)\times z$ &
$0.65 \le z \le 1.50$ \\
& & \\
$(B-K)=$ &$(3.526\pm 0.001) - (2.161\pm 0.001)\times z$ &
$ 0 < z \le 0.5$ \\
$(B-K)=$ &$(2.849\pm 0.002) - (0.794\pm 0.004)\times z$ &
$0.51 < z \le0.7$\\
$(B-K)=$ &$(2.255\pm 0.001) + (0.045\pm 0.001)\times z$ &
$0.71<z\le1.5$ \\
& & \\
$(B-R)=$ &$(0.595\pm 0.001) - (0.702\pm 0.001)\times z$ &
$0 <  z\le 0.47$ \\
$(B-R)=$ &$(0.024\pm 0.001) + (0.559\pm 0.001)\times z$ &
$0.48\le z\le 1.10$ \\
\enddata

\tablecomments{Colors are calculated using the median AGN SED of the
  optical/radio selected quasars from Elvis et al. (1994). }

\end{deluxetable}


\begin{table}
\label{tab:color_z} 
\end{table}


\clearpage
\begin{figure}
\plotfiddle{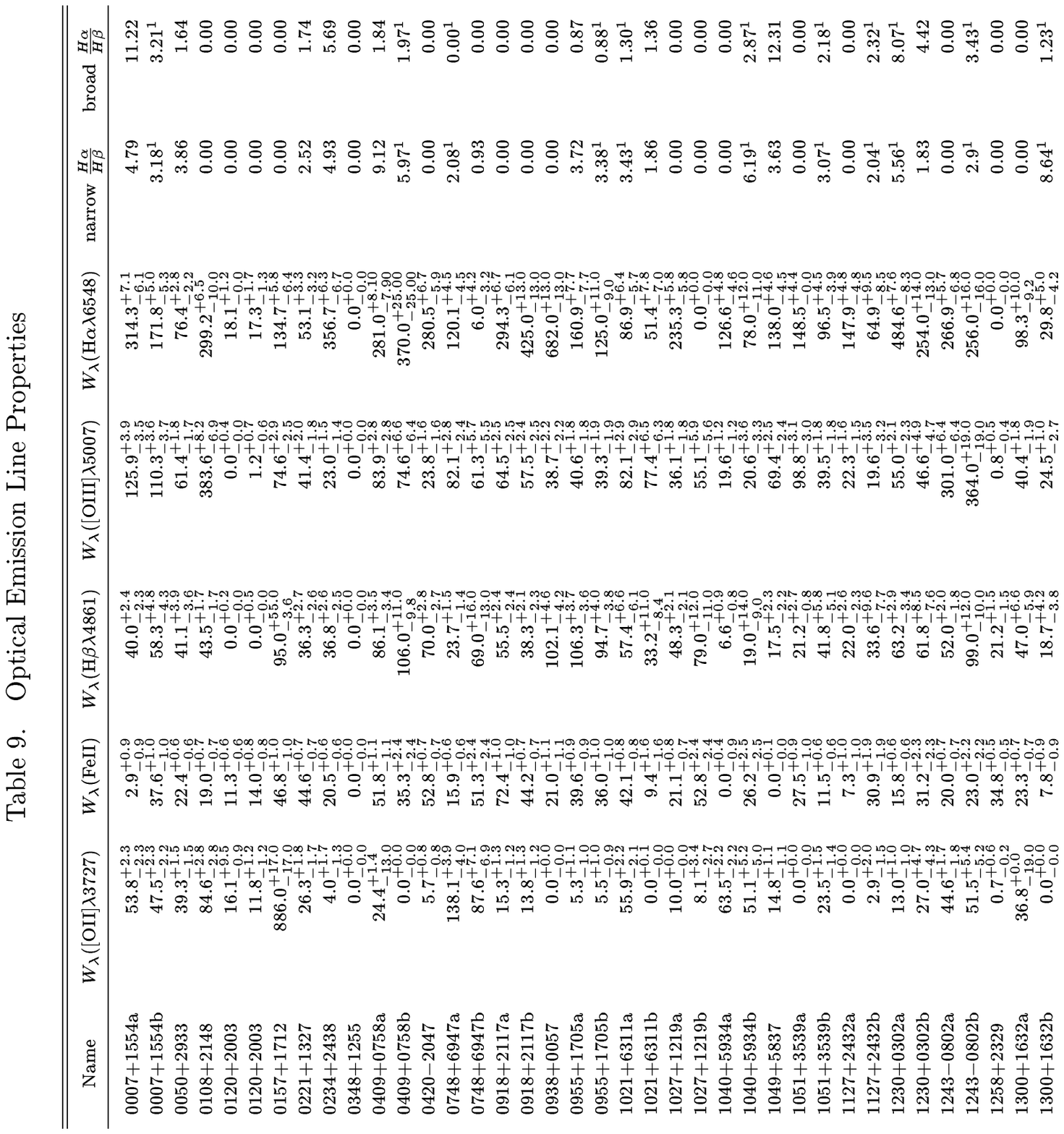}{6in}{0.}{90}{90}{-240}{-250} 
\label{tab:lines} 
\end{figure}

\clearpage
\begin{figure}
\plotfiddle{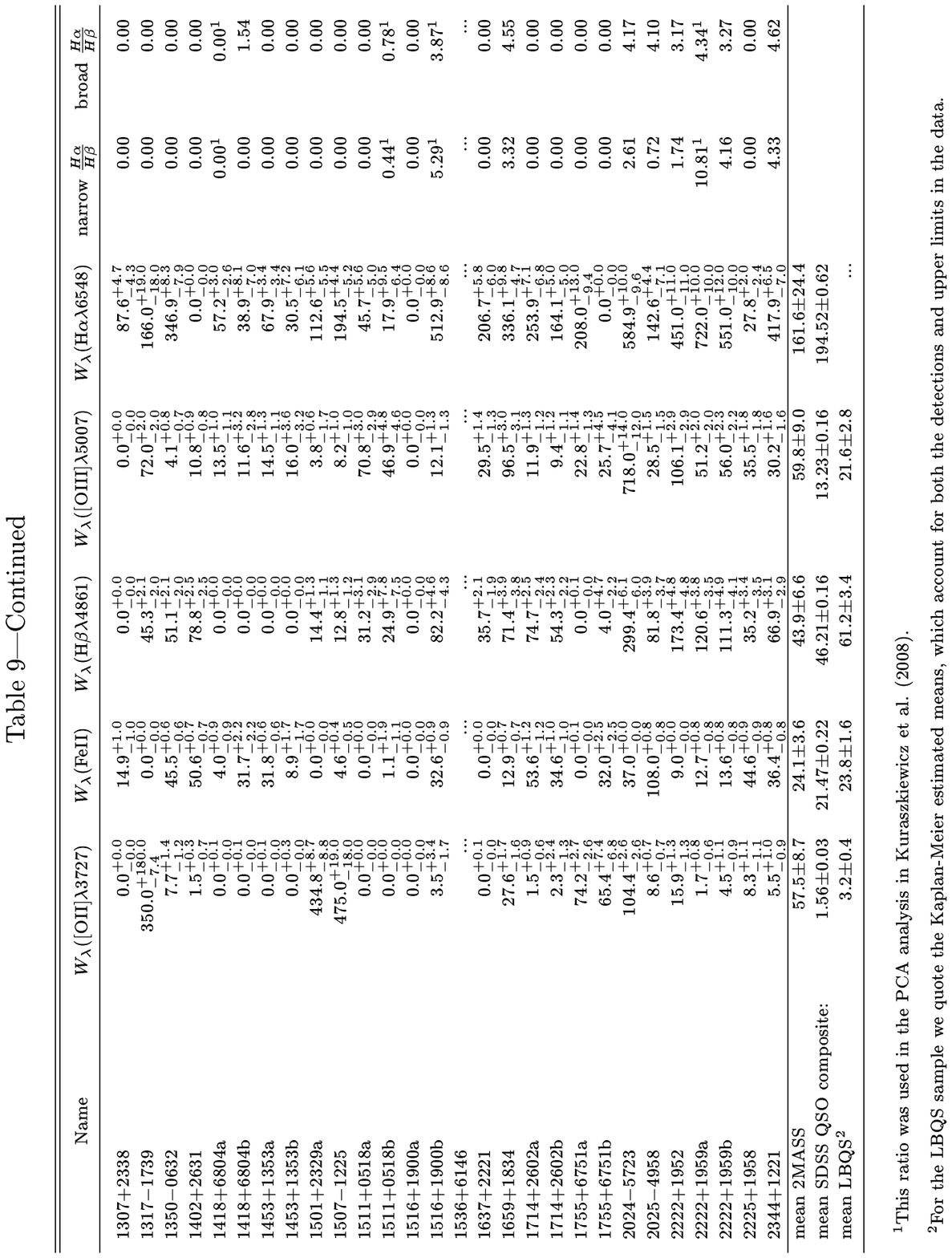}{6in}{0.}{90}{90}{-240}{-250} 
\end{figure}

\end{document}